\documentclass[useAMS,usenatbib]{mn2e}
\usepackage{times}

\usepackage{rotating}
\usepackage{subfigure}
\usepackage{epsfig}
\usepackage{psfrag}

\usepackage{multirow}

\def\up[#1]{\ensuremath{^{\mathrm{#1}}}}
\def\d{\dagger}
\def\na{\multicolumn{2}{c}{-}}
\def\uplim[#1]{\multicolumn{2}{c}{$<$#1}}

\def\gal{\psfrag{Q}{GPS}}
\def\qso{\psfrag{Q}{}}

\def\<{\ensuremath{<}}
\def\>{\ensuremath{>}}

\title[]{Observations and properties of candidate high frequency GPS radio sources in the AT20G survey}
\author[Hancock et al.]{
\parbox[t]{\textwidth}
{Paul J. Hancock$^{1,2}$,
Elaine M. Sadler$^{1}$,
Elizabeth K. Mahony$^{1,2}$,
Roberto Ricci$^{3}$
\vspace*{6pt} \\
\small $^{1}$Sydney Institute for Astronomy (SIfA), School of Physics, University of Sydney, NSW 2006, Australia \\
\small $^{2}$Australia Telescope National Facility, CSIRO, PO Box 76, Epping, NSW 1710, Australia\\
\small $^{3}$INAF-Istituto di Radioastronomia, Bologna, Via P. Gobetti, 101, 40129 Bologna, Italy\\
}}

\begin{document}
\date{}
\pagerange{\pageref{firstpage}--\pageref{lastpage}} \pubyear{}
\maketitle
\label{firstpage}

\begin{abstract}
We used the Australia Telescope Compact Array (ATCA) to obtain 40\,GHz and 95\,GHz observations of a number of sources that were selected from the Australia Telescope Compact Array 20\,GHz (AT20G) survey . The aim of the observations was to improve the spectral coverage for sources with spectral peaks near 20\,GHz or inverted (rising) radio spectra between 8.6\,GHz and 20\,GHz. We present the radio observations of a sample of 21 such sources along with optical spectra taken from the ANU Siding Spring Observatory 2.3m telescope and the ESO-New Technology Telescope (NTT). We find that as a group the sources show the same level of variability as typical GPS sources, and that of the 21 candidate GPS sources roughly 60\% appear to be genuinely young radio galaxies. Three of the 21 sources studied show evidence of being restarted radio galaxies. If these numbers are indicative of the larger population of AT20G radio sources then as many as 400 genuine GPS sources could be contained within the AT20G with up to 25\% of them being restarted radio galaxies.
\end{abstract}

%%%%%%%%%%%%%%%%%%%%%%%%%%%%%%%%%%%%%%%%%%%%%%%%%%%%%%%%%%%%%%%%%%%%%%%%%%%%

\section{Introduction}
Gigahertz Peaked Spectrum (GPS) radio sources are important to our understanding of the evolution of radio galaxies as they are thought to be the progenitors of large radio galaxies. The youth scenario described in \citet{odea_compact_1998} relates the spectral turnover frequency to the age of the radio source, with the youngest sources having the highest intrinsic turnover frequencies. 

As noted by \citet{torniainen_radio_2007} and others, it is important to distinguish between the different causes for a peaked radio spectrum. Nearly all QSO-type objects that have been observed to have a peaked spectrum have done so either due to non-simultaneous observations of variable flat spectrum sources, or to beamed components that are observed in a flaring state. Such sources show a peaked spectrum only for a short period of time and are not related to the evolution of radio galaxies, so are a major source of contamination to any sample of GPS radio sources. Monitoring of sources to exclude variability as a source of the spectral peak is essential, as is optical spectroscopy to identify QSO like sources.

To investigate the properties of the youngest radio galaxies, radio and optical observations were carried out on 21 candidate high frequency peaking GPS sources. The motivation for the radio observations is to identify variability and the high frequency radio properties, including the turnover frequency of each source. The goal of this paper is to investigate the candidate high-frequency GPS sources, and to classify them as being either genuine GPS galaxies, possible GPS galaxies and non GPS galaxies.

Genuine high-frequency GPS galaxies are characterised by: 
\begin{itemize}
\item A convex radio spectrum with a peak above 5\,GHz.
\item A compact radio morphology, with a low level of extended flux. Restarted radio sources can show jets and hot-spots from previous epochs of activity. The cores of these sources may still be considered to blue genuine GPS galaxies.
\item A typical total flux variability of $\sim$10\% or less in a year.
\item Little or no variability in spectral shape over time.
\item An optical spectrum indicative of AGN activity with broad or narrow emission lines or stellar absorption lines, but not the flat, featureless continuum shown by many blazars.
\end{itemize}

%\begin{sidewaystable}
\begin{table*}
 \centering
 %\footnotesize %smaller size to fit the table on the page
 \begin{tabular}{ccc r@{\,$\pm$\,}l r@{\,$\pm$\,}l r@{\,$\pm$\,}l r@{\,$\pm$\,}l cc r@{\,$\pm$\,}l r@{\,$\pm$\,}l r@{\,$\pm$\,}l r@{\,$\pm$\,}l}
 %\small
\hline
   (1) & \multicolumn{2}{c}{(2)}            &  \multicolumn{2}{c}{(3)}      & \multicolumn{2}{c}{(4)}      & \multicolumn{2}{c}{(5)}      &\multicolumn{2}{c}{(6)}      & (7)                 & (8)                 & \multicolumn{2}{c}{(9)}    &\multicolumn{2}{c}{ (10) }  & \multicolumn{2}{c}{(11)}      & \multicolumn{2}{c}{(12)}  \\ 
 Name  & \multicolumn{2}{c}{AT20G position} &  \multicolumn{2}{c}{$S_{95}$} & \multicolumn{2}{c}{$S_{40}$} & \multicolumn{2}{c}{$S_{20}$} &\multicolumn{2}{c}{$S_{20}$} & $V_{\mathrm{rms}}$  & $\Delta$ t      & \multicolumn{2}{c}{$S_8$}  &\multicolumn{2}{c}{ $S_5$ }& \multicolumn{2}{c}{$S_{1.4}$} & \multicolumn{2}{c}{$S_{0.8}$}  \\ 
 AT20G & \multicolumn{2}{c}{J2000}          &  \multicolumn{2}{c}{mJy}      & \multicolumn{2}{c}{mJy}      & \multicolumn{2}{c}{mJy}      &\multicolumn{2}{c}{mJy}      &   \%                & \tiny{Years}              & \multicolumn{2}{c}{ mJy }  &\multicolumn{2}{c}{   mJy }& \multicolumn{2}{c}{   mJy   } & \multicolumn{2}{c}{   mJy   }  \\
                       \multicolumn{3}{c}{ }&  \multicolumn{2}{c}{ Epoch 2} & \multicolumn{2}{c}{ Epoch 2} & \multicolumn{2}{c}{ Epoch 2} &\multicolumn{2}{c}{ Epoch 1} &                     &                     & \multicolumn{2}{c}{Epoch 1}&\multicolumn{2}{c}{Epoch 1}& \multicolumn{2}{c}{NVSS}      & \multicolumn{2}{c}{SUMSS}\\
\hline
J002616-351249 & 00:26:16.40 & $-$35:12:49.3 & 434 &    38 &  1150 & 25  &   1156 & 79 & 1123 & 43\up[c] &\<6 & 2.92 &   357 & 18 &  136 &  7 &    24.6 &   0.9 &    12.2 &   1.5 \\
J004417-375259 & 00:44:17.06 & $-$37:52:59.2 & 105 &     9 &   136 &  2  &     94 &  6 &  114 &  5\up[c] &7.7 & 2.92 &    86 &  4 &   66 &  3 &    18.4 &   0.7 &    45.6 &   1.8 \\
J004905-552110 & 00:49:05.62 & $-$55:21:10.5 & 128 &    11 &   204 &  4  &    142 &  9 &   89 &  5\up[b] &22  & 1.92 &    76 &  4 &   59 &  3 &       \na       &    11.8 &   1.0 \\
J005427-341949 & 00:54:27.92 & $-$34:19:49.2 &  69 &     6 &    90 &  1  &     82 &  5 &  110 &  4\up[c] &13  & 2.92 &    90 &  5 &   80 &  4 &    51.3 &   2.2 &    32.9 &   1.9 \\
J010333-643907 & 01:03:33.63 & $-$64:39:07.4 & 279 &    24 &   437 &  9  &    311 & 21 &  323 & 16\up[b] &\<6 & 1.92 &   235 & 12 &  187 &  9 &       \na       &   144.7 &   4.5 \\
J011102-474911 & 01:11:02.93 & $-$47:49:11.3 & \uplim[87]  &    82 &  1  &     72 &  4 &   83 &  4\up[c] &\<6 & 2.92 &    64 &  3 &   30 &  2 &       \na       &    10.5 &   0.9 \\
J012714-481332 & 01:27:14.87 & $-$48:13:32.1 & 187 &    16 &   276 &  6  &    174 & 12 &  237 & 12\up[c] &14  & 2.92 &   181 &  9 &  140 &  7 &       \na       &    80.3 &   2.6 \\
J012820-564939 & 01:28:20.38 & $-$56:49:39.7 & 120 &    10 &   222 &  4  &    131 &  9 &  189 & 10\up[b] &17  & 1.92 &   141 &  7 &  130 &  7 &       \na       &    82.6 &   2.6 \\
J180859-832526 & 18:08:59.53 & $-$83:25:26.8 & \uplim[128] &    93 &  2  &     80 &  5 &   95 &  5\up[a] &6.1 & 0.92 &    71 &  4 &   68 &  4 &       \na       &   106.8 &   3.3 \\
J181225-712006 & 18:12:25.07 & $-$71:20:06.3 & \uplim[97]  &    25 &  1  &     31 &  2 &   43 &  2\up[a] &15  & 0.92 &    39 &  2 &   26 &  2 &       \na       &     \uplim[6]   \\
J181857-550815 & 18:18:57.99 & $-$55:08:15.0 & \uplim[99]  &    71 &  1  &     55 &  3 &   75 &  4\up[a] &14  & 0.92 &    54 &  4 &   42 &  5 &       \na       &   124.4 &   4.0 \\
J203540-694407 & 20:35:40.43 & $-$69:44:07.9 & \uplim[104] &   106 &  2  &    157 & 10 &  160 &  8\up[b] &\<6 & 1.92 &   236 & 12 &  188 &  9 &       \na       &    16.4 &   1.0 \\
J203958-720655 & 20:39:58.06 & $-$72:06:55.0 & 151 &    13 &   266 &  5  &    332 & 22 &  417 & 21\up[b] &9.7 & 1.92 &   244 & 12 &  186 &  9 &       \na       &    99.1 &   3.1 \\
J205503-635207 & 20:55:03.83 & $-$63:52:07.0 & \uplim[91]  &    28 &  1  &     43 &  2 &   44 &  2\up[a] &\<6 & 0.92 &    29 &  2 &   12 &  1 &       \na       &     \uplim[6]   \\
J212222-560014 & 21:22:22.81 & $-$56:00:14.6 & \uplim[106] &    64 &  1  &     56 &  3 &   58 &  3\up[b] &\<6 & 1.92 &    34 &  2 &   28 &  2 &       \na       &     \uplim[6]   \\
J212402-602808 & 21:24:02.96 & $-$60:28:08.9 & \uplim[98]  &    60 &  1  &     89 &  6 &  135 &  7\up[a] &20  & 0.92 &  (125 &  6)\up[e]&  (81 &  4)\up[e]&       \na       &    33.0 &   1.3 \\
J213622-633551 & 21:36:22.08 & $-$63:35:51.2 & 224 &    19 &   354 &  7  &    372 & 25 &  411 & 21\up[c] &\<6 & 2.92 &  (281 & 14)\up[d]& (198 & 10)\up[d]&       \na       &   154.5 &   4.7 \\
J214447-694654 & 21:44:47.42 & $-$69:46:54.9 & 104 &     9 &   143 &  3  &    124 &  8 &  119 &  6\up[b] &\<6 & 1.92 &    94 &  5 &   81 &  4 &       \na       &    23.5 &   1.7 \\
J230737-354828 & 23:07:37.27 & $-$35:48:28.7 &  98 &     8 &   110 &  2  &     88 &  6 &  190 &  9\up[c] &36  & 2.92 &   186 &  9 &  112 &  7 &    58.7 &   2.2 &    93.2 &   3.1 \\
J233159-381147 & 23:31:59.43 & $-$38:11:47.3 & 132 &    11 &   316 &  6  &    443 & 30 &  523 & 25\up[c] &\<6 & 2.92 &   747 & 37 &  592 & 30 &   543.2 &  16.3 &   429.7 &  13.0 \\
J234743-494627 & 23:47:43.68 & $-$49:46:27.8 & 275 &    24 &   411 &  9  &    403 & 27 &  284 & 12\up[c] &16  & 2.92 &   251 & 13 &  211 & 11 &       \na       &   107.7 &   3.3 \\
\hline
 \end{tabular}
 \caption{Candidate high-frequency GPS sources. Data in columns 1, 2, 6, 9, and 10 are taken from the AT20G catalog. Columns 11 and 12 are from the NVSS and SUMSS surveys. Columns 3, 4, and 5 are new observations from this paper. The 20\,GHz variability index in column 7 is defined by $V_\mathrm{rms}$ in eq \ref{eqn:variability}. Column 8 shows the time in years between the Epoch 1 and Epoch 2 20\,GHz observations. The year of observation for Epoch 1 is denoted by: a - 2004.75, b - 2005.75, c - 2006.75, d - 2006.83, e - 2007.83. Sources with an upper limit in column 12 are present within the SUMSS survey maps, but are below the 6\,mJy detection limit of the catalog.}
 \label{tab:src_selection}
 \end{table*}
%\end{sidewaystable}

\section{Target Selection}
Sources were selected from the Australia Telescope 20\,GHz \citep[AT20G,][]{murphy_australia_2010} survey's list of confirmed sources available as of August 2007, and were restricted to 18h \< RA \< 01h30m to obtain the best visibility during the allotted ATCA observing time. Sources that showed a rising ($\alpha_{4.8}^{20} >+0.2$) spectrum that either remained inverted, or appeared to be peaking above 10\,GHz were identified as potential high frequency GPS sources. In total there were 148 sources that met this criteria, however not all were able to be observed due to timing constraints as discussed in section \ref{sec:atca_obs}. Previous observations by \citet{sadler_extragalactic_2008} have shown that fewer than 2\% of all sources with rising a spectrum ($\alpha_{8.6}^{20}>0$) continued to rise at 95\,GHz, and thus we expect to see a turnover in nearly all the sources selected.

\section{Optical identification}
Optical identification was performed via cross match with the SuperCOSMOS Sky Survey on-line database\footnote{http://surveys.roe.ac.uk/ssa/xmatch.html} \citep{hambly_supercosmos_2001}. The nearest object within a radius of 2.5\,arcsec was taken to be the optical counterpart. \citet{sadler_properties_2006}, using Monte Carlo simulations, find that fewer than 3\% of correlations made in this manner will be the result of chance. All but two of the sources were assigned an optical ID in this way. The source AT20G\,J203540-694407 has an optical ID that is offset by more than the 2.5\,arcsec cutoff, as an inspection of optical radio overlays revealed that there were no other nearby optical candidates, and the radio centroid was within the extent of the optical source. Table \ref{tab:optical_id} shows the SuperCOSMOS optical counterparts for the sample sources. The B, R, and I magnitudes are taken from the survey catalog, and sources with ``\dots'' have a corresponding magnitude that is fainter than the survey limit ($\sim23$, $\sim22$ and $\sim20$ mag respectively). The classification of objects as either stellar (Q) or galaxy like (G) is based solely on the optical morphology. The source J012714-481332 has no optical counterpart within 2.5 arcsec.

\begin{table*}
 \begin{tabular}{cccccccc}
\hline
Name  & \multicolumn{2}{c}{Optical Position} & Separation &\multicolumn{3}{c}{--- Magnitude ---} & Class \\
AT20G &             RA   &               DEC &  arcsec    &  B$_J$ & R$_F$ & I$_C$               &   ID   \\
\hline
J002616-351249 & 00:26:16.37 & -35:12:48.72  & 0.7      & 22.0 & 19.8 & 19.4 & G\\		  
J004417-375259 & 00:44:17.03 & -37:52:58.98  & 0.4      & 20.3 & 19.1 & 18.9 & G\\		  
J004905-552110 & 00:49:05.48 & -55:21:10.76  & 1.2      & 16.6 & 15.7 & 15.1 & G\\		  
J005427-341949 & 00:54:27.91 & -34:19:49.33  & 0.3      & 18.0 & 17.0 & 16.6 & G\\		  
J010333-643907 & 01:03:33.68 & -64:39:07.53  & 0.4      & 18.6 & 17.5 & 16.9 & G\\		  
J011102-474911 & 01:11:02.90 & -47:49:10.87  & 0.5      & 18.6 & 17.5 & 17.0 & G\\		  
J012714-481332 & \multicolumn{7}{c}{\dots}\\	       			  						  
J012820-564939 & 01:28:20.51 & -56:49:39.46  & 1.2      & 16.5 & 15.7 & 15.0 & G\\		  
J180859-832526 & 18:08:59.75 & -83:25:26.29  & 0.6      & 21.6 & \dots & \dots & G\\		  
J181225-712006 & 18:12:24.94 & -71:20:07.40  & 1.0      & 18.4 & 16.8 & 16.4 & G\\		  
J181857-550815 & 18:18:58.01 & -55:08:14.94  & 0.6      & 15.9 & 15.0 & 14.3 & G\\		  
J203540-694407 & 20:35:40.86 & -69:44:07.24  & 2.9$^\d$ & 17.2 & 16.3 & 16.1 & Q\\	  
J203958-720655 & 20:39:58.07 & -72:06:56.15  & 1.1      & 17.1 & 16.6 & 16.3 & Q\\		  
J205503-635207 & 20:55:03.78 & -63:52:06.46  & 0.7      & 20.2 & 19.0 & 18.3 & Q\\		  
J212222-560014 & 21:22:22.88 & -56:00:14.27  & 0.8      & 16.1 & 15.0 & 14.4 & G\\		  
J212402-602808 & 21:24:02.97 & -60:28:08.85  & 0.7      & 17.3 & 16.9 & 16.5 & Q\\		  
J213622-633551 & 21:36:22.03 & -63:35:50.87  & 0.6      & 19.8 & 19.0 & \dots & Q\\		  
J214447-694654 & 21:44:47.51 & -69:46:54.85  & 0.1      & 21.4 & 20.0 & \dots & Q\\		  
J230737-354828 & 23:07:37.24 & -35:48:28.24  & 0.7      & 20.9 & 19.8 & 20.0 & G\\		  
J233159-381147 & 23:31:59.47 & -38:11:47.61  & 0.7      & 17.3 & 16.8 & 16.4 & Q\\		  
J234743-494627 & 23:47:43.66 & -49:46:27.81  & 0.2      & 19.8 & 18.2 & 18.9 & Q\\             
\hline
 \end{tabular}
 \caption[Optical identification of the sources from Table \ref{tab:src_selection}]{Optical identification of the sources from table \ref{tab:src_selection}. Optical positions, magnitudes and classes are as reported in the SuperCOSMOS sky survey catalog. Magnitudes listed as ``\dots'' are fainter than the detection limit of the survey. Class ID is G - Galaxy, Q - Star-like. Notes: $\dagger$ - This identification was accepted despite being more than 2.5\,arcsec from the radio centroid and is discussed in section \ref{sec:J203540-694407}.}
 \label{tab:optical_id}
\end{table*}

\section{Observations}
\subsection{ATCA 20, 40, and 95\,GHz}
\label{sec:atca_obs}
Sources from the GPS candidate sample were observed in October 2007 using the ATCA in its hybrid H75 configuration. The H75 configuration has a maximum baseline of 82m, and has been shown to provide robust phases at 95\,GHz \citep{sadler_extragalactic_2008}. The observations were carried out with a bandwidth of 256\,MHz ( 2$\times$128\,MHz -- the maximum available at the time), and the resulting synthesized beam sizes were 31, 15, and 7\,arcsec FWHM at 20, 40, and 95\,GHz respectively. The observing dates are shown in Table \ref{tab:obs_dates}. The 95\,GHz observations were conducted using the 3mm system that had been recently installed on the ATCA \citep{moorey_cryogenically_2008}, and were done on the same day as the 20\,GHz observations, which were included to detect variability. The 40\,GHz observations were carried out on a separate day to avoid losing time moving the feed translator needed to bring the 7mm feed into the telescope focus. As there is no 3mm receiver on the 6km antenna, the observations were made using five of the compact array antennas in the H75 configuration, with a goal of 3 separate observations at different hour angles to maximize the available $(u,v)$ coverage for each source.

The raw visibility data were reduced using the astronomical software package {\sc miriad} \citep{sault_retrospective_1995}. Fluxes were extracted from the calibrated data using the {\sc miriad} task {\sc calred}, which computes the triple product amplitude \citep[Ricci, Sault \& Ekers (in preparation);][]{cornwell_imaging_1995,murphy_australia_2010}. Compared to measuring the flux densities from images, as is typical, this method of measuring flux densities is robust to the effects of phase decorrelation. The band-passes were calibrated using the strong ATCA high-frequency calibrator PKS\,1921-293. The new observations are included in Table \ref{tab:src_selection} along with previous AT20G observations and data from the Sydney University Molonglo Sky Survey \citep[SUMSS,][]{mauch_sumss:wide-field_2003} and NRAO-VLA Sky Survey \citep[NVSS,][]{condon_nrao_1998} catalogs when available.

\subsection{SSO 2.3m spectra}
Spectroscopic observations for three high-frequency peaking (HFP) sources were taken during the nights of 1-5 May 2008 with the ANU 2.3m telescope at Siding Springs, using the 158R grating in the blue arm of the spectrometer, without a dichroic. This hybrid setup allows for a large spectral window of 3500-11000\AA, with a resolution of $\sim 15$\AA. Fringing toward the blue end of the spectrum and a reduced sensitivity in the red end reduce the effective spectral window to 4000-8000\AA\,in most cases. Three sources of interest to this paper were observed during this time. The data were reduced using the Interactive Data Reduction and Analysis Facility \citep[IRAF,][]{valdes_interactive_1984}, and spectral features were identified with the aid of the program {\sc runz} \citep{saunders_improvements_2004}, which was developed for the 6 degree field Galaxy Redshift Survey  \citep[6dFGS,][]{jones_6df_2004,jones_6df_2009}. The resulting spectra and redshifts are shown in the Appendix.

\subsection{ESO NTT imaging and spectroscopy}
A subset of sources from table \ref{tab:src_selection} were included in a backup observing schedule as part of a program by Dick Hunstead and Helen Johnston, on the 3.58m ESO New Technology Telescope (NTT). Imaging and spectroscopy of the target sources was carried out on the nights of 24 - 29 of August 2008. For each source a short exposure acquisition image was observed to align the slit of the spectrograph, and was followed by longer exposure of 1000sec for spectroscopy. The spectra were reduced using IRAF and redshifts were identified using {\sc runz}. A byproduct of the observing program is the set of acquisition images for each of the sources. In each case the radio ID was a single, isolated galaxy, with the exception of AT20G\,J181225-712006 which is discussed further in section \ref{sec:J181225-712006}.

\begin{table}
\caption[The observation dates of the data for this paper]{The observation dates of the data for this paper.}
\label{tab:obs_dates}
\centering
\begin{tabular}{ccc}
\hline
Telescope & $\nu$/$\lambda$ & Observation Date\\
\hline
ATCA     & 20\,GHz       & 18-19 October 2007\\
ATCA     & 40\,GHz       & 18-19 October 2007\\
ATCA     & 95\,GHz       & 20-21 October 2007\\
SSO 2.3m & 4000-8000\AA  & 1-5 May 2008\\
ESO NTT  & 4000-9000\AA  & 24-29 August 2008\\
\hline
\end{tabular}
\end{table}

\begin{table*}\centering
 \begin{tabular}{cccccc}
  \hline
 Name  & Name & Redshift & Ref. & Class &  Spectrum\\
 AT20G & Alt. &          &      &  ID   &   Figure / Type \\
  \hline
J002616-351249 & PMN J0026-3512           & $0.6_{-0.4}^{+0.9}$ & est  & G &  \\
J004417-375259 & PMN J0044-3752           & 0.483               & NTT  & G &\ref{fig:J004417-375259} em I \\
J004905-552110 & SUMSS J004905-552111     & 0.0626              & NTT  & G &\ref{fig:J004905-552110}  em+abs\\
J005427-341949 & SUMSS J005428-341949     & 0.11                & 2dF  & G &\ref{fig:J005427-341949_2df}  em+abs\\
J010333-643907 & PMN J0103-6438           & 0.163               & NTT  & G &\ref{fig:J010333-643907_spectrum}  em I\\
J011102-474911 & SUMSS J011101-474931     & 0.154               & 2dF  & G &\ref{fig:J011102-474911_2df}  em+abs II\\
J012714-481332 & PMN J0127-4813           &\multicolumn{4}{c}{\dots}          \\
J012820-564939 & PMN J0128-5649           & 0.066               & D03  & G &  \\
J180859-832526 & SUMSS J180857-832526     & $0.5_{-0.3}^{+0.8}$ & est  & G &  \\
J181225-712006 & 2MASX J18122510-7120065  & 0.199               & NTT  & G &\ref{fig:J181225-712006_spectra}  abs \\
J181857-550815 & PMN J1818-5508           & 0.073               & 6dF  & G &  abs\\
J203540-694407 & PMN J2035-6944           & 0.875               & SSO  & Q &\ref{fig:specJ203540-694407}  em I\\
J203958-720655 & PMN J2039-7207           & 1.00                & SSO  & Q &\ref{fig:specJ203958-720655}  em I\\
J205503-635207 & SUMSS J205516-635325     & [1]                 & est  & Q &  \\
J212222-560014 & 2MASX J21222292-5600143  & 0.0518              & 6dF  & G &  abs+em\\
J212402-602808 & SUMSS J212402-602808     & [1]                 & est  & Q &  \\
J213622-633551 & PMN J2136-6335           & [1]                 & est  & Q &  \\
J214447-694654 & SUMSS J214447-694652     & [1]                 & est  & Q &  \\
J230737-354828 & NVSS J230737-354828      & $0.4_{-0.2}^{+0.6}$ & est  & G &  \\
J233159-381147 & PMN J2332-3811           & 1.20                & J78  & Q &  \\
J234743-494627 & PMN J2347-4946           & 0.643               & H08  & Q &  em I\\
  \hline
 \end{tabular}
 \caption[List of redshifts for the sources in table \ref{tab:src_selection}]{List of redshifts for the sources in table \ref{tab:src_selection}. Redshifts come from: 2dF - \citet{colless_2df_2001}; D03 - \citet{drake_radio-excess_2003}; 6dF - \citet{jones_6df_2004}; J78 - \citet{jauncey_redshifts_1978}; H08- \citet{healey_cgrabs:all-sky_2008}; NTT - Observations by RWH/HJ; SSO - Observations by PJH/EM; est - approximate redshift estimated from B magnitude using equation \ref{eq:bhzb} in section \ref{sec:redshift_estimation} for galaxies, and as z$\sim$1 for quasars. The final column lists the figure number for spectra that are shown in this paper as well as the type of spectrum:em - emission lines present, abs - absorption lines present, I - AGN Type I spectrum, II - AGN Type II spectrum.}
 \label{tab:redshifts}
\end{table*}

\section{Results}

\subsection{Redshift Estimation}
\label{sec:redshift_estimation}
\begin{figure}
 \centering
 \epsfig{file=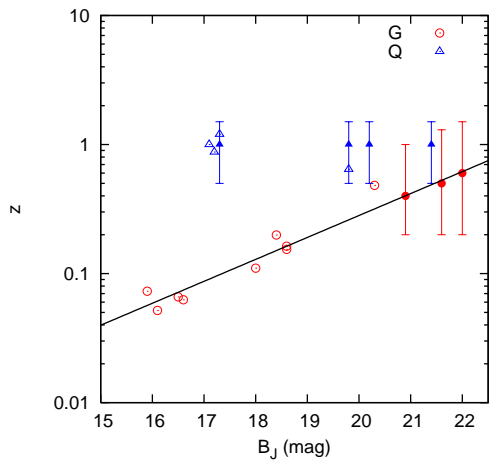,bb=210 55 555 440, height=0.8\linewidth, angle=-90,clip=}
 \caption{Redshift as a function of B$_J$. The red circles are sources identified as galaxies (G) by SuperCOSMOS, and the blue triangles are those identified as stellar (Q). The solid line represents equation \ref{eq:bhzb}. The galaxies are in good agreement with the model fit. The solid markers with error bars are the sources that have had their redshifts estimated.}
 \label{fig:bz_comp}
\end{figure}

For sources that remained without a redshift after observations and searching of the literature, redshifts were estimated. Following the work of \citet{burgess_molonglo_2006}, the redshifts of the sources were estimated based on their B magnitudes using:
\begin{equation}
\log_{10} z = (-3.95\pm 0.18)+(0.17\pm 0.01)B_J
\label{eq:bhzb}
\end{equation}
\citet{burgess_molonglo_2006} also used R magnitudes to estimate the redshift of objects within their sample, however comparison with known redshifts from this sample showed that the B magnitude based estimate was more accurate. Figure \ref{fig:bz_comp} shows that the model of \citet{burgess_molonglo_2006} is a good estimator for the redshift for the sources identified as galaxies by SuperCOSMOS. The estimated redshifts are included in table \ref{tab:redshifts}. The median redshift for the galaxies, including those with estimated redshifts is z=0.18, whilst the median spectroscopic redshift of the quasars is z=1.0. The sample of sources used to determine equation \ref{eq:bhzb} are exclusively galaxies, and thus the stronger blue emission of quasars will result in their redshifts being underestimated. For this reason quasars of unknown redshift were assigned a value of z=1. Only quasars with a measured spectroscopic redshift are used in the analysis of this paper.

\subsection{Radio Variability at 20GHz}
The AT20G catalog contains 4.8, 8.6, and 20\,GHz flux densities whilst the new observations span 20, 40, and 95\,GHz. The overlap at 20\,GHz allows for a measurement of variability. Following \citet{barvainis_radio_2005}, who use a de-biased variability index which takes into account the uncertainties in measurement, the fractional variability $V_{\mathrm{rms}}$ is taken to be
\begin{equation}
V_{\mathrm{rms}} = \frac{100}{\langle S\rangle}\sqrt{\frac{\Sigma[S_i-\langle S\rangle]^2 - \Sigma\sigma_i^2}{N}}
\label{eqn:variability}
\end{equation}
where $S_i$ are the individual flux density measurements for the same sources, and $\sigma_i$ is the error on each measurement, N is the number of measurements, and $\langle S\rangle$ is the mean flux density. When the amount of variability is less than can be measured with the given uncertainties, the radicand becomes negative, and the variability index is then considered to also be negative. \citet{sadler_properties_2006} analyzed the distribution of positive and negative variability indices calculated in this way and found that the minimum amount of variability that could be detected within their data was 6\%, and listed all sources with a variability index below this limit to be \<6\%. This approach was adopted for this data, and 6\% is assumed to be the minimum detectable variability. 

\begin{figure}
 \centering
 \subfigure[]{\epsfig{file=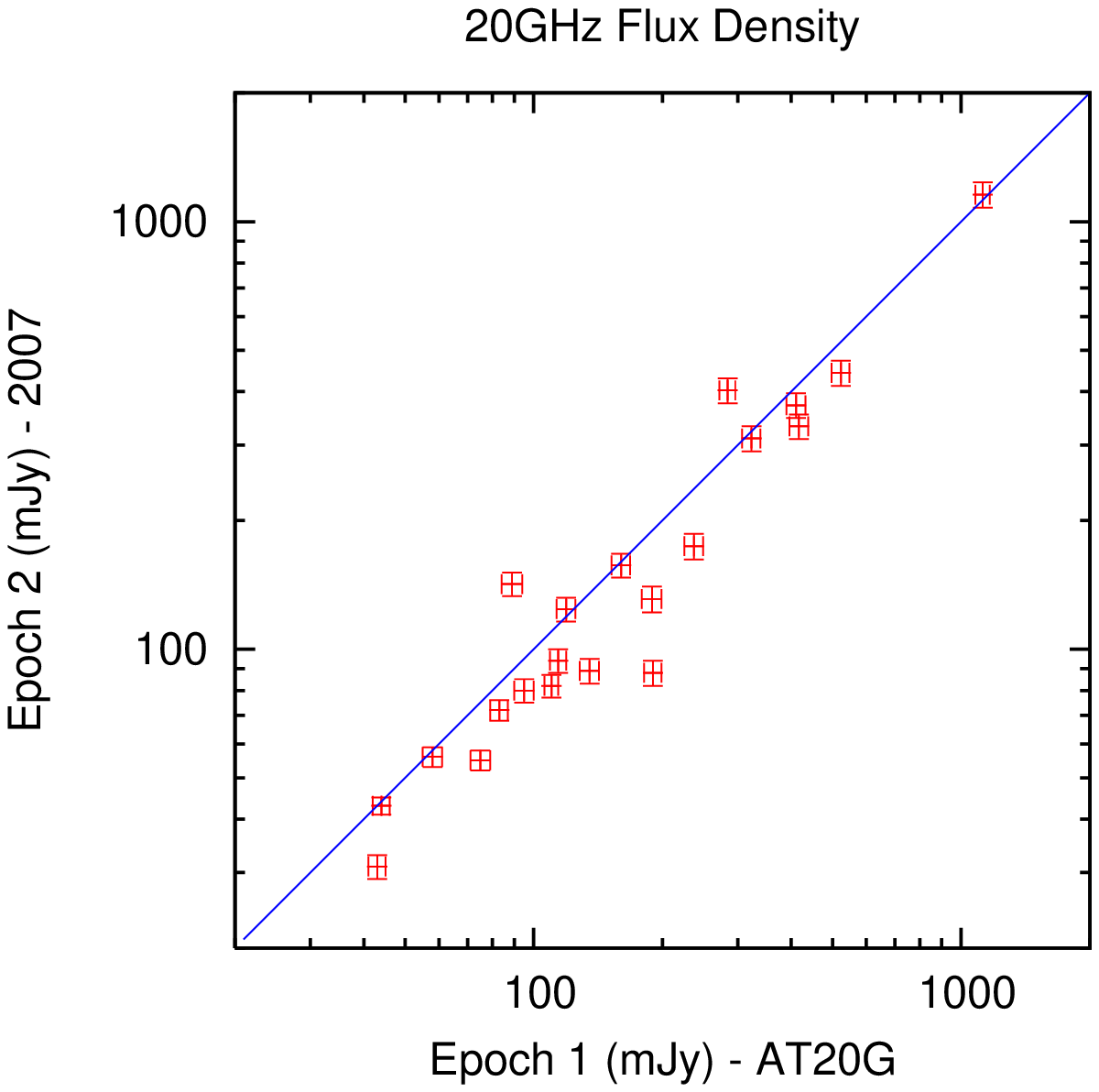,bb= 210 55 555 405, height=0.8\linewidth, angle=-90}}
 \subfigure[]{\epsfig{file=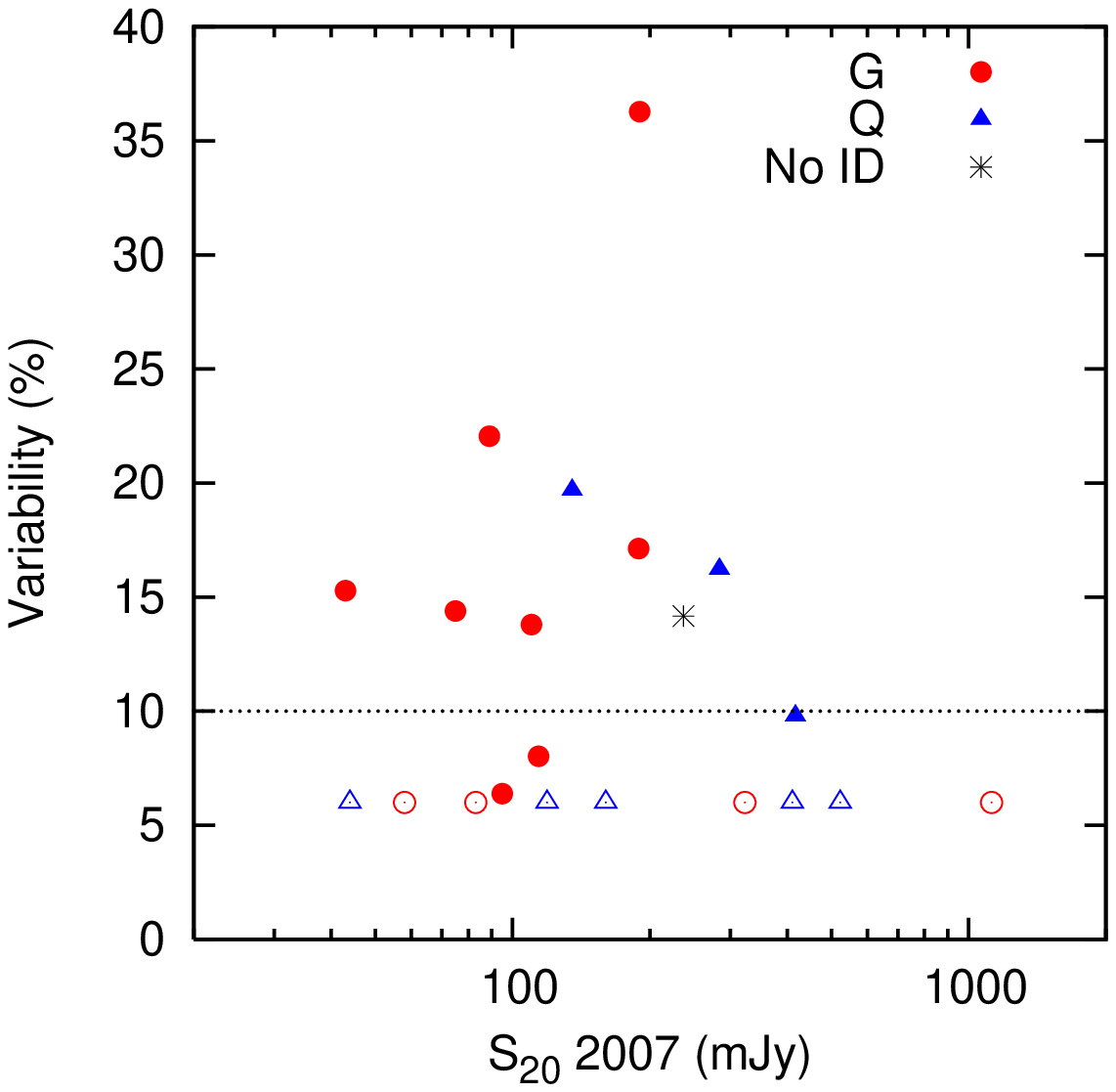,bb= 210 55 555 405, height=0.8\linewidth, angle=-90}}
 \caption[Variability of the candidate GPS sources]{Variability of the candidate GPS sources. (a) The 20\,GHz flux density at each of the two epochs of observation. (b) The percentage variability between the different epochs of observation as a function of the 2007 (Epoch 2) flux density. Galaxies (G) are in filled red circles, star like objects (Q) are in filled blue triangles, the black star is the source J012714-481332 which has no optical ID. Sources that varied by less than 6\% are shown as upper limits (empty circles and triangles). The horizontal line at 10\% is a typical level of variability seen in genuine GPS sources.}
 \label{fig:variability}
\end{figure}

The 20\,GHz variability (and the timescale over which it has been measured) is listed in columns 7 and 8 of table \ref{tab:src_selection} for the candidate GPS sources. The Epoch 2 20\,GHz flux density is that measured along with the 40 and 95\,GHz fluxes in 2007, whilst the Epoch 1 flux density is that measured during the follow up observations of the AT20G. GPS sources are generally the least variable class of radio sources selected at centimeter wavelengths, with a typical variation of $\sim10$\% over a year \citep{odea_compact_1998}. Sources that are not related to the evolutionary model described by the youth scenario may appear to have a peaked spectrum during phases of outburst and are thus contaminants to samples of GPS and HFP sources. The level of variability is shown in Figure \ref{fig:variability}(b). For eleven of the twenty sources in the sample the variability is less than 10\% on timescales of 1--3 years. The remaining nine sources have higher variability which is an indication that they may not be genuine GPS sources. It is interesting that of the sources that vary more than 10\%,  6/9 have optical IDs that are galaxies. Figure \ref{fig:variability}(a) shows a direct comparison between the 20\,GHz fluxes seen at different epochs. All but two of the sources are observed at either the same or a lower flux level during the second epoch of observations. This behaviour is typical for variable sources in a flux limited sample, with the higher activity phase being more easily detectable than the low activity phase, resulting in more sources being detected in a state of high activity, and subsequent observations being at lower activity. Another consideration is that in general, well monitored AGN show only small flux variations on 1-2 year timescales, with larger outbursts only occurring roughly once every 6 years \citep{hovatta_statistical_2007,tornikoski_long-term_2009}.

\subsection{Radio Spectra}
Following the work of \citet{snellen_new_1998} the radio spectra were fitted using:
\begin{equation}
F(\nu) = \frac{S_m}{1-e^{-1}} \left( \frac{\nu}{\nu_m}\right)^k \left(1-e^{-\left(\frac{\nu}{\nu_m}\right)^{(l-k)}} \right)
\end{equation}
where $k,\,l$ are the optically thick and thin spectral indices, and $S_m,\,\nu_m$ are the flux density and frequency at which the spectral peak occurs. The SUMSS and NVSS fluxes were not used in the fitting of the spectrum. Figures A\ref{fig:radio_sed:a}-\subref{fig:radio_sed:c} show the radio spectra for each of the 21 sources observed. For sources with low (\<10\%) variability the SED fit is quite robust, however for sources with larger variability the fit is less well defined. In the cases where the radio spectra remain inverted at 95\,GHz or are of no particular shape, a linear fit was used in place of the above equation. The spectra of individual sources are discussed in the Appendix.

The median value of the turnover frequency is $84.5\pm 8.9$\,GHz for SuperCOSMOS galaxies, and only $20.5\pm 9.4$\,GHz for the quasars (SuperCOSMOS stellar objects). As there are not enough accurate redshift measurements to convert all the spectra into the rest-frame, the data in table \ref{tab:turnover} are still in the observed frame. The large difference in median turnover frequency between the galaxy and quasar population may be partly due to the difference in redshift.

\begin{table*}
 \centering
 \begin{tabular}{ccc|ccc}
  \hline
   Name & $\nu_m$ & $\alpha_{\mathrm{low}}$ & Name & $\nu_m$ & $\alpha_{\mathrm{low}}$ \\
  AT20G &   GHz   &   &  AT20G &   GHz\\
  \hline
\multicolumn{3}{c|}{galaxy ID}    &\multicolumn{3}{c}{stellar ID}\\
J002616-351249  &  29 & +1.47 & J203540-694407  &   5 & +2.81 \\       
J004417-375259  &  93 & +0.32 &	J203958-720655  &  26 & +0.68 \\       
J004905-552110  &  79 & +0.66 &	J205503-635207  &  15 & +1.54 \\       
J005427-341949  &  53 & +0.21 &	J212402-602808  &  14 & +0.83 \\       
J010333-643907  &  84 & +0.40 &	J213622-633551  &  30 & +0.61 \\       
J011102-474911  &   6 & +3.44 &	J214447-694654  &  82 & +0.29 \\       
J012820-564939  &  85 & +0.26 &	J233159-381147  &  7  & +1.63 \\       
J180859-832526  &$\d$ & +0.16 & J234743-494627  &  73 & +0.35 \\
J181225-712006  &  13 & +0.94 &\\
J181857-550815  &$\d$ & +0.23 &\multicolumn{3}{c}{no ID}\\
J212222-560014  &$\d$ & +0.39 & J012714-481332  &  89 & +0.30 \\
J230737-354828  &$\d$ & -0.10 &\\
 \hline
 \end{tabular}
 \caption[The fitted observed-frame turnover frequency, $\nu_m$, and optically thick spectral index,$\alpha_{\mathrm{low}}$]{The fitted observed-frame turnover frequency, $\nu_m$, and optically thick spectral index, $\alpha_{\mathrm{low}}$. Sources in the left hand column are identified as galaxies within SuperCOSMOS, whilst those in the right hand column are identified as stellar. J012714-481332 has no optical ID.  Note: Sources marked with $\dagger$ remain inverted at 95\,GHz and so peak above 95\,GHz.}
 \label{tab:turnover}
\end{table*}

\subsection{Optical Colours}
The spectral energy distribution of galaxies is sufficiently regular that it is possible to estimate their stellar populations from the optical colours \citep{eisenstein_spectroscopic_2001,burgess_molonglo_2006}. Using the data of \citet{fukugita_galaxy_1995} the SuperCOSMOS B$_J$ and R$_F$ magnitudes were converted to Johnson-Cousins B and R$_C$ magnitudes. The conversion was done using the B$_J$-B and R$_F$-R$_C$ colours of -0.16 and -0.11 for elliptical galaxies, and resulted in a conversion scheme of: B-R$_C$ = B$_J$-R$_F$ + 0.05\,mag, and R$_C$-I$_C$= R$_F$-I$_C$ +0.11\,mag. The colours were then compared to the models of \citet{fukugita_galaxy_1995} for a elliptical and S0 galaxies at different redshifts. \citet{eisenstein_spectroscopic_2001} use a similar method to show that the redshift loci for different SEDs are degenerate, with different galaxy types following the same locus as a function of redshift but with differing initial points.

\begin{figure}
 \centering
 \epsfig{file=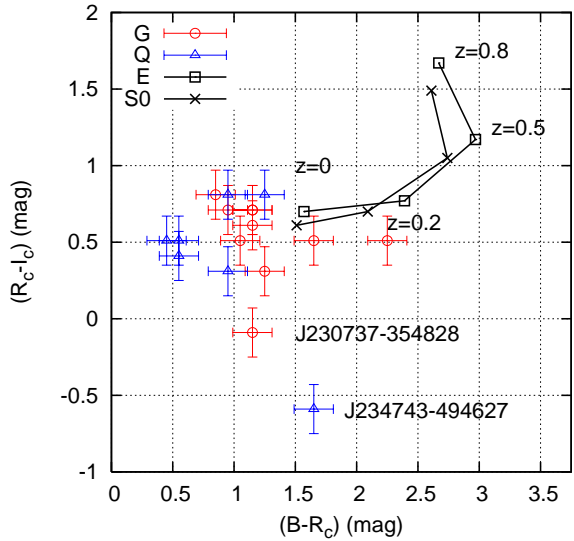,bb= 210 55 555 420, height=0.9\linewidth, angle=-90, clip=}
 \caption[Colour B-R$_c$ vs R$_c$-I$_c$]{Colour B-R$_c$ vs R$_c$-I$_c$ for each of the sources in the sample that were detected in all three bands. Galaxies (G) and star-like (Q) objects are shown as red circles and blue triangles respectively. Error bars are $\sigma_{B-R,R-I} = 0.16$\,mag which is typical for objects with B$_J$=20. The black boxes and crosses represent the model of \citet{fukugita_galaxy_1995} for elliptical and S0 galaxies at various redshifts. Note that the B, R, and I magnitudes were not measured simultaneously, so may be affected by variability.}
 \label{fig:cc_plot}
\end{figure}

Figure \ref{fig:cc_plot} shows the B-R$_c$ and R$_c$-I$_c$ colours of sources from the candidate GPS sample with the models of an elliptical and S0 galaxy from \citet{fukugita_galaxy_1995} shown. The three colour magnitudes within the SuperCOSMOS database are taken from observations that are typically separated by at least a decade. The galaxies that were used by \citet{fukugita_galaxy_1995} to construct the elliptical and S0 models are not active galaxies. The separation between the galaxies in this sample and the E/S0 models is in the negative B-R$_c$ direction, indicating that the candidate GPS sources are bluer than the models. The increased amount of blue emission can be attributed to either a non-stellar continuum from the AGN or a young stellar population as a result of recent star formation. The R$_c$-I$_c$ colour is less sensitive to AGN activity as the red part of the spectrum is still dominated by stellar emission. There is an agreement between the colour of the candidate GPS galaxies and the low redshift end of the E/S0 models which is consistent with the measured and estimated redshifts being less than $\sim$0.5. Some of the sources identified as star like in the SuperCOSMOS database are grouped with galaxies around the z=0 end of the galaxy evolution model. The object identified as a galaxy with B-R$_c$=1.15 and R$_c$-I$_c$=-0.09 is AT20G J230737-354828, and is also the most variable source at 20\,GHz. The colours of this object are likely affected by variability, and this source is probably a quasar.

\subsection{Radio and Optical luminosities}
The 20\,GHz radio luminosities of the candidate GPS sources are shown in table \ref{tab:radiolum}. The luminosity distances were calculated using the cosmology calculator of \citet{wright_cosmology_2006}, using H$_0$ = 71 km/s/Mpc, $\Omega_{M}=0.27$, and $\Omega_{vac}=0.73$. The absolute B magnitudes of the objects were calculated without making any K-corrections.

High-frequency GPS sources are very compact ( $<\sim$ 10-50\,pc) at frequencies of 20\,GHz and higher. It is therefore likely that the optical and compact radio emission from the AGN is synchrotron radiation that is being emitted from the same volume and with the same amount of Doppler boosting. This is on contrast to the steep spectrum lobes/jets that are seen in more classical radio galaxies and steep spectrum quasars, where the emission is dominated by lobes that are not sensitive to the core luminosity. If the radio and optical emission is indeed synchrotron then the radio and optical luminosities would be expected to be correlated at frequencies above the self absorbed peak. Stellar emission from the host galaxy would be expected to dominate the optical luminosity up to the point at which the AGN becomes radio loud, and then the optical luminosity would increase with radio power. 

\begin{figure}
 \centering
 \epsfig{file=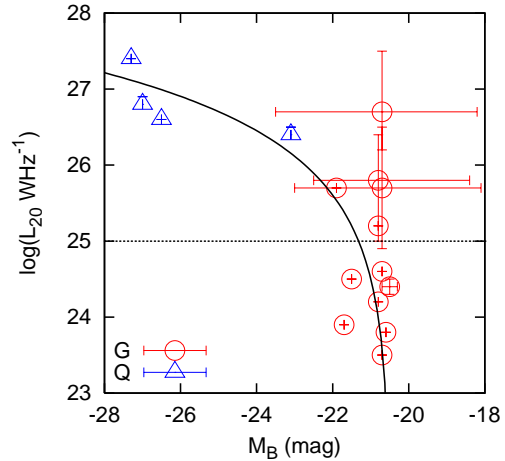, bb= 260 50 555 375, height=0.8\linewidth, angle=-90, clip=}
 \caption{The 20\,GHz radio luminosity as a function of absolute B magnitude for the sample. Galaxies are shown as red circles and star like objects as blue triangles. The horizontal line is the radio power which separates FR-I and FR-II galaxies. The solid curve is the model of equation \ref{eq:model}.}
 \label{fig:radioloud}
\end{figure}

Figure \ref{fig:radioloud} shows the radio and optical luminosities of the sample GPS sources, as well as a model which describes the above scenario. The model illustrated is:
\begin{equation}
\log(\mathrm{L}_{20}) = 25.2 +\log(-\mathrm{M}_B-20.5)
\label{eq:model}
\end{equation}
where the parameters have been chosen to fit the data. A similar correlation was noted previously by \citet{sadler_extragalactic_2008} at 95\,GHz, for a sample of inverted spectrum ($\alpha_8^{20}>0$) sources, nearly all of which fit the selection criteria of this paper.

\begin{table}
 \centering
 \begin{tabular}{clllc}
\hline
Name & $\log(L_{20G})$  & \multicolumn{1}{c}{M$_B$} & \multicolumn{1}{c}{z} & Class \\%&Name & $\log(L_{20G})$  & M & z \\
     & WHz$^{-1}$       &                       &                       &   ID  \\%&     & WHz$^{-1}$       &   &   \\
\hline
\multicolumn{5}{c}{Spectroscopic z}\\
J004417-375259  & $25.7$                & $-21.9$                 & 0.483                & G\\ 
J004905-552110  & $23.8$                & $-20.6$                 & 0.0626               & G\\ 
J005427-341949  & $24.4$                & $-20.5$                 & 0.11                 & G\\ 
J010333-643907  & $25.2$                & $-20.8$                 & 0.163                & G\\ 
J011102-474911  & $24.6$                & $-20.7$                 & 0.154                & G\\ 
J012820-564939  & $24.2$                & $-20.8$                 & 0.066                & G\\ 
J181225-712006  & $24.5$                & $-21.5$                 & 0.199                & G\\ 
J181857-550815  & $23.9$                & $-21.7$                 & 0.073                & G\\ 
 J203540-694407  & $26.6$                & $-26.5$                 & 0.875               & Q\\
 J203958-720655  & $26.8$                & $-27.0$                 & 1.00                & Q\\
J212222-560014  & $23.5$                & $-20.7$                 & 0.0518               & G\\ 
 J233159-381147  & $27.4$                & $-27.3$                 & 1.20                & Q\\
 J234743-494627  & $26.4$                & $-23.1$                 & 0.643               & Q\\
\multicolumn{5}{c}{Estimated z}\\
J002616-351249  & $26.7^{+ 0.8}_{-0.5}$ & $-20.7^{+ 2.5}_{-2.8}$  & $0.6_{-0.4}^{+0.9}$  & G\\               
J180859-832526  & $25.7^{+ 0.8}_{-0.8}$ & $-20.7^{+ 2.6}_{-2.3}$  & $0.5_{-0.3}^{+0.8}$  & G\\     
J230737-354828  & $25.8^{+ 0.6}_{-0.8}$ & $-20.8^{+ 2.4}_{-1.7}$  & $0.4_{-0.2}^{+0.6}$  & G\\          
\hline											   
 \end{tabular}
\caption{The 20\,GHz radio luminosity and absolute magnitude of each of the candidate GPS sources. Redshift and class ID are as in table \ref{tab:redshifts}. Quasars with estimated redshifts are not shown.}
\label{tab:radiolum}
\end{table}

\section{Discussion}
The initial optical identification of the sample of 21 sources was carried out by cross matching with the SuperCOSMOS database. A morphological identification of galaxy or star-like object was based on the SuperCOSMOS catalog.

\begin{table}
 \centering
 \begin{tabular}{cccc}
\hline
Name           & Class & Revised & Notes\\
AT20G          &  ID   &  Class  &      \\
 (1)           &  (2)  &   (3)   &  (4) \\
\hline
J002616-351249 &   G   &    x     & Likely beamed source\\		  
J004417-375259 &   G   &   GPS?   & BLRG?\\
J004905-552110 &   G   &    x     & \\
J005427-341949 &   G   &   GPS    & Possibly restarted\\
J010333-643907 &   G   &   GPS    & Possibly restarted\\
J011102-474911 &   G   &   GPS    & \\
J012714-481332 &   -   &   GPS?   & Need more information\\
J012820-564939 &   G   &   GPS?   &\\
J180859-832526 &   G   &   GPS    &\\  
J181225-712006 &   G   &   GPS    & Merging components?\\  
J181857-550815 &   G   &   GPS    & Restarted\\	  
J203540-694407 &   Q   &    x     &\\
J203958-720655 &   Q   &    x     &\\  
J205503-635207 &   Q   &   GPS?   & New radio source\\  
J212222-560014 &   G   &   GPS    & Cluster member\\  
J212402-602808 &   Q   &    x     & No optical spectrum\\
J213622-633551 &   Q   &    x     &\\
J214447-694654 &   Q   &   GPS    & No optical spectrum\\		  
J230737-354828 &   G   &    x     & Large 20\,GHz variability\\
J233159-381147 &   Q   &    x     & Previously well studied\\
J234743-494627 &   Q   &    x     & FSRQ\\
\hline
 \end{tabular}
 \caption[Original and revised classification of candidate GPS sources]{Original and revised classification of candidate GPS sources. Column 2 shows the classification from SuperCOSMOS as either Galaxy (G) or star-like (Q). Column 3 shows the revised class as a genuine or likely GPS galaxy or as unlikely to be a GPS galaxy (x). Notes: BLRG - Broad Line Radio Galaxy, FSRQ - Flat Spectrum Radio Quasar.}
 \label{tab:classification}
\end{table}

Based on optical colours, spectra, variability and multi-frequency information from the literature, sources were then reclassified as being: genuine, possible or unlikely GPS galaxies. Sources with an initial identification of galaxy were reclassified as genuine or likely GPS galaxies 75\% (9/12) of the time, and sources with an initial identification of star like were classified as unlikely to be a GPS galaxy 75\% (6/8) of the time. The single source that had no SuperCOSMOS ID is a possible GPS source. Thus if a sample of GPS sources is to be found, removal of sources classified as star-like within the SuperCOSMOS database is an efficient first step removing 75\% of the undesirable candidates whilst rejecting only 17\% of the likely GPS sources. Table \ref{tab:classification} shows the original and revised classification for each of the 21 sources. Notes on individual sources are given in the Appendix.

The optical colours of the source sample were compared to those of normal E or S0 galaxy models. In each instance there was an excess of blue emission suggesting that an AGN or recent star formation was present. Where available the optical spectra indicate that an AGN is the most likely cause of the excess emission.

The 20\,GHz variability of the candidate sources was measured over a period of 1-3 years. Although the variability was measured on three different time scales there is no significant difference between the variability index for sources in each of the sub-groups, the average being 10-12\% over 1-3 years. As a group the sources show the same level of variability as typical GPS sources, however it is interesting to note that in this sample the optically identified galaxies show a slightly higher amount of variability ($13\pm2.5\%$) than the star like objects ($9.5\pm 2.0\%$). The source with the highest variability is classified as a galaxy within the SuperCOSMOS database, but was not considered to be a genuine GPS source. 

The radio and optical luminosities of the sample sources are consistent with a scenario in which the two are correlated in the radio loud regime, and unrelated in the radio quiet, for both quasars and galaxies.

Three (25\%) of the 12 confirmed GPS galaxies show evidence of being restarted. Such sources exhibit a radio core that is typical of a genuine GPS source, but also show extended emission, hot spots or jets. These restarted GPS sources are interesting as they show more than one epoch of radio galaxy evolution within a single source \citep{saikia_recurrent_2010}. The evolution of these sources may differ from that of GPS sources that are beginning their radio evolution for the first time. If a merger or star formation history of these sources could be determined then the causes of the quenching and restarting process can be investigated.

If these numbers are indicative of the larger population of radio sources then as many as 400 genuine GPS sources could be contained within the AT20G with up to 25\% of them being restarted.

The four sources with the lowest measured redshift in this sample have spectra which peak above $\sim80$\,GHz, and are all associated with galaxies. This is very promising for future efforts to identify the High Frequency Peaking GPS sources from the AT20G survey.

\section{Conclusions and Future work}
A sample of 21 radio sources with inverted spectra in the AT20G catalog were observed at 40 and 95\,GHz to determine their high-frequency spectral shape and turnover frequency. A second epoch of 20\,GHz observations was included to measure the variability of the sources. Redshift information was taken from the literature when possible and observations were carried out when no redshift was known. In 7/21 cases the redshifts were estimated based on optical colours.

Detailed study of the 21 GPS candidates reveals that 8 are likely GPS galaxies and 4 are possible GPS galaxies with further investigation required in each case. Nine of the 21 GPS candidates were considered not to be genuine GPS galaxies and do not require further work. Three of the 12 surviving GPS candidates show evidence of previous activity, indicating that, whilst GPS sources are some of the youngest radio sources, a significant number of them have been recently restarted. If these numbers are indicative of the full AT20G sample then as many as 400 genuine GPS sources could be contained within the AT20G with up to 25\% of them being restarted.

Further work needs to be done to increase the redshift coverage of the GPS population and to expand upon the number of HFP sources with peaks above 20\,GHz.
\vspace{-2mm}
\section*{Acknowledgments}
The authors would like to thank Dick Hunstead and Helen Johnston for the use of their ESO-NTT backup observing time, and for their help in reducing the data of same. The authors also thank Jim Condon for his insightful comments as to the interpretation of the radio and optical luminosity relation.

This research has made use of the NASA/IPAC Extragalactic Database (NED) which is operated by the Jet Propulsion Laboratory, California Institute of Technology, under contract with the National Aeronautics and Space Administration. 
\vspace{-2mm}
\appendix 
\section{}
This appendix includes the radio spectra for the 21 sources observed at 40 and 95\,GHz as well at notes on the individual sources. 

For each of the sources the following information is reported, (where available or notable): The optical identification and colours, radio observations from source other than the AT20G and this paper, the radio variability, the radio spectrum, the radio morphology, and other notes of interest. The revised classification of each object will also be given as either genuine GPS or probably QSO based on the criteria set out in the introduction.

The Parkes-MIT-NRAO 4.85\,GHz survey \citep[PMN,][]{griffith_parkes-mit-nrao_1993} is used as an independent measure of flux and to determine source variability. The primary beam of the PMN survey is 4.2 arcmin at 4.8\,GHz whilst the AT20G has a primary beam of 2.2 and 0.53 arcmin at 4.8 and 20\,GHz respectively. The AT20G does not contain any baselines shorter than 30m and has only limited sensitivity to flux on scales larger than $\sim$30\,arcsec \citep{murphy_australia_2010}. The AT20G beam is often elliptical at 4.8 and 8.6\,GHz and results in elongated radio contours, that appear to be due to extended emission, even when the source is unresolved. The observation dates for the PMN survey sources cannot be determined more accurately than to within the year of 1990, as the survey observations spanned many months within this year.

The SSO and NTT spectra that were observed for this project are included within the individual source notes.

\clearpage
\begin{figure*}
 \centering
\subfigure{
     \epsfig{file=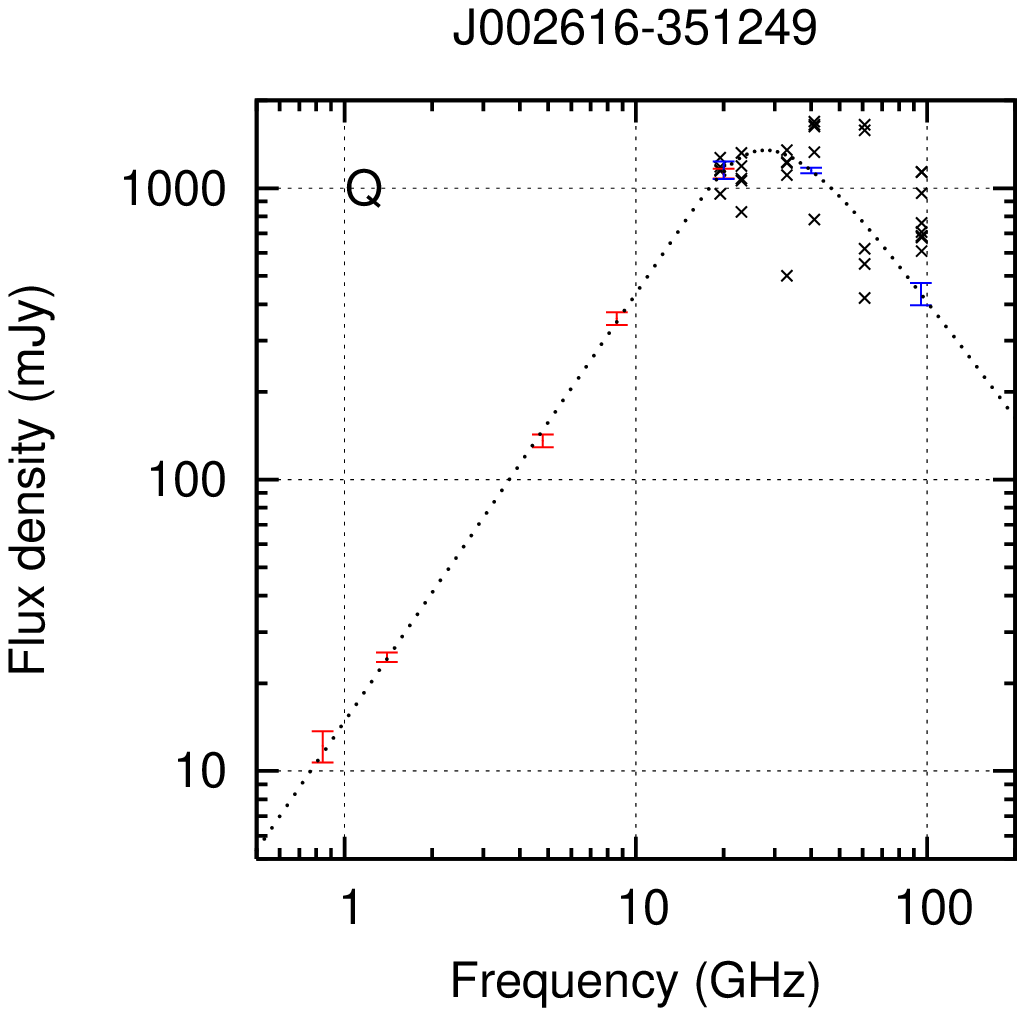,  width=0.32\linewidth, bb=265 55 540 75, angle=-90,clip=} %flux label
\qso \epsfig{file=C1392/J002616-351249.eps, height=0.32\linewidth, bb=265 75 540 350, angle=-90,clip=}
\gal \epsfig{file=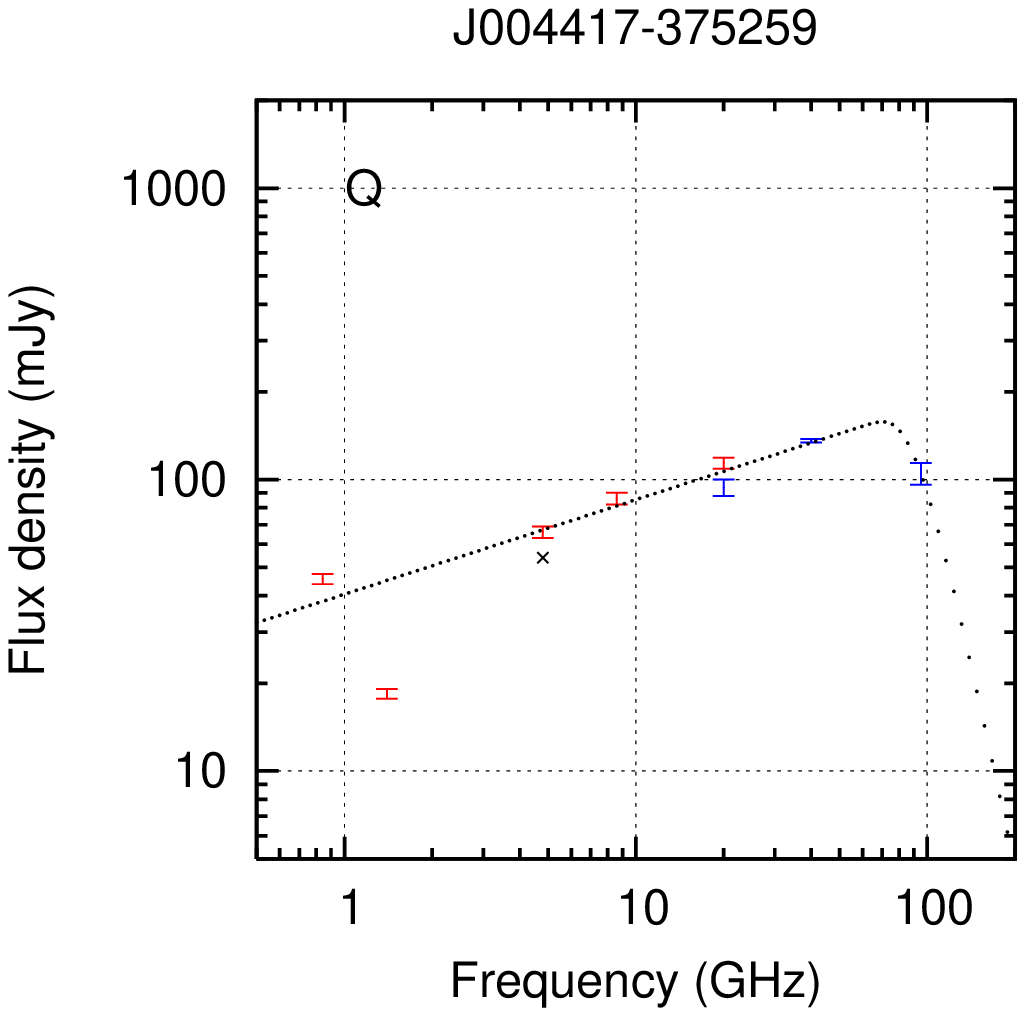, height=0.32\linewidth, bb=265 75 540 350, angle=-90,clip=}
\qso \epsfig{file=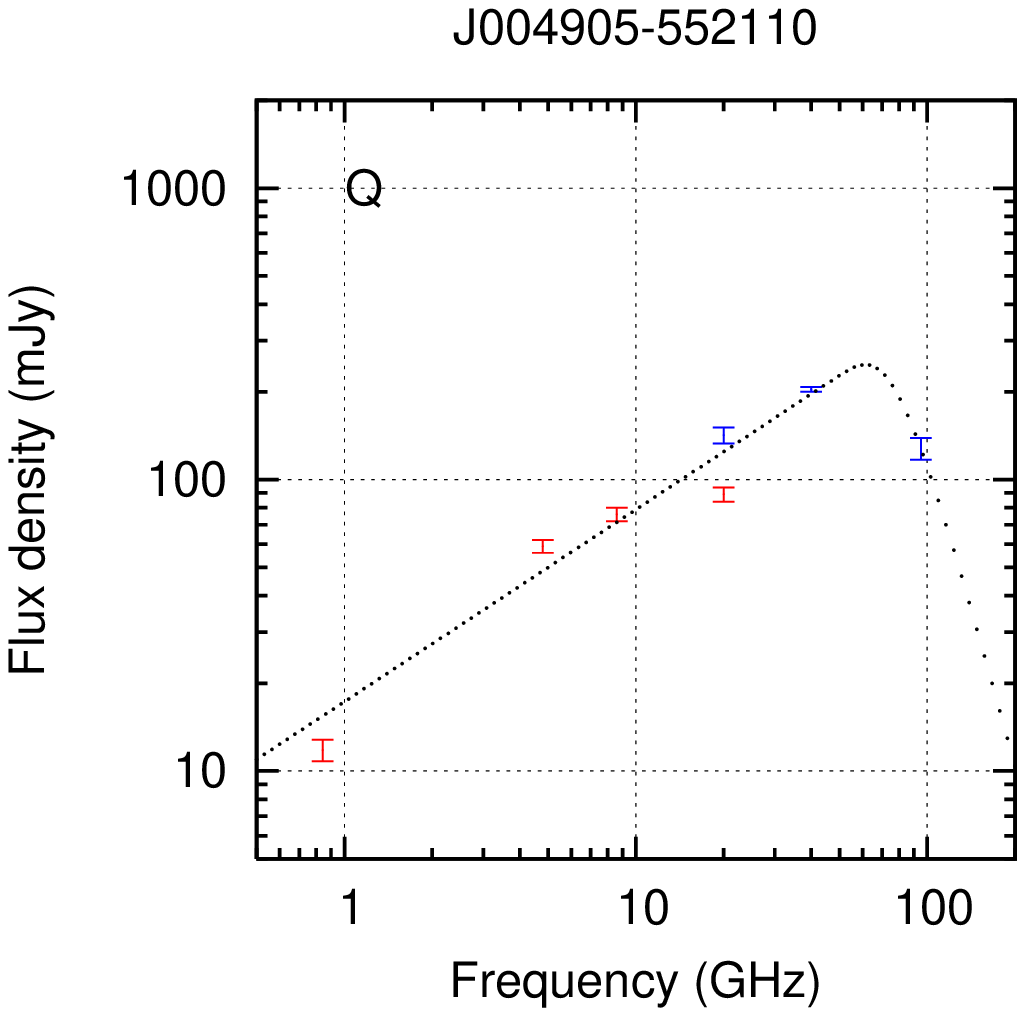, height=0.32\linewidth, bb=265 75 540 350, angle=-90,clip=}
}
\\
\subfigure{
     \epsfig{file=C1392/J002616-351249.eps,  width=0.32\linewidth, bb=265 55 540 75, angle=-90,clip=} %flux label
\gal \epsfig{file=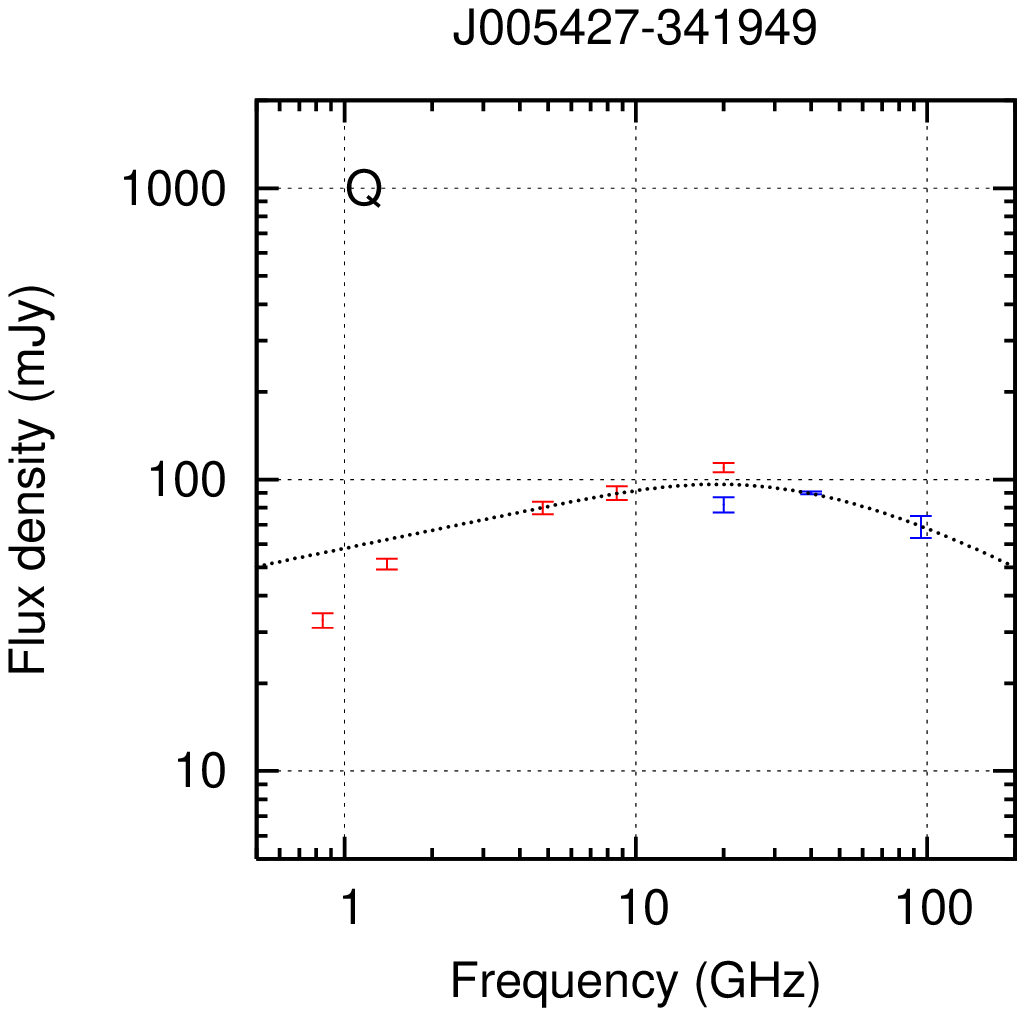, height=0.32\linewidth, bb=265 75 540 350, angle=-90,clip=}
\gal \epsfig{file=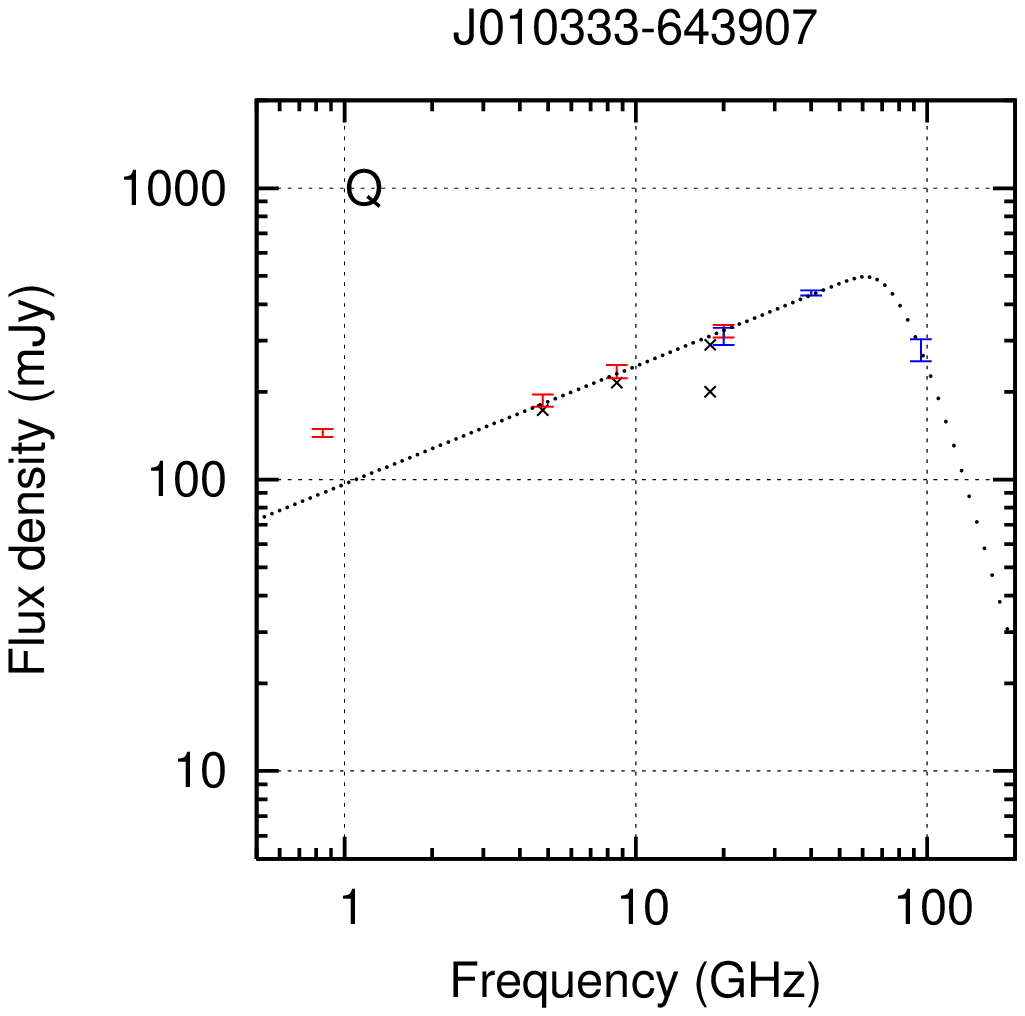, height=0.32\linewidth, bb=265 75 540 350, angle=-90,clip=}
\gal \epsfig{file=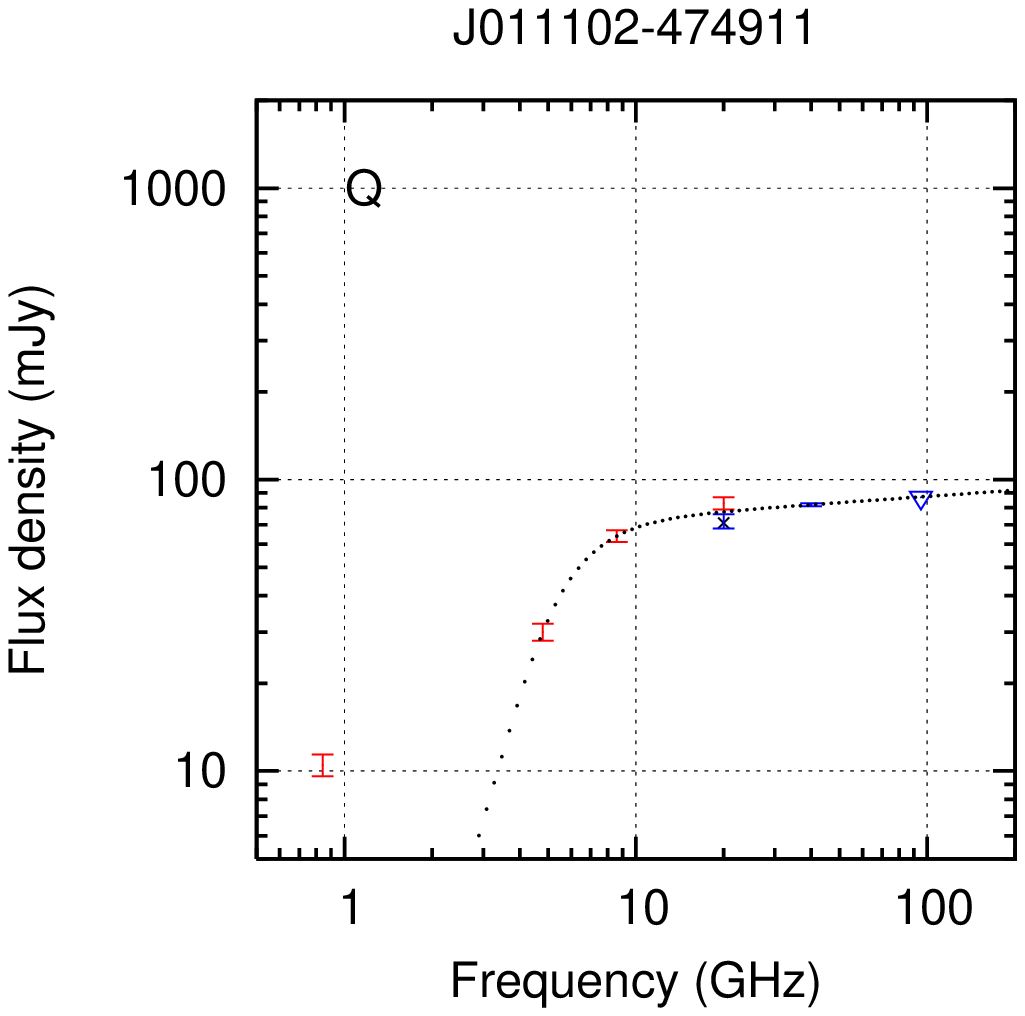, height=0.32\linewidth, bb=265 75 540 350, angle=-90,clip=}
}
\\
\subfigure{
     \epsfig{file=C1392/J002616-351249.eps,  width=0.32\linewidth, bb=265 55 540 75, angle=-90,clip=} %flux label
\gal \epsfig{file=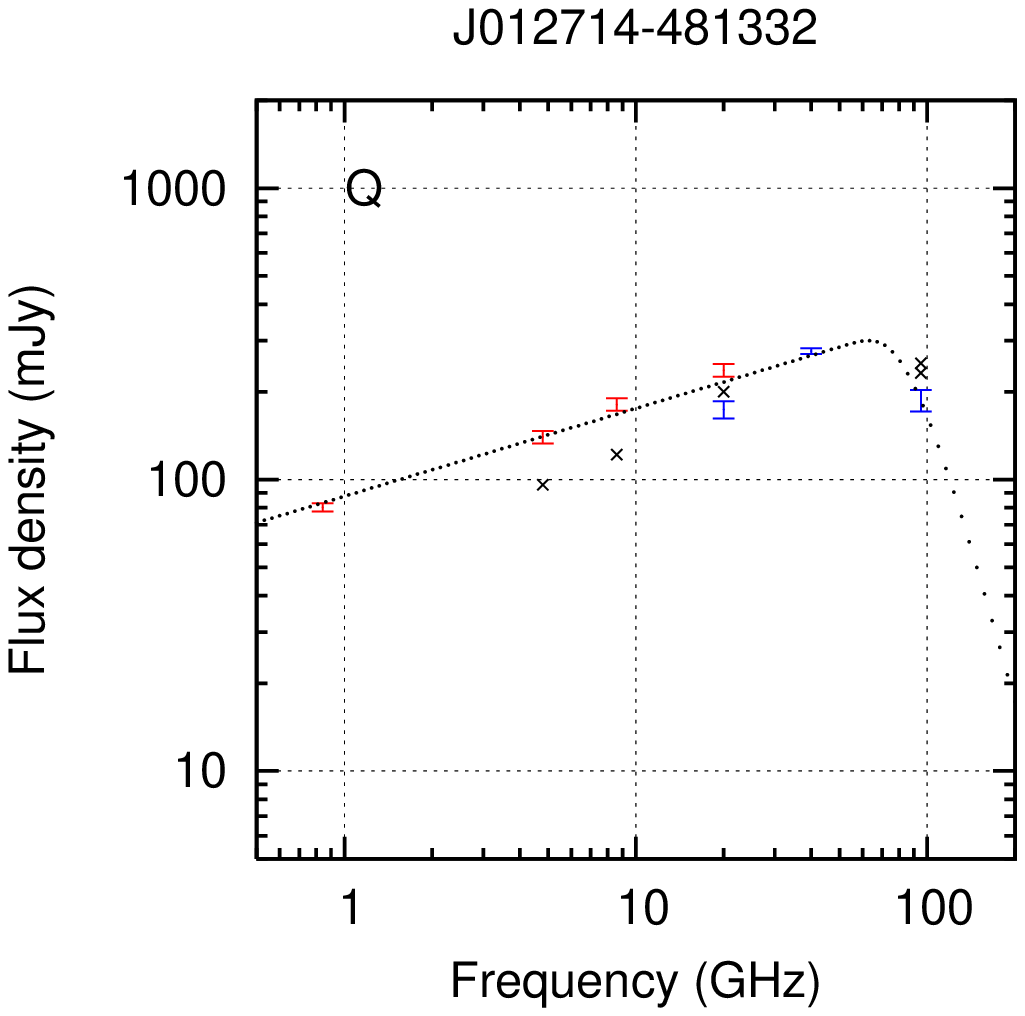, height=0.32\linewidth, bb=265 75 540 350, angle=-90,clip=}
\gal \epsfig{file=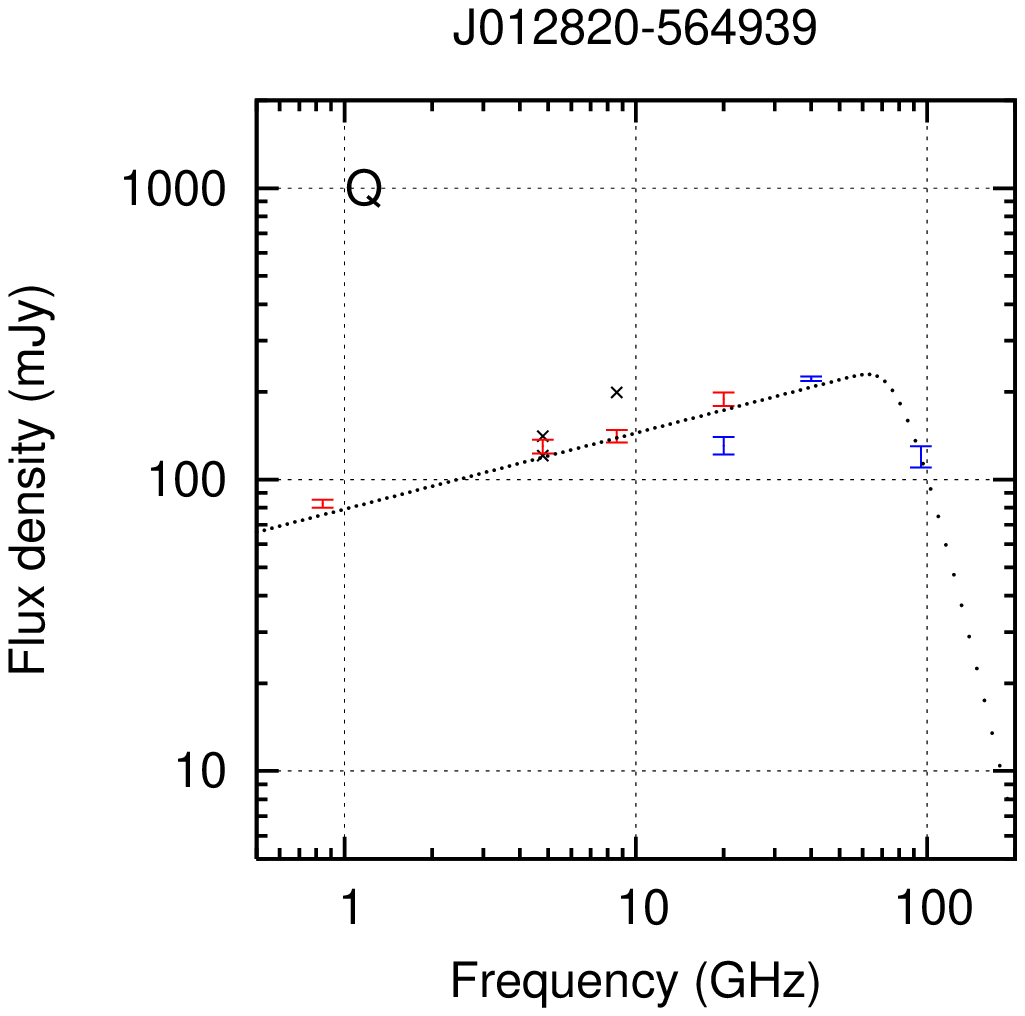, height=0.32\linewidth, bb=265 75 540 350, angle=-90,clip=}
\gal \epsfig{file=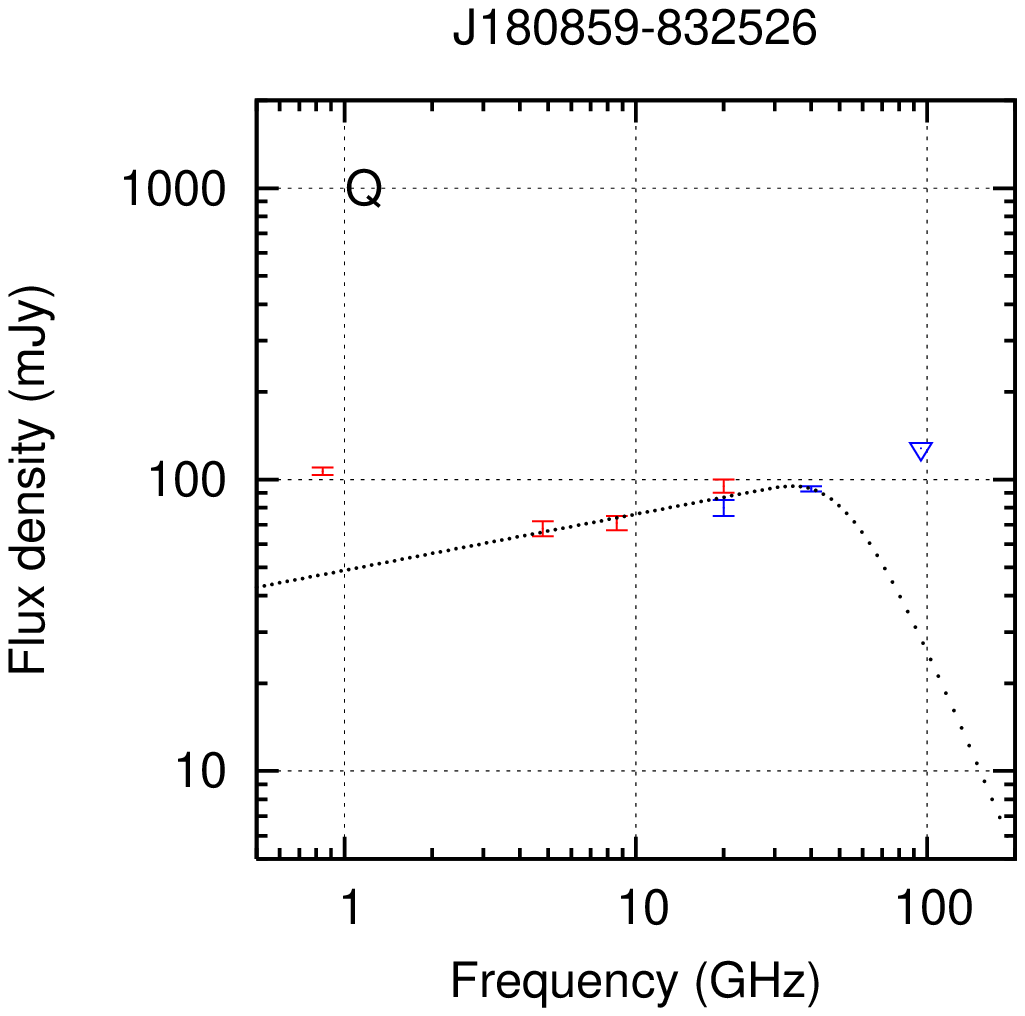, height=0.32\linewidth, bb=265 75 540 350, angle=-90,clip=}
}
\\
\addtocounter{subfigure}{-3}
\subfigure[]{
\label{fig:radio_sed:a}
\epsfig{file=C1392/J002616-351249.eps,  width=0.025\linewidth, bb=265 55 285 75, angle=-90,clip=} %empty square	  
\epsfig{file=C1392/J012714-481332.eps, height=0.32\linewidth, bb=540 75 555 350, angle=-90,clip=} %frequecny label
\epsfig{file=C1392/J012820-564939.eps, height=0.32\linewidth, bb=540 75 555 350, angle=-90,clip=} %frequecny label
\epsfig{file=C1392/J180859-832526.eps, height=0.32\linewidth, bb=540 75 555 350, angle=-90,clip=} %frequecny label
}
\label{fig:radio_sed}
\caption{Spectral energy distributions for the 21 sources observed. Red data are from SUMSS (843\,MHz), NVSS (1.4\,GHz), and the AT20G catalog (4.8, 8.6 and 20\,GHz). Blue data is from this paper (20, 40, and 95\,GHz). Open triangles represent 95\,GHz flux density upper limits. Small black crosses are data from other surveys mentioned in the discussion of each source. The dotted line is a fit to the data of this paper and the AT20G using either the function of Snellen et al. (1998), or a powerlaw.}
\end{figure*}

\addtocounter{figure}{-1}
\begin{figure*}
 \addtocounter{subfigure}{1}
 \centering
 \subfigure{
     \epsfig{file=C1392/J002616-351249.eps,  width=0.32\linewidth, bb=265 55 540 75, angle=-90,clip=}%flux label
\gal \epsfig{file=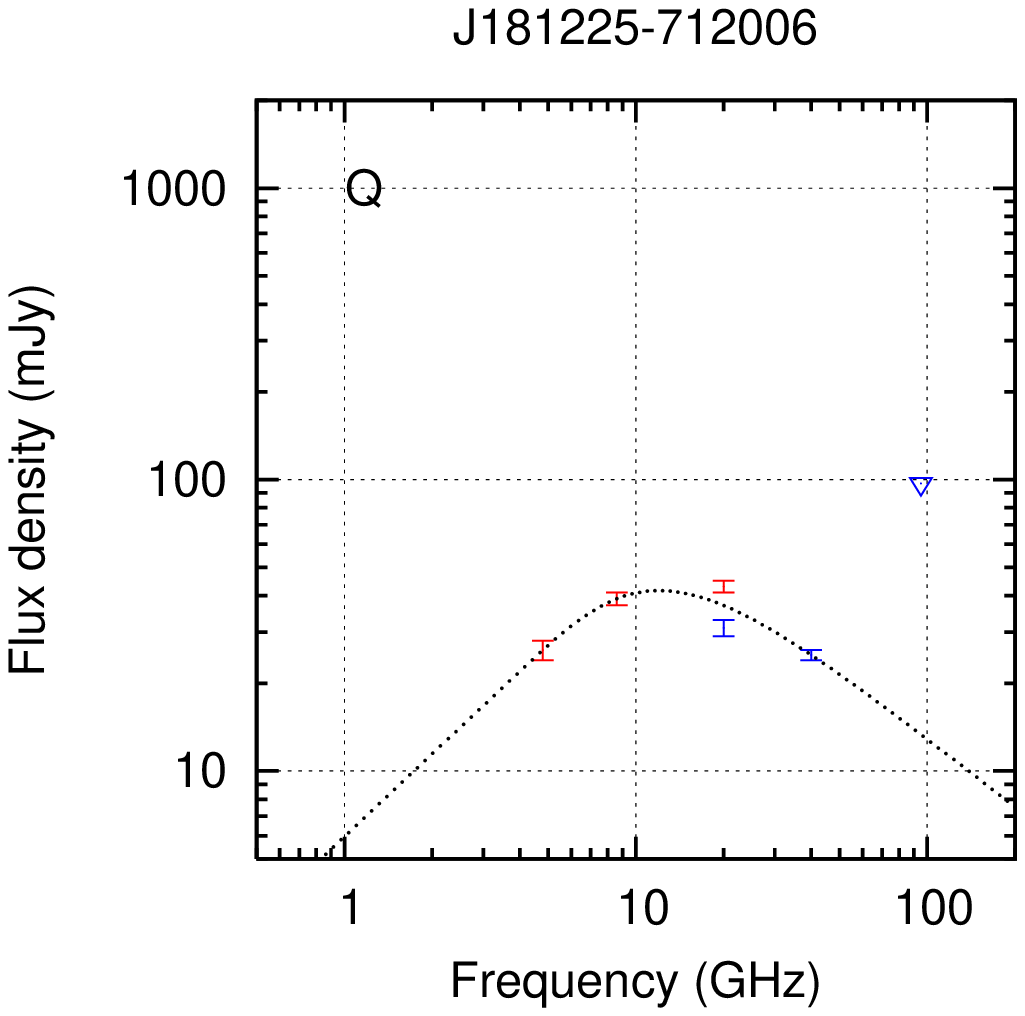, height=0.32\linewidth, bb=265 75 540 350, angle=-90,clip=}
\gal \epsfig{file=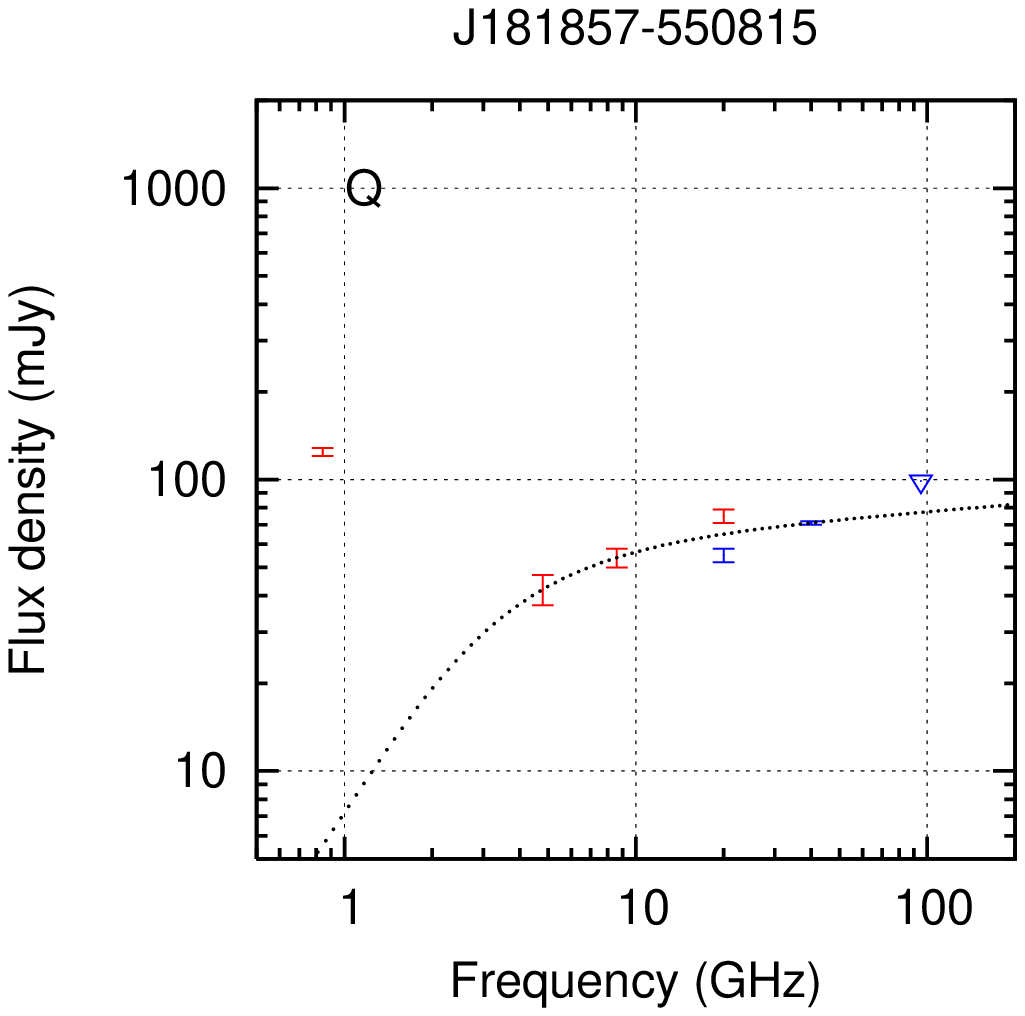, height=0.32\linewidth, bb=265 75 540 350, angle=-90,clip=}
\qso \epsfig{file=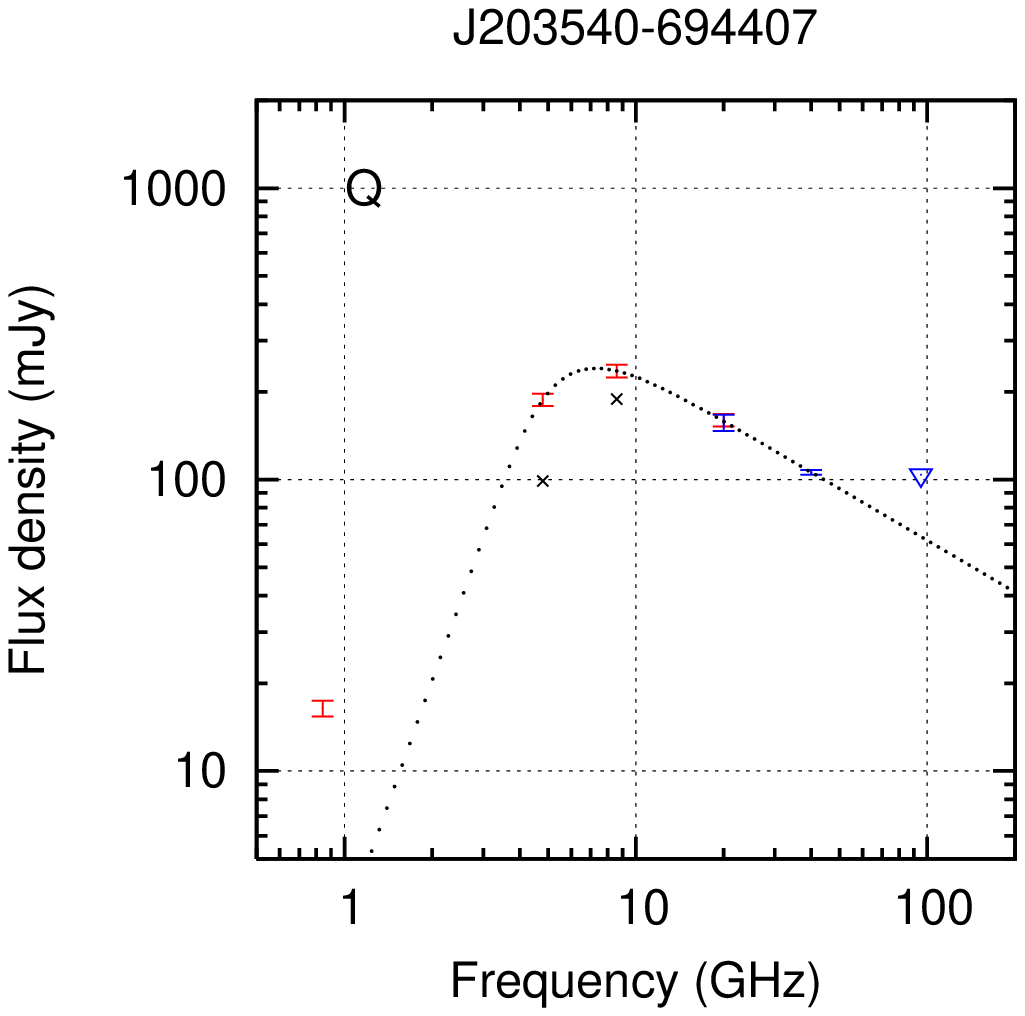, height=0.32\linewidth, bb=265 75 540 350, angle=-90,clip=}
}
\\									    
\subfigure{
     \epsfig{file=C1392/J002616-351249.eps,  width=0.32\linewidth, bb=265 55 540 75, angle=-90,clip=}%flux label
\qso \epsfig{file=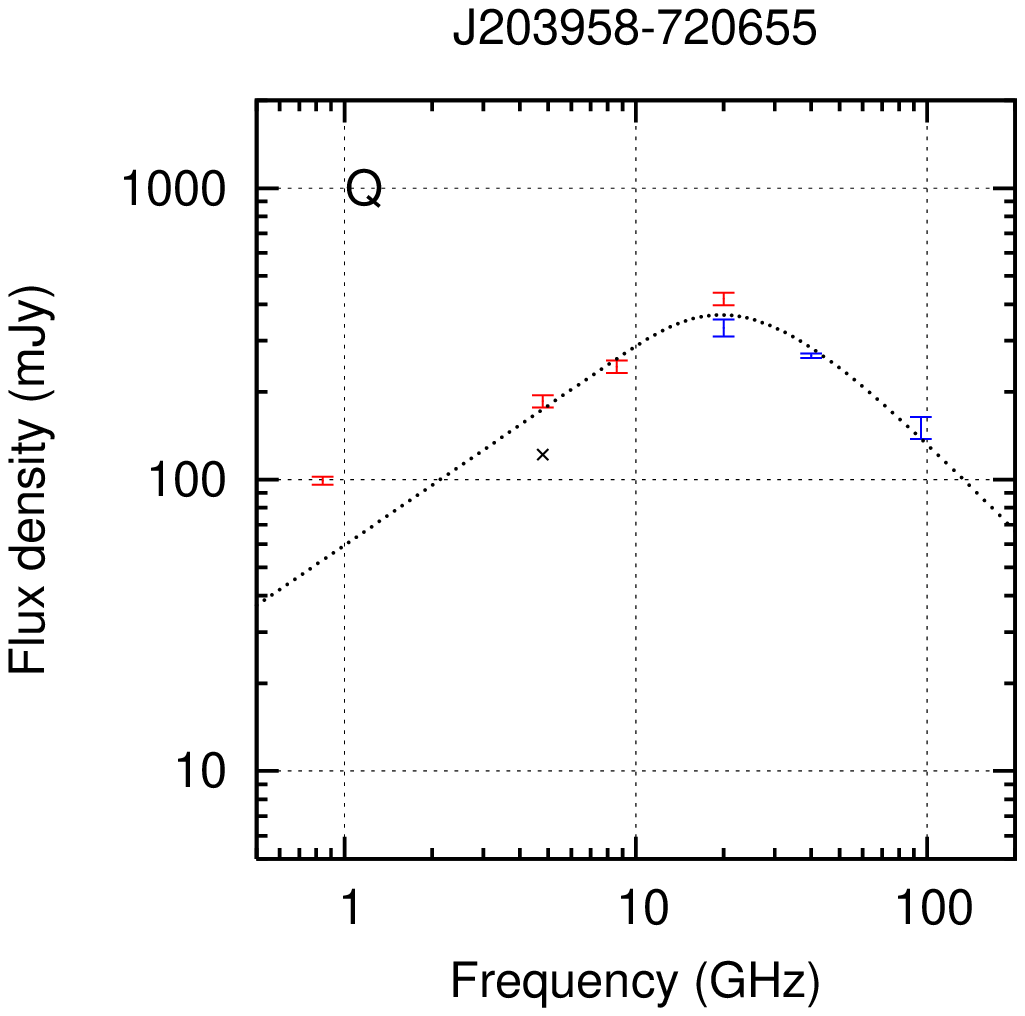, height=0.32\linewidth, bb=265 75 540 350, angle=-90,clip=}
\gal \epsfig{file=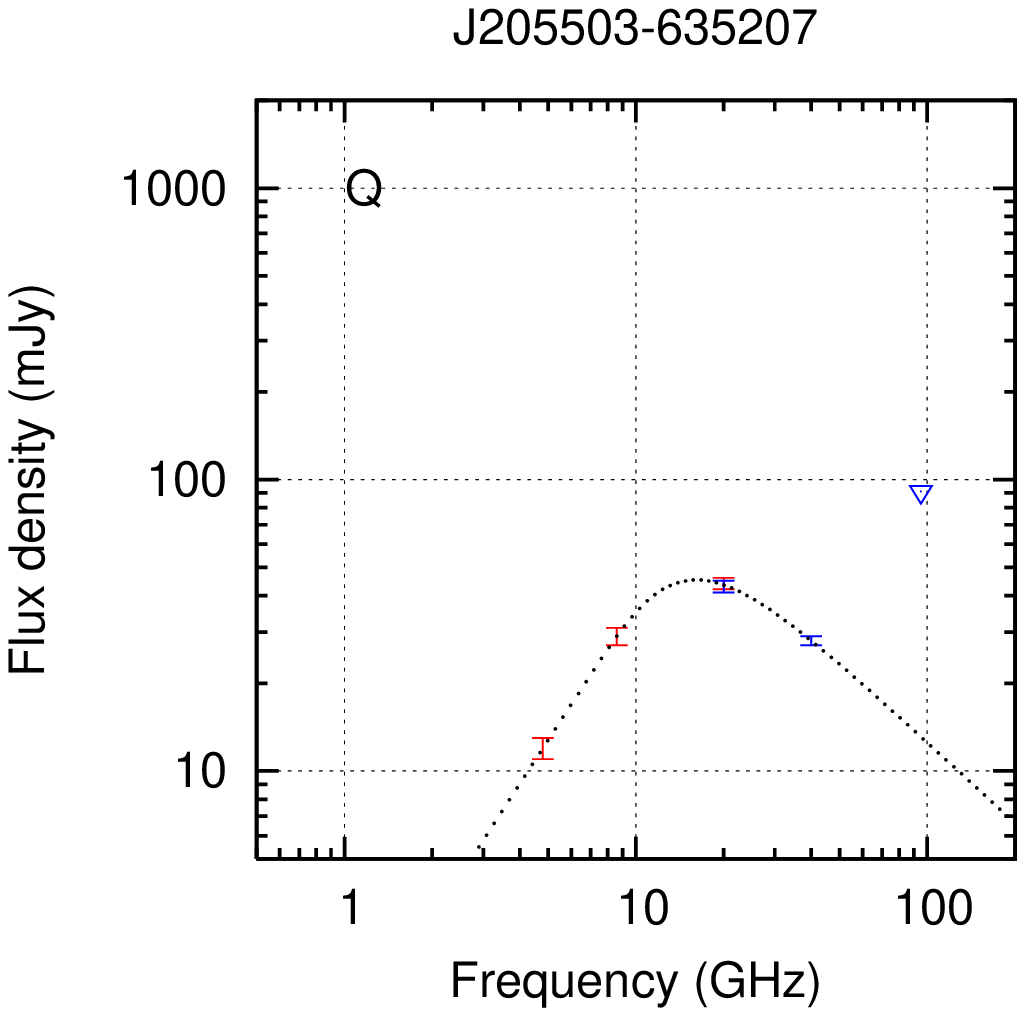, height=0.32\linewidth, bb=265 75 540 350, angle=-90,clip=}
\gal \epsfig{file=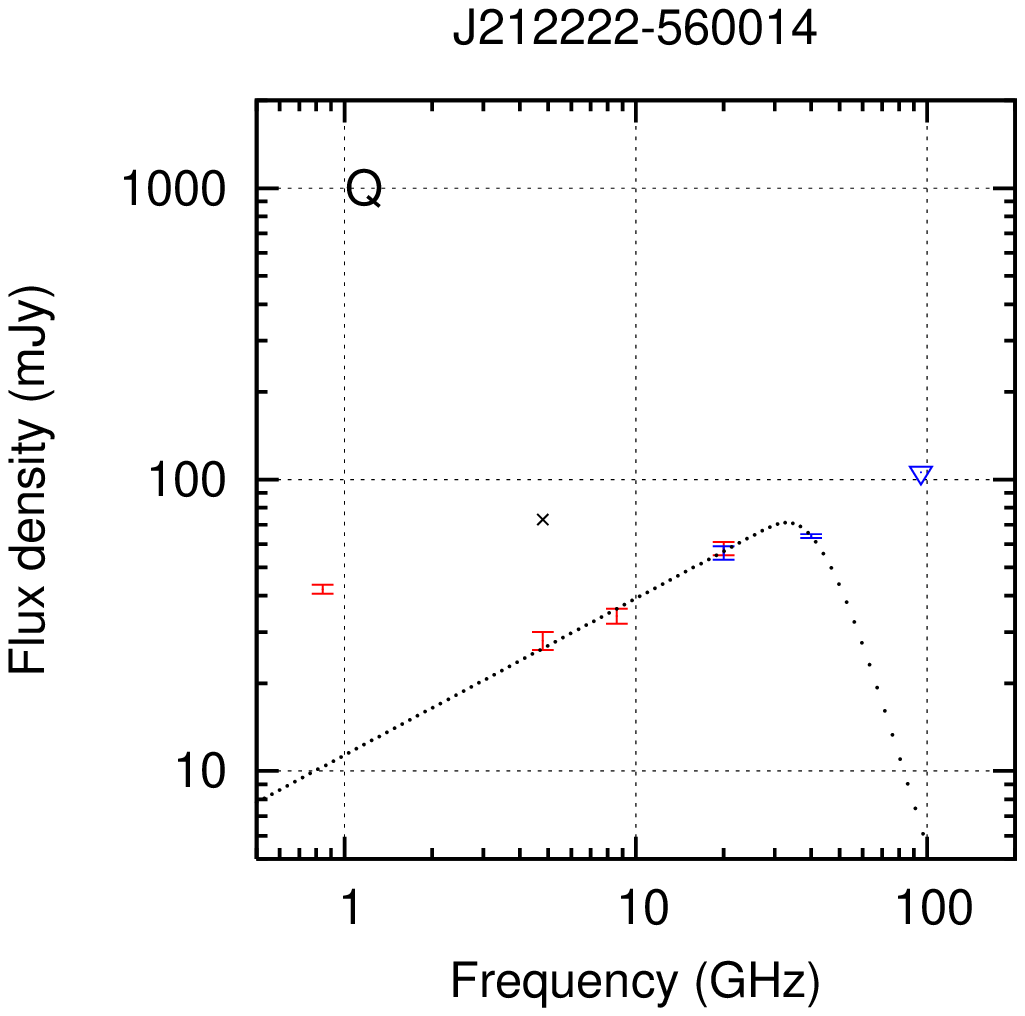, height=0.32\linewidth, bb=265 75 540 350, angle=-90,clip=}
}
\\		
\subfigure{
     \epsfig{file=C1392/J002616-351249.eps,  width=0.32\linewidth, bb=265 55 540 75, angle=-90,clip=}%flux label
\qso \epsfig{file=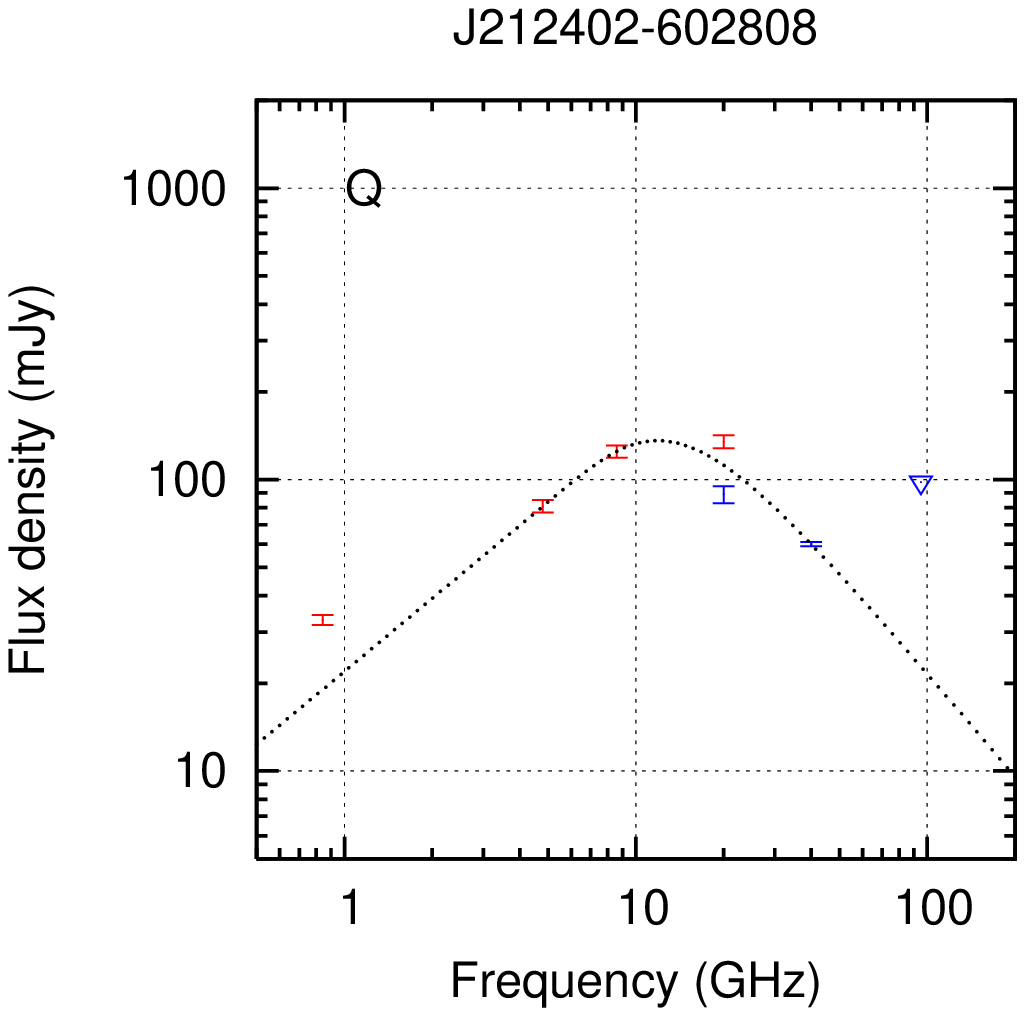, height=0.32\linewidth, bb=265 75 540 350, angle=-90,clip=}
\qso \epsfig{file=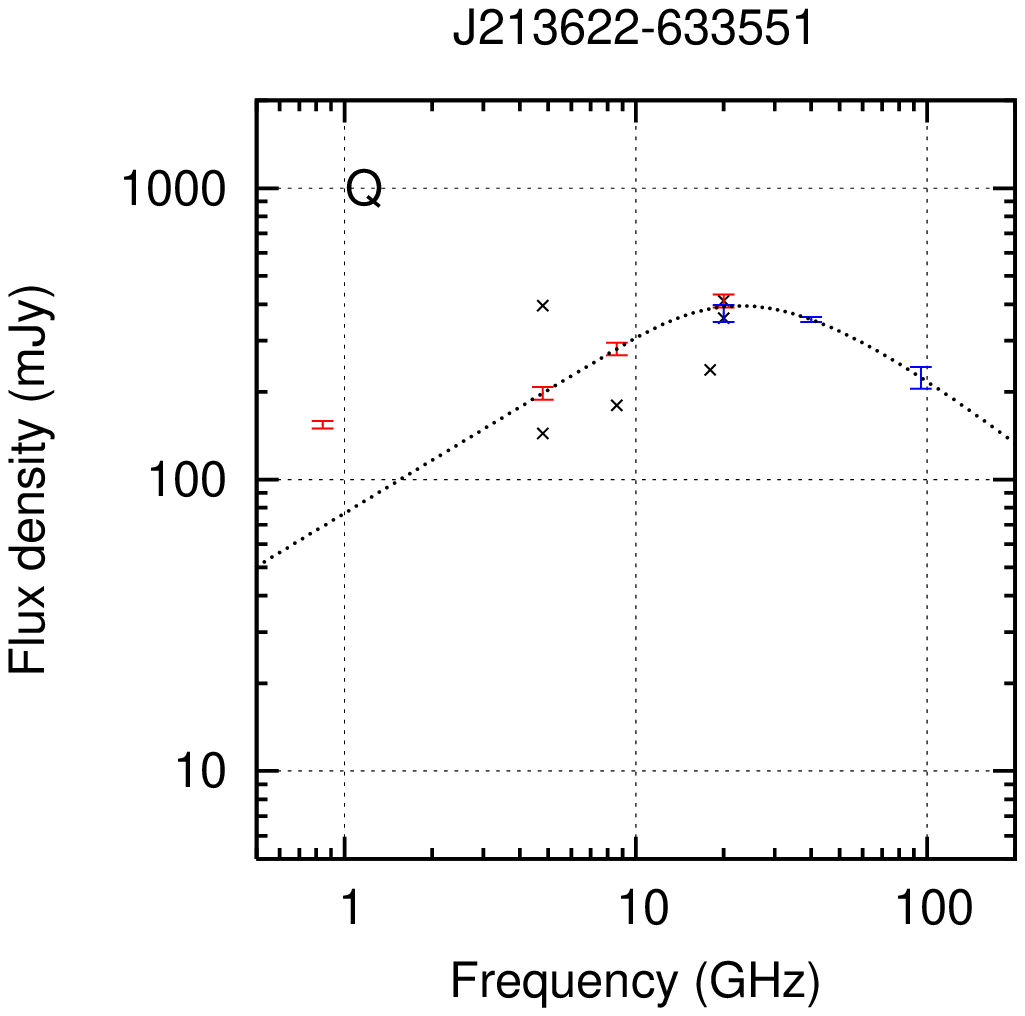, height=0.32\linewidth, bb=265 75 540 350, angle=-90,clip=}
\gal \epsfig{file=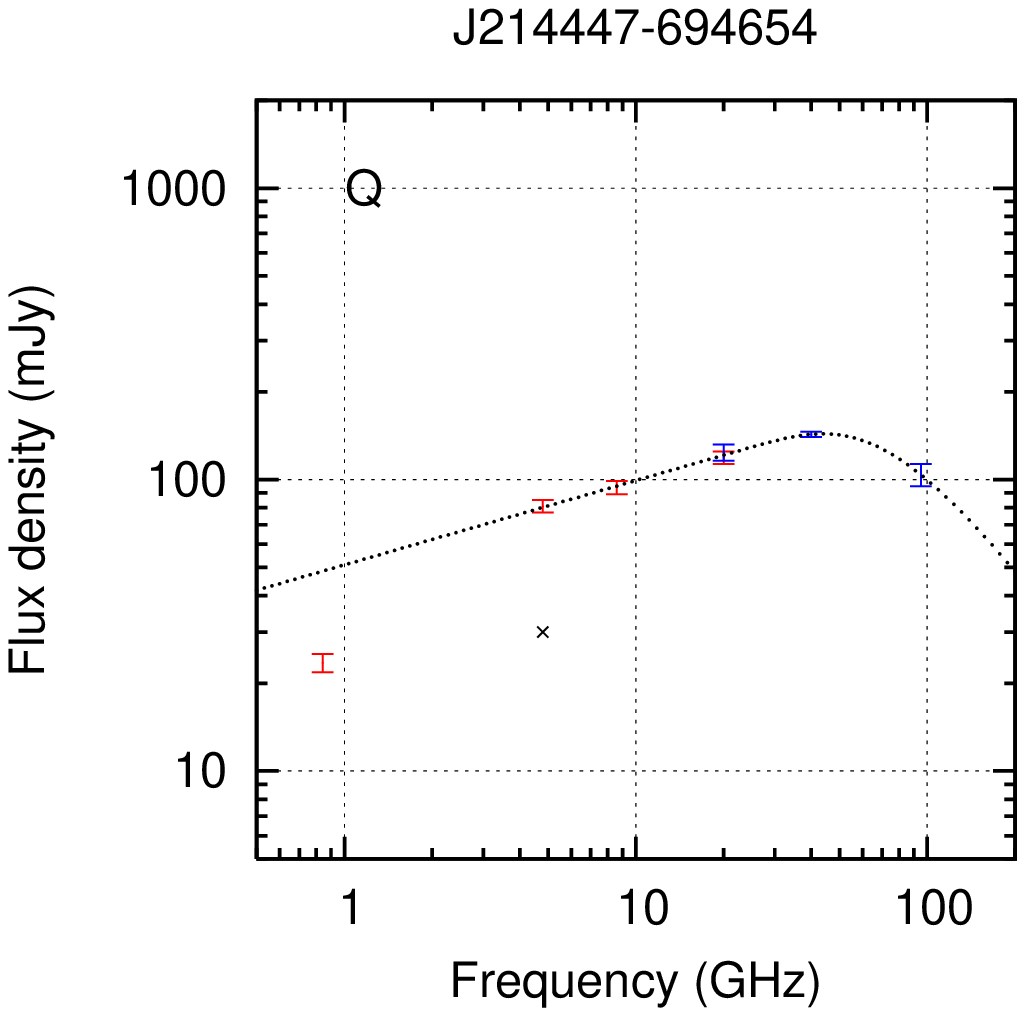, height=0.32\linewidth, bb=265 75 540 350, angle=-90,clip=}
}
\\
\subfigure{
     \epsfig{file=C1392/J002616-351249.eps,  width=0.32\linewidth, bb=265 55 540 75, angle=-90,clip=} %flux label
\qso \epsfig{file=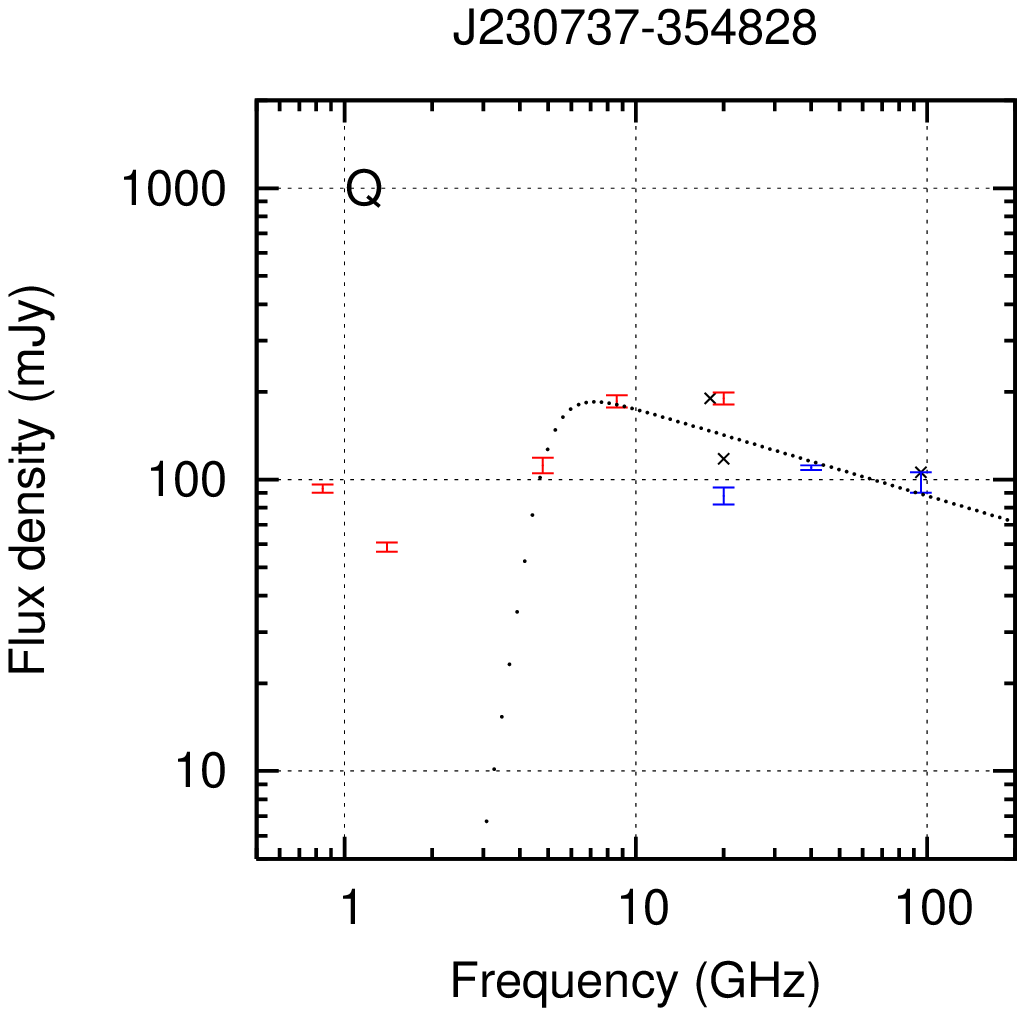, height=0.32\linewidth, bb=265 75 540 350, angle=-90,clip=}
\qso \epsfig{file=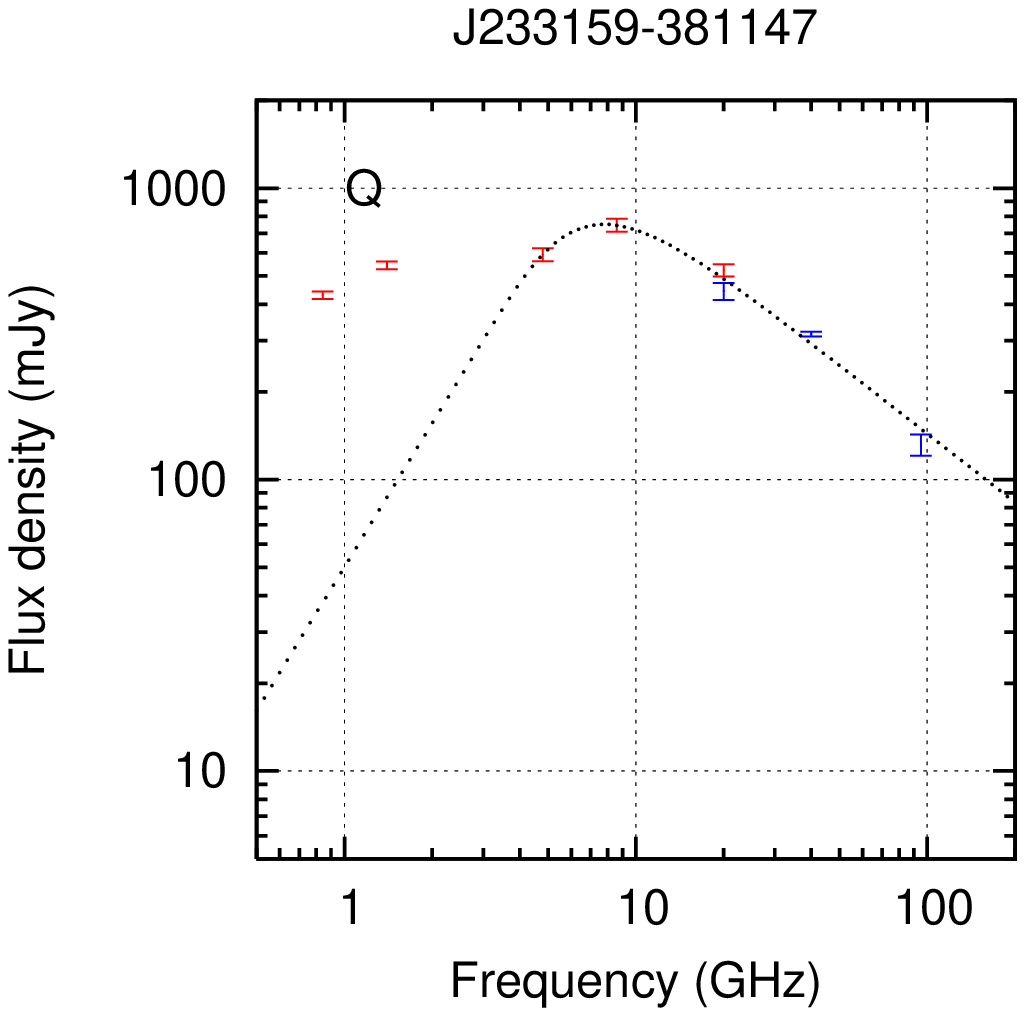, height=0.32\linewidth, bb=265 75 540 350, angle=-90,clip=}
\qso \epsfig{file=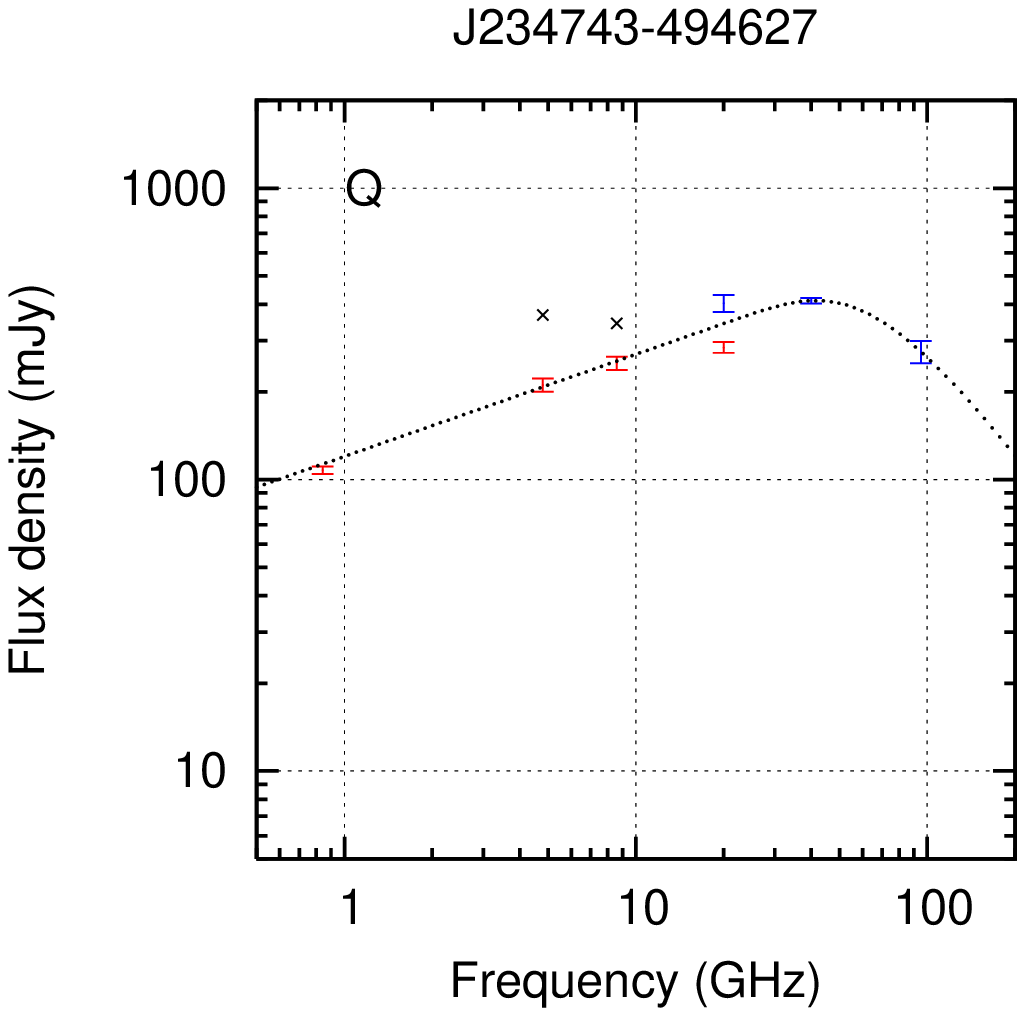, height=0.32\linewidth, bb=265 75 540 350, angle=-90,clip=}
}
\addtocounter{subfigure}{-4}
\subfigure[]{
\label{fig:radio_sed:c}
\label{fig:radio_sed:b}
\epsfig{file=C1392/J002616-351249.eps,  width=0.025\linewidth, bb=265 55 285 75, angle=-90,clip=} %empty square	  
\epsfig{file=C1392/J012714-481332.eps, height=0.32\linewidth, bb=540 75 555 350, angle=-90,clip=} %frequecny label
\epsfig{file=C1392/J012820-564939.eps, height=0.32\linewidth, bb=540 75 555 350, angle=-90,clip=} %frequecny label
\epsfig{file=C1392/J180859-832526.eps, height=0.32\linewidth, bb=540 75 555 350, angle=-90,clip=} %frequecny label

}
\caption[Radio spectral energy distributions for the 21 sources observed (10-18)]{Continued}
\end{figure*}

%\addtocounter{figure}{-1}
%\begin{figure*}
% \centering
% \subfigure{
%     \epsfig{file=C1392/J002616-351249.eps,  width=0.32\linewidth, bb=265 55 540 75, angle=-90,clip=} %flux label
%\qso \epsfig{file=C1392/J230737-354828.eps, height=0.32\linewidth, bb=265 75 540 350, angle=-90,clip=}
%\qso \epsfig{file=C1392/J233159-381147.eps, height=0.32\linewidth, bb=265 75 540 350, angle=-90,clip=}
%\qso \epsfig{file=C1392/J234743-494627.eps, height=0.32\linewidth, bb=265 75 540 350, angle=-90,clip=}
%}
%\addtocounter{subfigure}{1}
%\subfigure[]{
%\label{fig:radio_sed:c}
%\epsfig{file=C1392/J002616-351249.eps,  width=0.025\linewidth, bb=265 55 285 75, angle=-90,clip=} %empty square	  
%\epsfig{file=C1392/J012714-481332.eps, height=0.32\linewidth, bb=540 75 555 350, angle=-90,clip=} %frequecny label
%\epsfig{file=C1392/J012820-564939.eps, height=0.32\linewidth, bb=540 75 555 350, angle=-90,clip=} %frequecny label
%\epsfig{file=C1392/J180859-832526.eps, height=0.32\linewidth, bb=540 75 555 350, angle=-90,clip=} %frequecny label
%}
%  \caption[Radio spectral energy distributions for the 21 sources observed (19-21)]{Continued}
%\end{figure*} 

\clearpage
\begin{table}
\centering
\caption{Data summary for J002616-351249}
\label{tab:data_J002616-35124}
\begin{tabular}{cccc}
\hline
\hline
Data Source & $\nu$ & Flux Density & Obs \\
            & GHz   & mJy          & Year\\ 
\hline
PMN         & 4.85  & $121\pm 12$ & 1990    \\
AT20G       & 4.8   & $136\pm 7$  & 2004.83 \\
\hline
VCS2        & 8.4   & \<200       & 2002.1 \\
AT20G       & 8.6   & $357\pm 18$ & 2004.83  \\
\hline
AT20G                             & 20 & $1123\pm 43$  & 2004.75  \\
\citet{sadler_extragalactic_2008} & 20 & $1273\pm 67$  & 2005.5\\
This paper                        & 20 & $1156\pm 79$  & 2007.83\\
\hline
This paper                        & 40 & $1150\pm 25$  & 2007.83\\
\hline
\citet{sadler_extragalactic_2008} & 95 & $1155\pm 136$ & 2005.5\\
                                  & 95 & $1119\pm 132$ & 2005.5\\
This paper                        & 95 & $434\pm 38$   & 2007.83\\
\hline
\end{tabular}
\end{table}
\begin{figure}
 \centering
 \epsfig{file=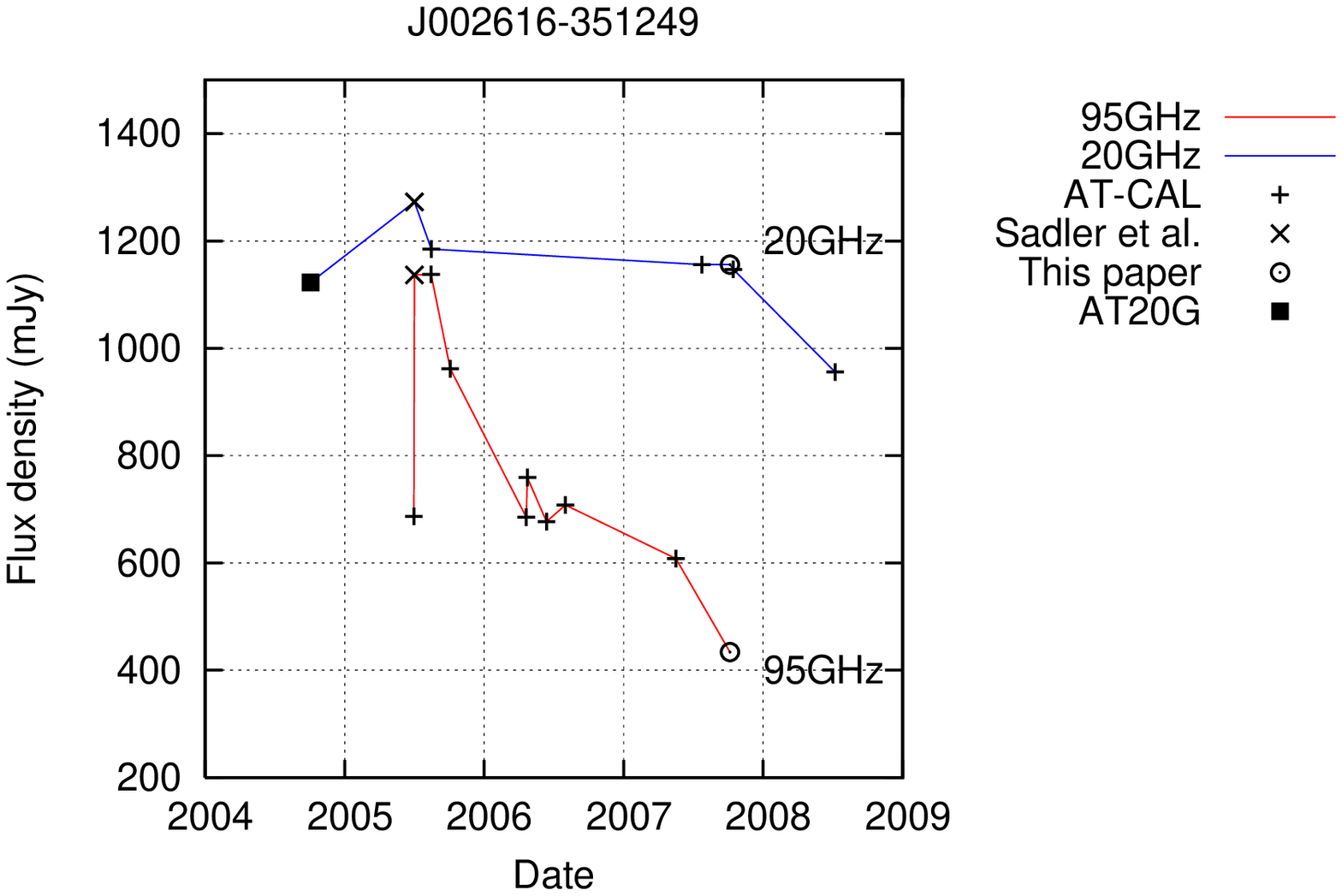, height=0.8\linewidth, angle=-90,clip=}
 \caption{Flux history at 20 and 95\,GHz, with fluxes from the AT20G,\citet{sadler_extragalactic_2008}, this paper, and the AT calibrator catalog. }
 \label{fig:J002616-351249_fluxhist}
\end{figure}

\begin{figure}
 \centering
 \includegraphics[bb=230 55 555 380, height=0.6\linewidth, angle=-90]{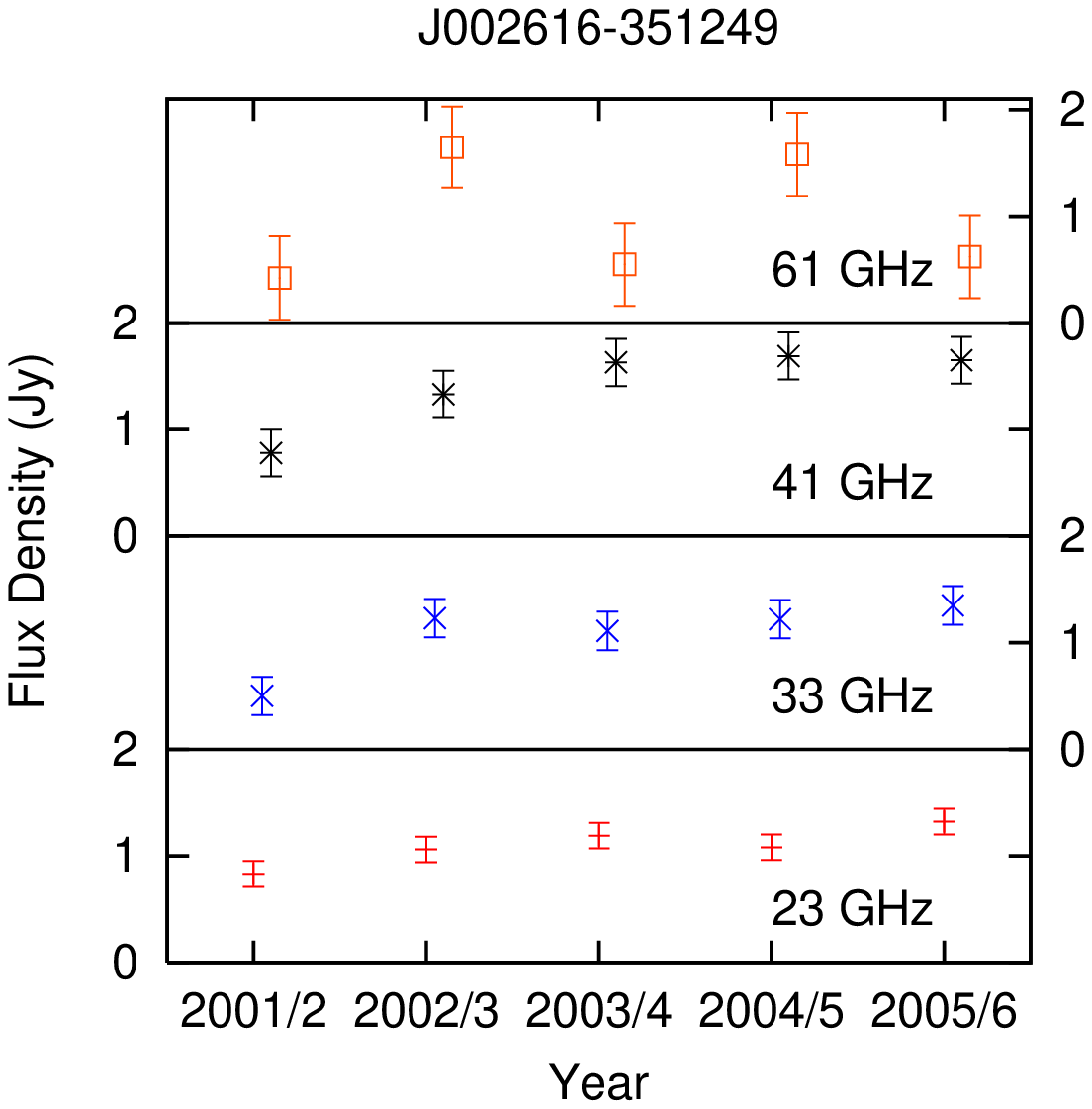}
 \caption{Flux history of J002616-351249 at 23, 33, 41, and 61\,GHz taken from the WMAP 5 year catalog. The increasing amount of variability above the spectral peak suggests that the emission is beamed.}
 \label{fig:J002616-351249_WMAP}
\end{figure}

\subsection{AT20G J002616-351249 - a $z\sim 0.6$ non-GPS galaxy}
This source is identified as a B$_J$ = 22.0\,mag galaxy within the SuperCOSMOS database. The optical colours are B-R$_C$ = 2.25\,mag and R$_C$-I$_C$= 0.51\,mag. The estimated redshift is 0.6 but the colours are bluer than expected for a normal E or S0 galaxy, suggesting that an AGN or recent star formation is present.

This source is in the PMN catalog \citep{wright_parkes-mit-nrao_1996}, the second VLA Calibrator Survey \citep[][VCS2]{fomalont_second_2003}, and was also observed by \citet{sadler_extragalactic_2008}. The details of these observations are summarised in table \ref{tab:data_J002616-35124}.

The 4.8\,GHz variability of this source is less than 6\% over a period of 14 years. There is good agreement between the AT20G and PMN fluxes. This source was added to the Australia Telescope calibrator catalog in 2005 and has thus been monitored at 20 and 95\,GHz since this time. Figure \ref{fig:J002616-351249_fluxhist} shows the flux history of this source with data from the AT calibrator catalog, the AT20G survey, \citet{sadler_extragalactic_2008}, and this paper. The 20\,GHz variability is $\sim$10\% over 4 years, whilst the 95\,GHz flux shows an outburst and decline in flux in mid 2005. Figure \ref{fig:J002616-351249_WMAP} shows a five year flux history for this source as seen by the WMAP mission. The variability over the 5 years of the WMAP mission is 10, 22, 18 and 38\% at 23, 33, 41, and 61\,GHz respectively. The SUMSS and NVSS fluxes for this source are in agreement with the fitted radio spectrum of the AT20G, 40, and 95\,GHz observations, which given the large amount of time between the SUMSS, NVSS and AT20G observations suggests that there is no significant variability below the spectral peak. 

The lack of variability below the spectral peak, and an increasing amount of variability above the peak, suggests that we are seeing beamed emission and that there are blobs of material moving in and out of the beamed component. Beamed emission and a large 95\,GHz variability is not indicative of a genuine GPS source. This object may be a beamed source rather than a genuine GPS galaxy.

\subsection{AT20G J004417-375259 - a $z=0.48$ non-GPS galaxy}
\begin{figure}
 \centering
 \subfigure[]{\epsfig{file=C1392/J0044-3752.ps,bb=80 30 580 720, height=0.95\linewidth,angle=-90,clip=}}
 \subfigure[]{\epsfig{file=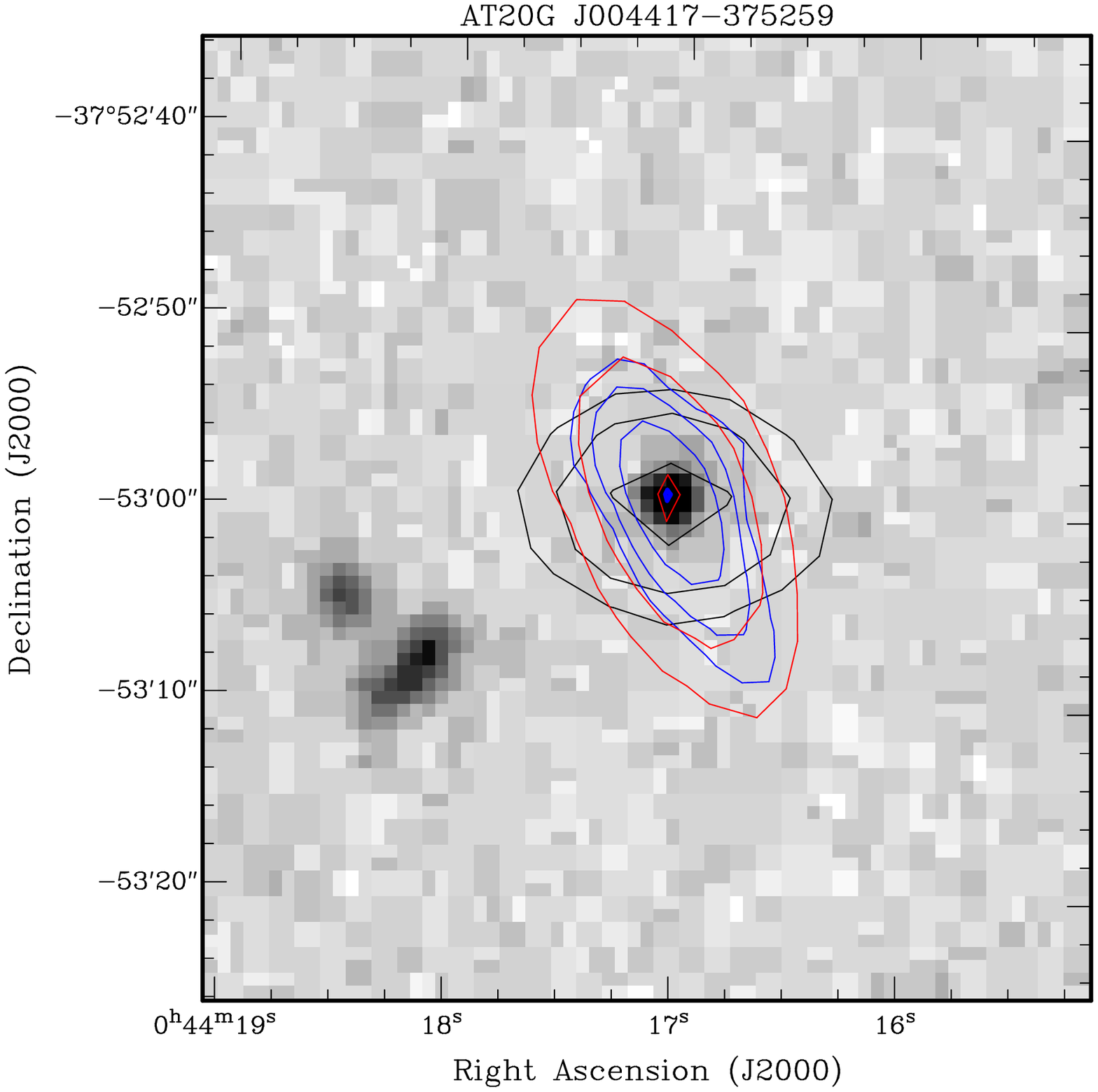, width=0.8\linewidth}}
 \caption{{\bf(a)} The ESO NTT spectrum of AT20G\,J004417-375259. Flux units are arbitrary. H$\beta$, O[III], and MgII are seen in emission, along with the high ionisation states of [NeIII] and [NeV]. The redshift is 0.483.{\bf (b)} SuperCOSMOS optical image of AT20G J004417-375259 overlaid with AT20G contours. Contours are: Black - 20\,GHz at 15, 30, and 60\,mJy/beam, Blue - 8.6\,GHz at 10, 20, 40, and 80\,mJy/beam, and Red - 4.8\,GHz at 15, 30 and 60\,mJy/beam. The radio source is unresolved at all frequencies.}
 \label{fig:J004417-375259_overlay}
 \label{fig:J004417-375259}
\end{figure}

This source is identified as a B$_J$ = 20.3\,mag galaxy in the SuperCOSMOS database. At a measured redshift of 0.48, the optical colours of B-R$_C$ = 1.25\,mag and R$_C$-I$_C$= 0.31\,mag, are much bluer than would be expected for a normal E or S0 galaxy. The presence of broad lines in the NTT spectrum (see fig \ref{fig:J004417-375259}(a)), indicates that the excess blue emission is due to an AGN. The broad emission lines and non-stellar blue continuum are typical of a QSO or broad-line radio galaxy.

J004417-37525 is present in the PMN catalog as PMN\,J0044-3752 with a 4.85\,GHz flux density of $54\pm9$\,mJy. This source shows no significant variability at 4.85\,GHz and is only 7.7\% variable at 20\,GHz over a period of 3 years. 

A radio-optical overlay is shown in Figure \ref{fig:J004417-375259_overlay}(b). There is no evidence for extended flux in any of the three frequencies within the AT20G. The QSO-like spectrum indicates that this source may not be a genuine GPS galaxy.

\subsection{AT20G J004905-552110 - a $z=0.062$ non-GPS galaxy}
\begin{figure}
 \centering
 \subfigure[]{\epsfig{file=C1392/J0049-5521.ps,bb=80 30 580 720, height=0.95\linewidth,angle=-90,clip=}}
 \subfigure[]{\epsfig{file=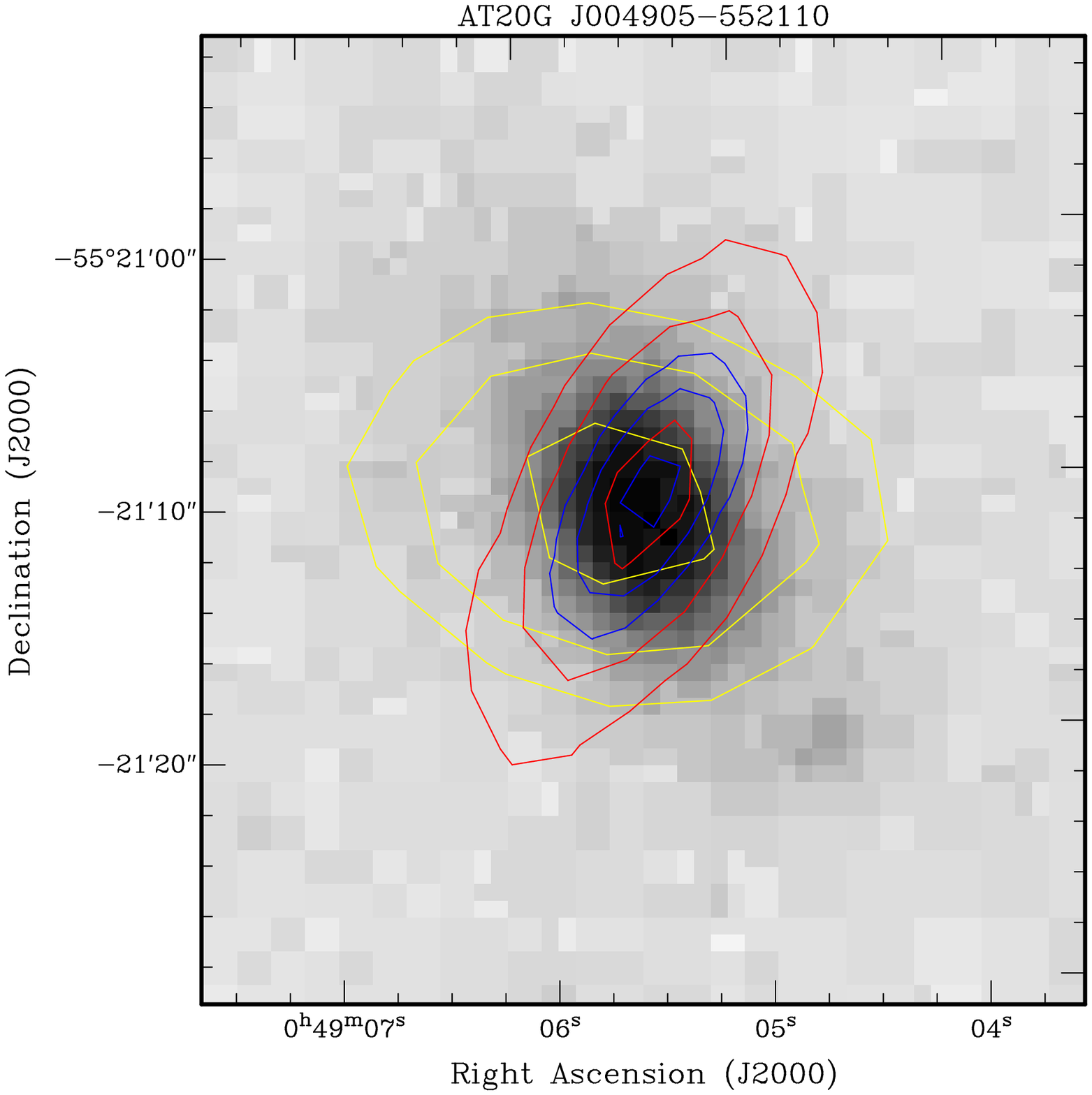, width=0.8\linewidth}}
 \caption{{\bf (a)} The ESO NTT spectrum of AT20G\,J004905-552110. Both emission and absorption are present, giving a redshift of 0.062. The spectrum is dominated by stellar features, indicative of a late type galaxy. The unlabeled absorption feature near 7630\AA is from atmospheric oxygen. {\bf (b)} SuperCOSMOS blue optical image of AT20G J004905-552110 overlaid with AT20G contours. Contours are: Yellow - 20\,GHz at 15, 30, and 60\,mJy/beam, Blue - 8.6\,GHz at 15, 30, and 60\,mJy/beam, and Red - 4.8\,GHz at 12, 25, and 50\,mJy/beam. }
 \label{fig:J004905-552110_overlay}
 \label{fig:J004905-552110}
\end{figure}

This source is identified as a B$_J$ = 16.6\,mag galaxy within the SuperCOSMOS database. The SuperCOSMOS blue image shows a bright core with some surrounding extended emission, consistent with a Seyfert type galaxy as shown in Figure \ref{fig:J004905-552110_overlay}(b). The ESO NTT spectrum in Figure \ref{fig:J004905-552110}(a) shows a redshift of 0.062 with strong emission lines superimposed on a strong-lined stellar continuum typical of an early-type galaxy. The optical colours of this galaxy are B-R$_C$ = 0.95\,mag and R$_C$-I$_C$= 0.71\,mag, which is consistent with a z$\sim$0 E or S0 galaxy with some excess blue emission due to either an AGN or recent star formation. In the SuperCOSMOS image in Figure \ref{fig:J004905-552110_overlay}(b) there is a possible bright spot towards the southwest of the image, which may be a satellite galaxy that is merging with the host galaxy and could contribute to the star formation or AGN activity.

With the exception of SUMSS there are no other radio observations of this source listed within any of the databases that are incorporated into NED or VizieR. There is no evidence for any extended emission or jet like structures at any of the AT20G frequencies. A 20\,GHz variability index of 22\% over two years suggests that this source may not be a genuine GPS source, and the radio SED (Figure \ref{fig:radio_sed}1(a)) shows some signs that the location of the spectral peak may also be variable. This source is probably not a genuine GPS galaxy.

\subsection{AT20G J005427-341949 - a $z=0.11$ GPS galaxy}
\begin{figure}
 \centering
 \subfigure[]{\epsfig{file=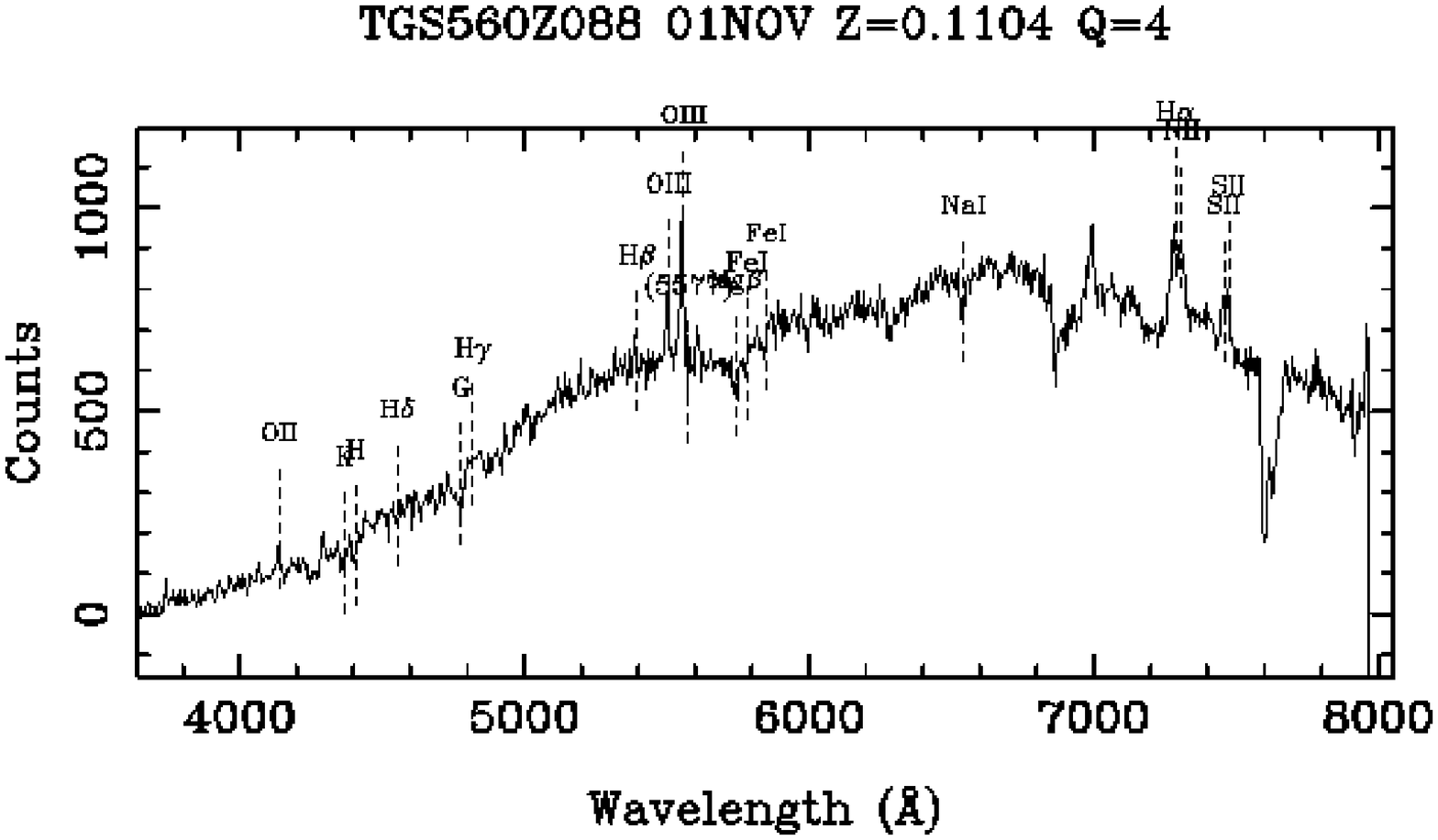, bb= -10 205 615 540, width=0.9\linewidth,clip=}}
 \subfigure[]{\epsfig{file=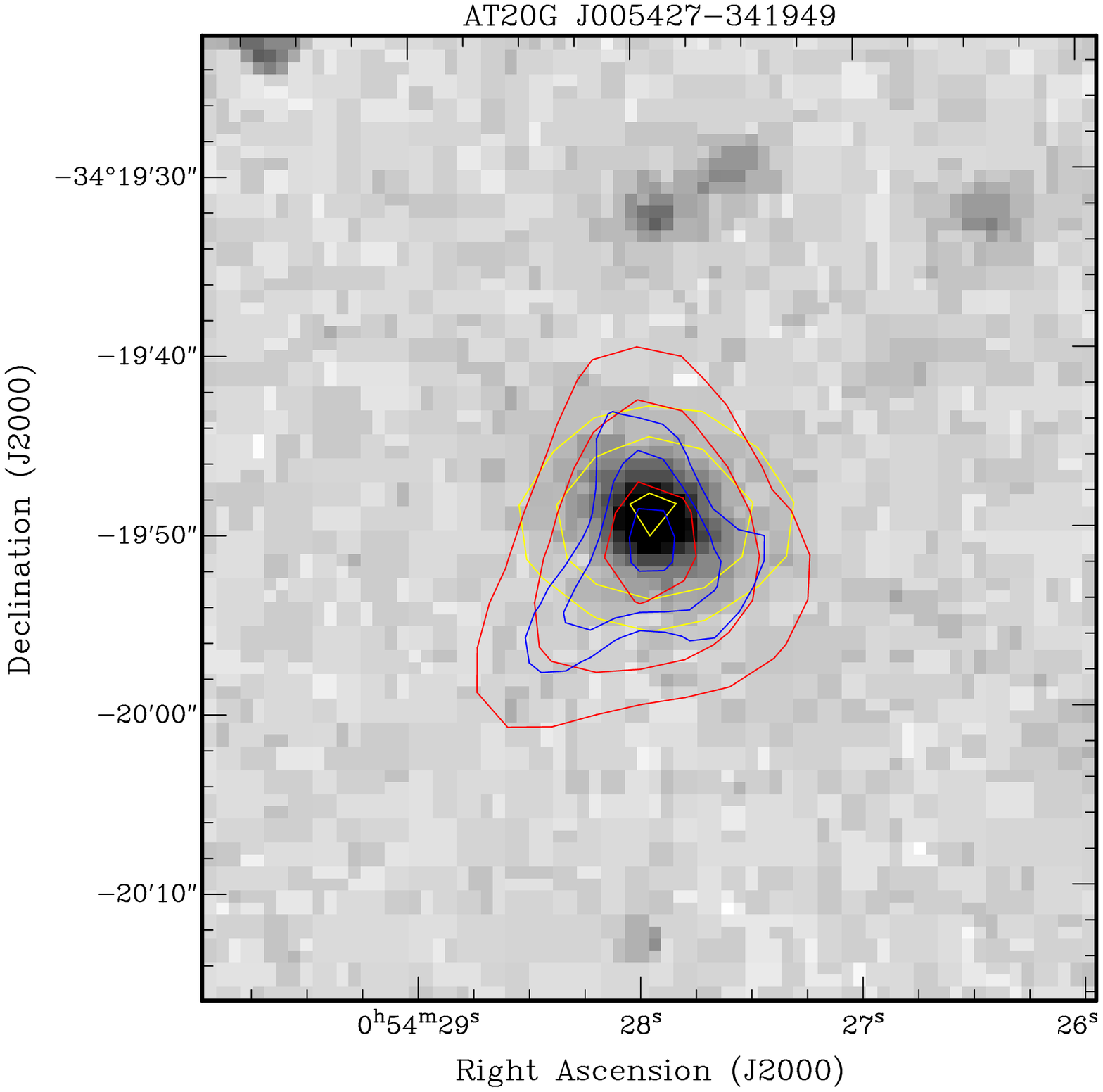, width=0.65\linewidth}}
 \subfigure[]{\epsfig{file=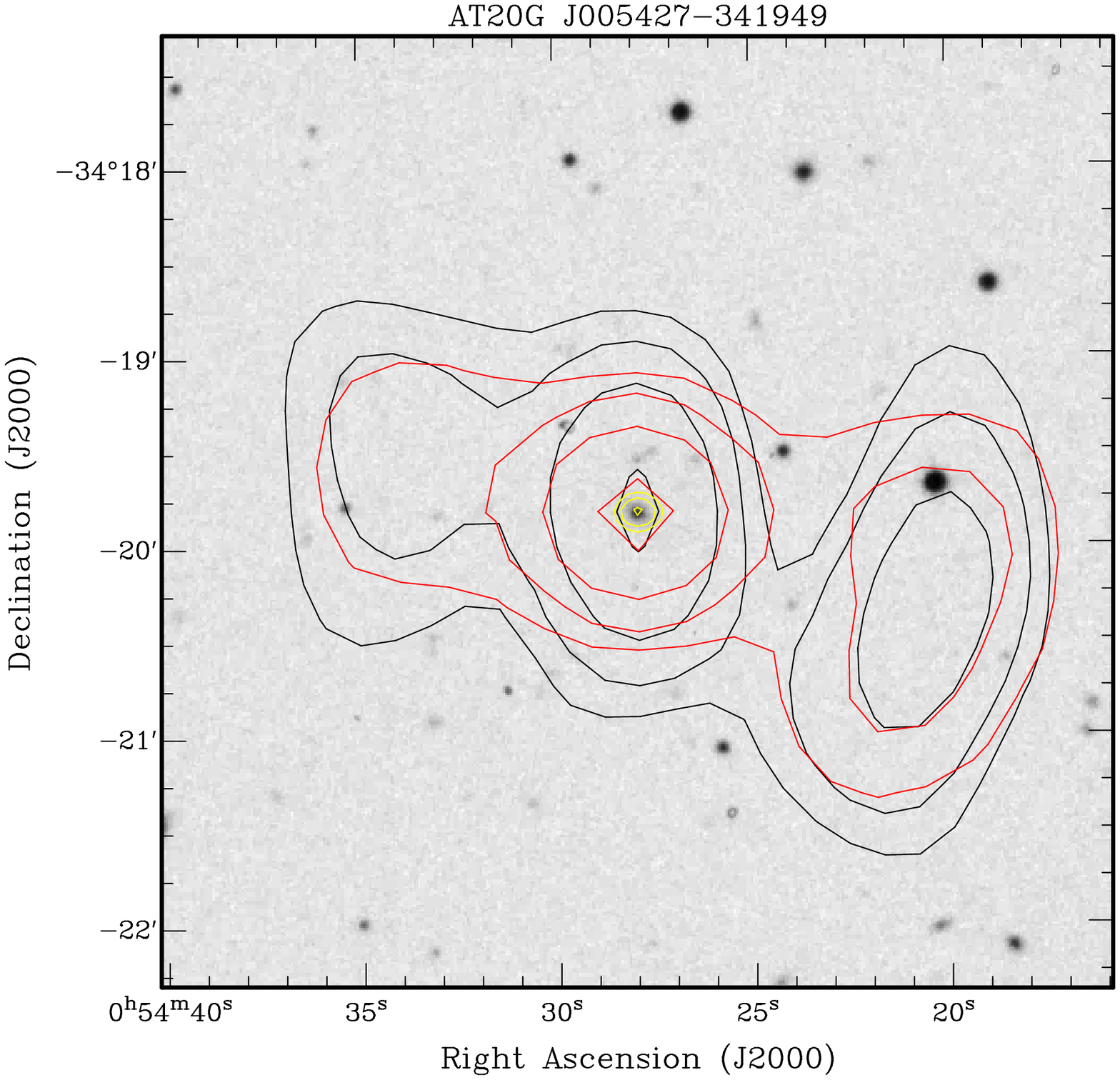, width=0.65\linewidth}}
 \caption{AT20G J005427-341949. {\bf (a)} 2dFGRS spectrum. H$\alpha$, [SII] and [OIII] are visible in emission, and H, K and Fe are seen in absorption. The spectrum is dominated by starlight. The redshift is z=0.11. {\bf (b)} SuperCOSMOS blue image overlaid with 15, 30, and 60\,mJy/beam AT20G contours at 4.8\,GHz (red), 8.6\,GHz (blue), and 20\,GHz (yellow). The 4.8 and 8.6\,GHz contours suggest that the source may have some extended emission. {\bf (c)} As in (b) with AT20G 20\,GHz contours in yellow, SUMSS 843\,MHz contours in black at 4, 8, 16, and 32\,mJy/beam, and NVSS 1.4\,GHz contours in red at 4, 8, 16, and 32\,mJy/beam.}
 \label{fig:J005427-341949_2df}
\end{figure}

This source is identified as a B$_J$ = 18.0\,mag galaxy in the SuperCOSMOS database. The optical colours are B-R$_C$ = 1.05\,mag and R$_C$-I$_C$= 0.51\,mag which are consistent with a low redshift E or S0 type galaxy with excess blue emission from an AGN or star formation. The 2dF spectrum of this object in Figure \ref{fig:J005427-341949_2df}(a) shows OIII, SII and H$\alpha$ in emission, but almost no H$\beta$. This source is at a redshift of 0.11 \citep{colless_2df_2001}.

J005427-341949 is detected as an X-ray source in the ROSAT All-Sky Faint Source Catalog \cite[RASS-FSC][]{voges_rosat_2000} at $3.51\times10^{-2}$\,ct/s. This source is identified as being within a cluster of galaxies by \citet{tago_clusters_2006}, so it is possible that the X-ray emission may be related to the cluster rather than the galaxy. \citet{turriziani_roxa:new_2007} attribute the X-ray flux to the galaxy and list this source as being borderline between a radio galaxy and a flat spectrum radio quasar.

There is some evidence for extended emission in the 4.8 and 8.6\,GHz AT20G images as shown in Figure \ref{fig:J005427-341949_2df}(b). The sparse $(u,v)$ coverage of the observations means that even though extended emission is able to be detected, the structure cannot be determined, and the contours of Figure \ref{fig:J005427-341949_2df}(b) are not indicative of the true extended structure. The 843\,MHz SUMSS and 1.4\,GHz NVSS contours show a triple component source, with the central component coincident with the optical ID and AT20G radio positions (figure \ref{fig:J005427-341949_2df}(c)). The core has a rising spectral index of $\alpha_{0.8}^{1.4} = +0.8$ whilst the lobes have a steep spectral index of $\alpha_{0.8}^{1.4}\sim -1.1$. The steep spectral index of the lobes, and the fact that they are not detected at 20\,GHz indicates that they are no longer being powered, and that this source is probably restarted.

At a redshift of z=0.11 the linear scale is 1.98\,kpc/arcsec, and the projected size of the lobes is $\sim$ 155 and $\sim$180\,kpc for the East and West components respectively. The total projected linear size is $\sim$355\,kpc.

A 20\,GHz variability of 13\% over 2 years is consistent with that of a genuine GPS galaxy. The extended emission is possibly due to either the galaxy cluster, or the restarted nature of this source. This source is probably a genuine GPS galaxy.

\begin{figure}
 \centering
\epsfig{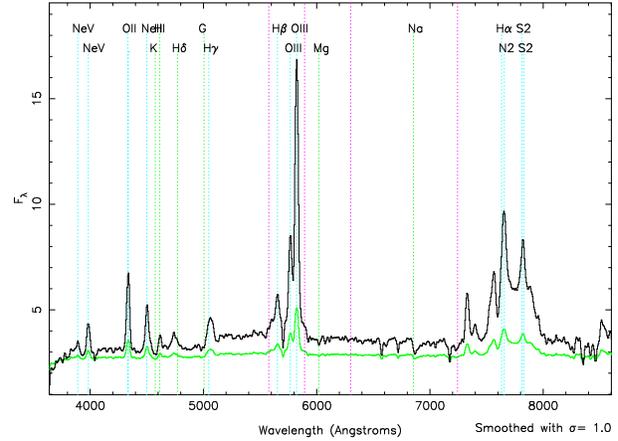}
 \caption[Spectrum of AT20G J010333-643907]{The ESO NTT spectrum of AT20G J010333-643907. The emission features are wide and strong and include higher ionisation states of [NeIII] and [NeV]. The unlabeled emission feature at around 7300\AA  is [OI]. The very wide emission feature of H$\alpha$ is suppressed by atmospheric absorption by oxygen at 7630\AA. This object is a redshift 0.163 emission line AGN.}
 \label{fig:J010333-643907_spectrum}
\end{figure}

\subsection{AT20G J010333-643907 - a $z=0.163$ GPS galaxy}
This source is identified as a B$_J$ = 18.6\,mag galaxy in the SuperCOSMOS database. The optical colours are B-R$_C$ = 1.15\,mag and R$_C$-I$_C$= 0.71\,mag which are consistent with a low redshift E or S0 type galaxy with excess blue emission from an AGN or star formation. The NTT spectrum of this object in Figure \ref{fig:J010333-643907_spectrum} and shows strong emission in H$\alpha$, [OII], [OIII], [OI], [NeV], and [NeIII]. This is an AGN at a redshift of 0.163. 

This source has been observed several times over the course of the AT20G survey and related projects as summarised in Table \ref{tab:010333-643907}. The variability is \<6-7\% over a period of 1-2 years at each of the frequencies. The radio spectrum of this source has a peak frequency near 84\,GHz.

\begin{figure}
 \centering
 \subfigure[]{\epsfig{file=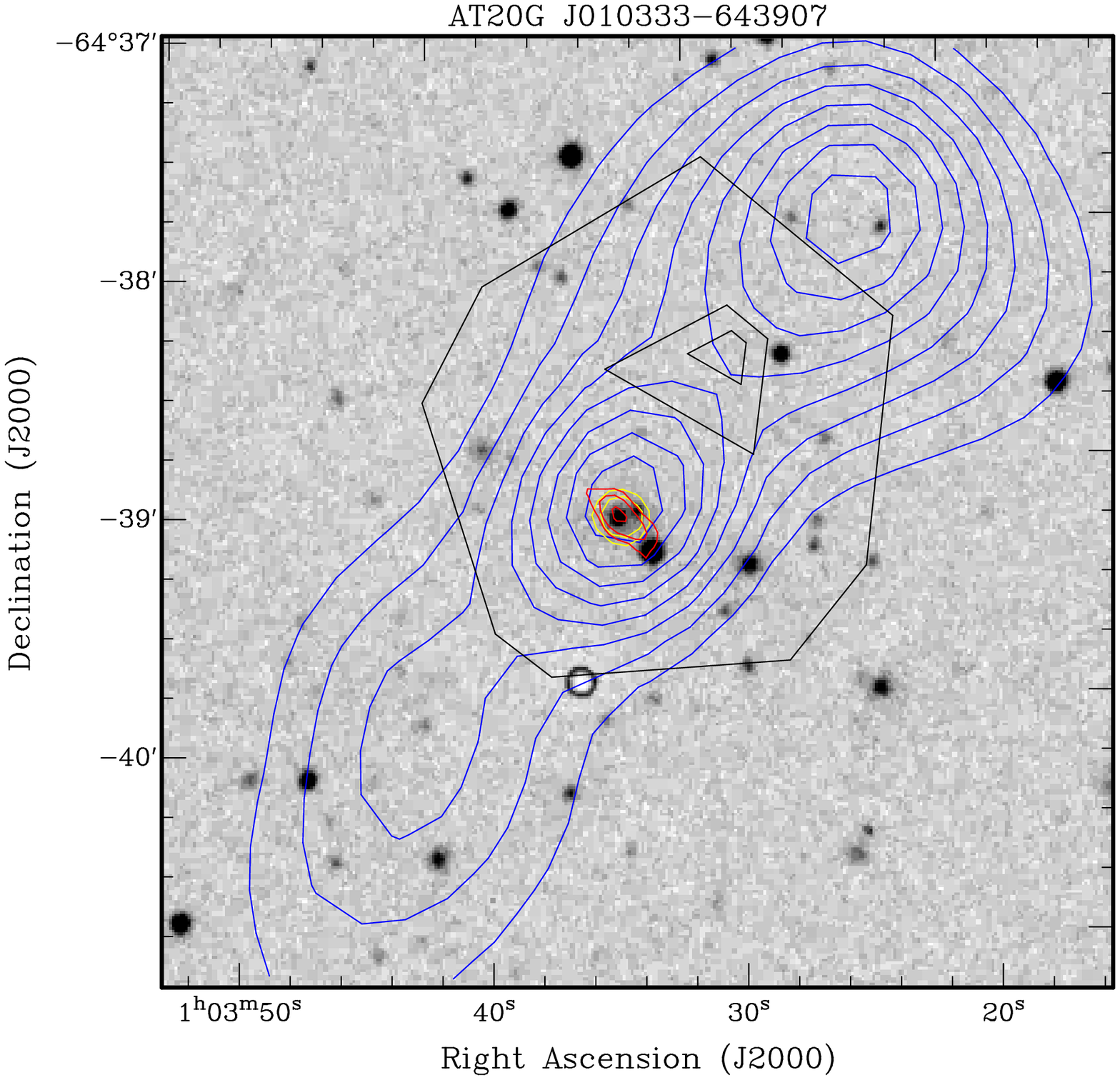, width=0.65\linewidth}}
 \subfigure[]{\epsfig{file=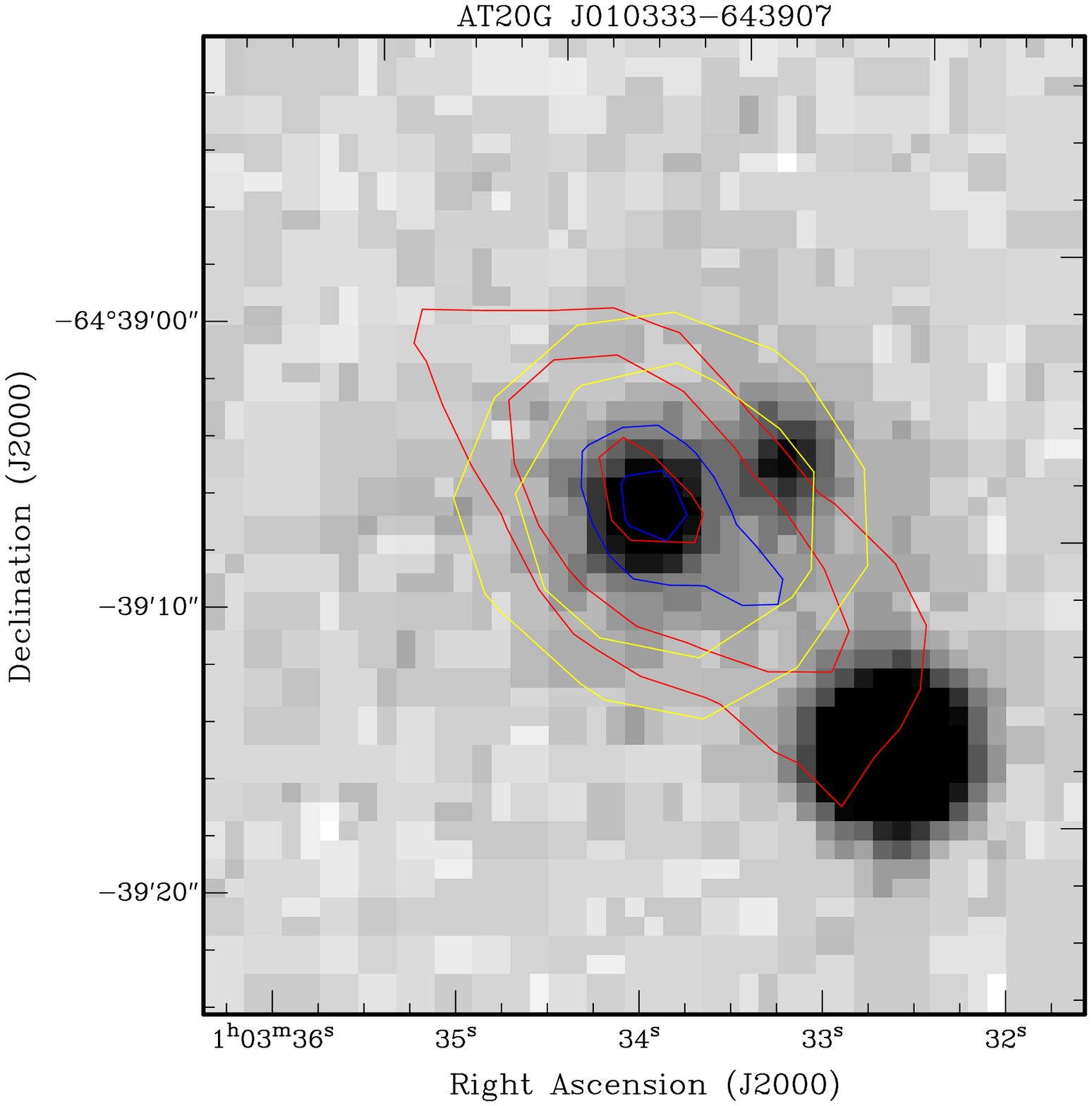, width=0.65\linewidth}}
 \caption[Radio optical overlays for AT20G J010333-643907]{SuperCOSMOS Radio-Optical overlays for AT20G J010333-643907. {\bf(a)} SUMSS contours in blue (33-150\,mJy/beam in steps of 16\,mJy/beam), PMN contours in black (300, 330 and 335\,mJy/beam), AT20G 4.8\,GHz in red (smaller contours, at 40, 80 and 160\,mJy/beam), and 20\,GHz in yellow (60 and 120\,mJy/beam). The PMN flux includes contributions from both the core and both lobes, whilst the AT20G fluxes represent only the core flux. {\bf(b)} Zoom of central region with AT20G contours only, contours as in (a) but with 8.6\,GHz contours in blue at 80 and 160\,mJy/beam. The elongated contours at 4.8 and 8.6\,GHz are due to an elliptical beam shape.}
 \label{fig:J010333-643907_overlay}
\end{figure}

Figure \ref{fig:J010333-643907_overlay}(a) shows a SUMSS-SuperCOSMOS radio optical overlay for this source. The SUMSS counterpart shows a strong radio core with two lobes. One of the lobes is equal in brightness to the core whilst the other is approximately half the power. The linear scale at z=0.163 is 2.77\,kpc/arcsec giving a core-to-lobe separation of 235 and 225 kpc for the fainter and brighter lobe respectively. The projected linear size is $\sim$460\,kpc. The PMN counterpart is $395\pm 22$\,mJy and is halfway between the core and strong lobe seen in the SUMSS image. Figure \ref{fig:J010333-643907_overlay}(b) shows an AT20G-SuperCOSMOS radio optical overlay. This source is unresolved in all of the AT20G images, and the elongation seen at 4.8 and 8.6\,GHz is due to an elliptical beam shape. 

There is no counterpart to either of the SUMSS lobes in the AT20G catalog. This lack of emission could indicate that the lobes seen at 843\,MHz are no longer being powered and the higher energy electron population has been depleted. A restarted radio galaxy would possess a young radio source which would explain the high frequency GPS nature of the core emission. A rest-frame spectral peak above 100\,GHz is also consistent with a young or recently re-started radio source. The possible companion to the North-West of this object, as seen in Figure \ref{fig:J010333-643907_overlay}(b), is a likely merger candidate which is triggering the current phase of radio emission. With a linear scale of 2.77\,kpc/arcsec the companion is at least 21kpc from the host galaxy, if it is at the same redshift. A spectrum of the possible companion would be very useful.

This source is a genuine GPS galaxy and is possibly a restarted radio galaxy.

\begin{table}
\centering
\caption{Data summary for J010333-643907}
\begin{tabular}{cccc}
\hline
\hline
Data Source & $\nu$ & Flux Density & Obs \\
            & GHz   & mJy          & Year\\ 
\hline
\citet{sadler_properties_2006} & 4.8 & $173\pm 5$  & 2003.92 \\
AT20G                          & 4.8 & $187\pm 9$  & 2005.83 \\
\hline
\citet{sadler_properties_2006} & 8.6 & $215\pm 9$  & 2003.92 \\
AT20G                          & 8.6 & $253\pm 12$ & 2005.83 \\
\hline
\citet{sadler_properties_2006} & 18  & $200\pm 14$ & 2003.83 \\
\citet{ricci_first_2004}       & 18  & $290\pm 58$ & 2002.83 \\
\hline
AT20G                          & 20  & $323\pm 16$ & 2005.75 \\
This paper                     & 20  & $311\pm 21$ & 2007.83 \\
\hline
\end{tabular}
\label{tab:010333-643907}
\end{table}

\subsection{AT20G J011102-474911 - a $z=0.154$ GPS galaxy}
\begin{figure}
 \centering
 \epsfig{file=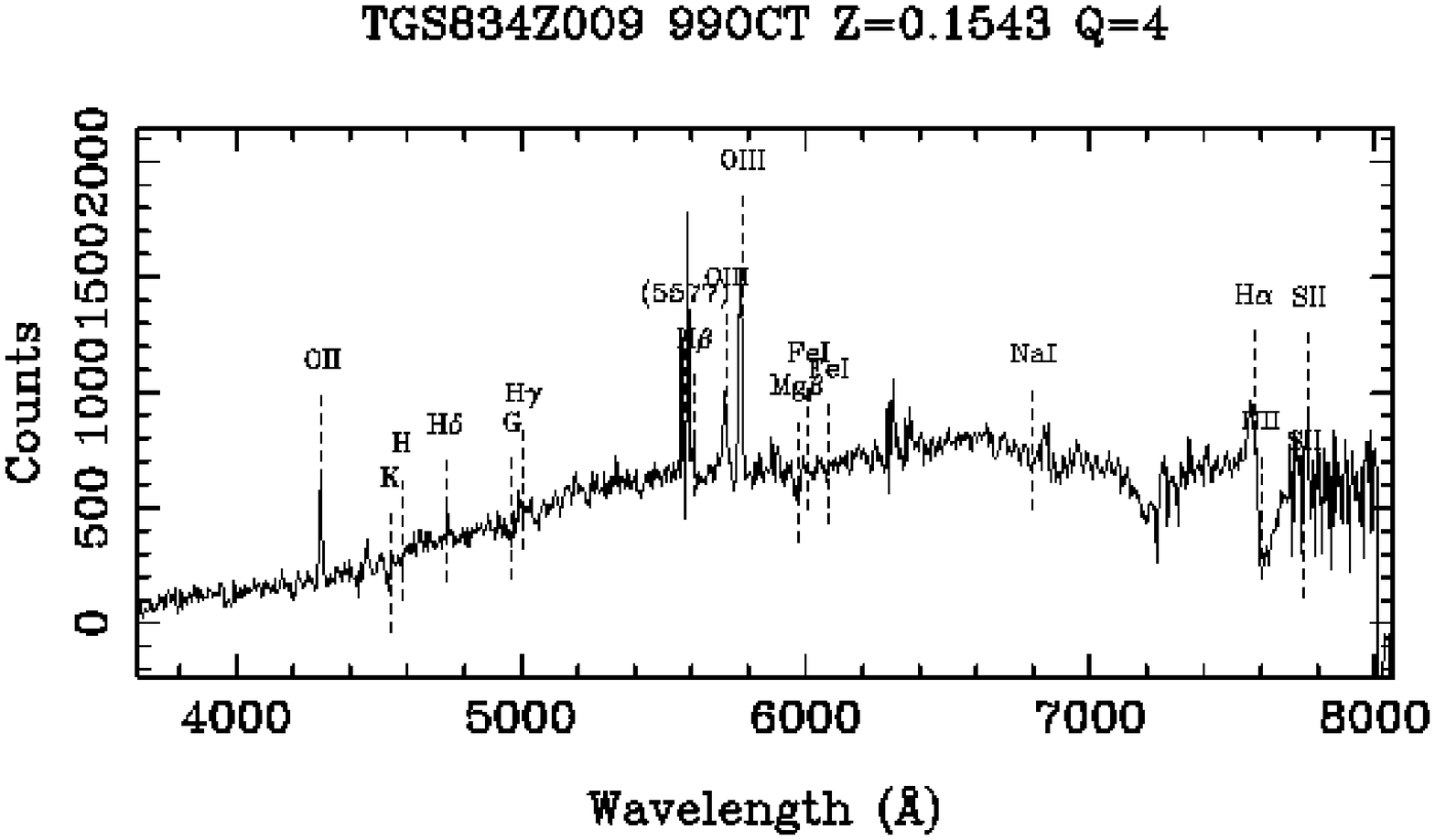,bb=-10 205 620 530, width=0.95\linewidth,clip=}
 \caption{2dFGRS spectrum of AT20G\,J011102-474911 showing emission in H$\alpha$, [OIII], and [OII] and absorption of Na, Fe, Mg. The redshift is 0.154}
 \label{fig:J011102-474911_2df}
\end{figure}

This source is identified as a B$_J$ = 18.6\,mag galaxy in the SuperCOSMOS database. The optical colours are B-R$_C$ = 1.15\,mag and R$_C$-I$_C$= 0.61\,mag which are consistent with a low redshift E or S0 type galaxy with excess blue emission from an AGN or star formation. The 2dFGRS spectrum of this source is shown in Figure \ref{fig:J011102-474911_2df} where strong emission in H$\alpha$, [OIII], and [OII] can be seen. These emission lines are typical of an emission-line AGN with a redshift of 0.154.

This source was not detected at 95\,GHz and thus it is not clear where the spectrum turns over. It is possible that this source remains flat above 95\,GHz as it does not require a steep $40-95$\,GHz spectral index to avoid detection.

In 2005 \citet{sadler_extragalactic_2008} observed this source twice at 95\,GHz on consecutive days and found the flux to be \<99\,mJy and \<64\,mJy. Observations at 20\,GHz within a few days of the 95\,GHz observations show a flux density of $71\pm 4$\,mJy. The 95\,GHz upper limits are consistent with the measurements of this paper, and the revised 20\,GHz variability is still \<6\% over 3 years. The new observations strengthen the evidence that this source is not variable. The lack of variability is consistent with a genuine GPS source.

\subsection{AT20G J012714-481332 - possible GPS galaxy}
\begin{figure}
 \centering
 \epsfig{file=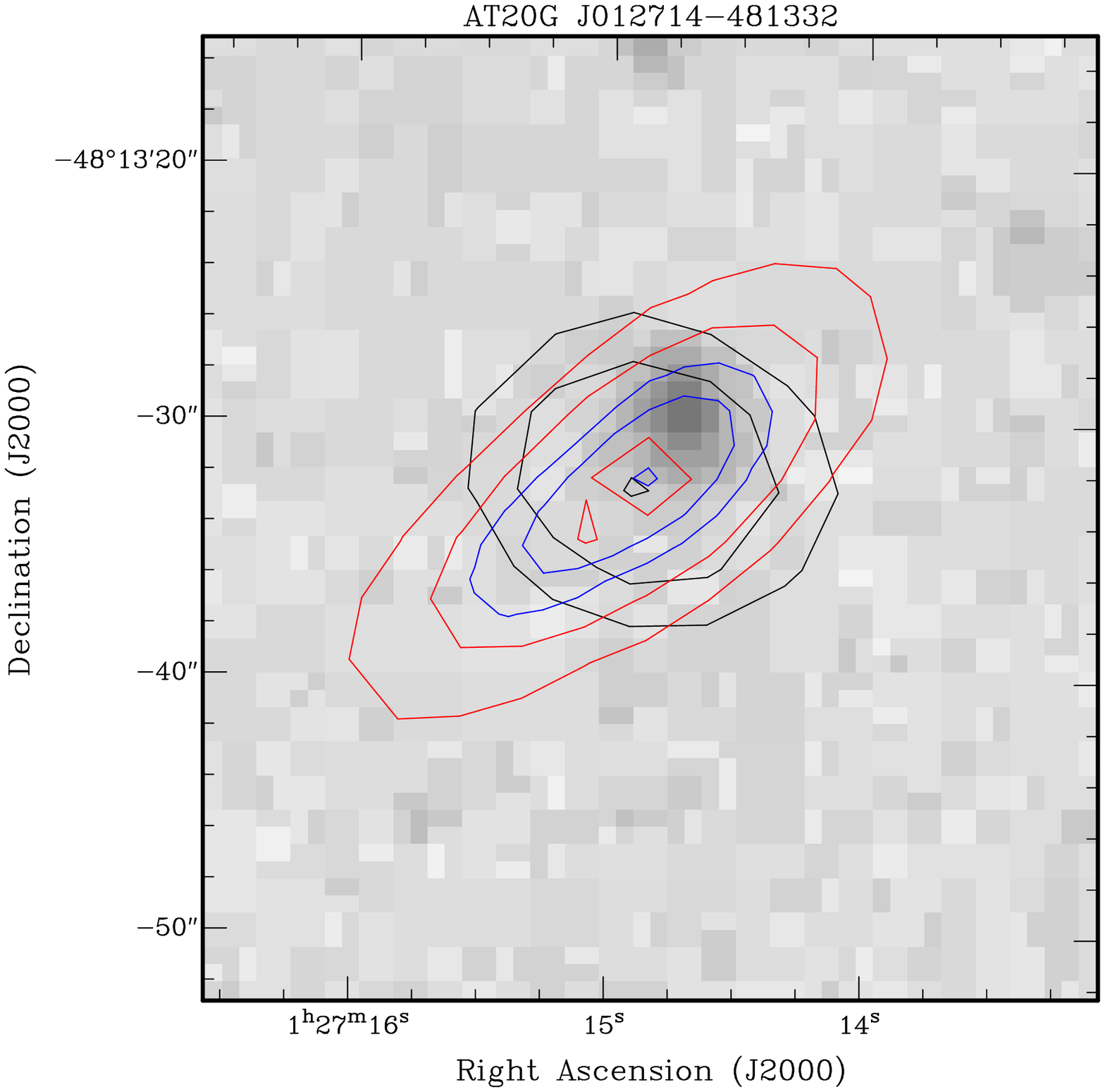, width=0.8\linewidth}
 \caption[Radio optical overlay for AT20G J012714-481332]{SuperCOSMOS blue image of J012714-481332 overlaid with AT20G radio contours at 4.8\,GHz (red at 20, 40, and 80\,mJy/beam), 8.6\,GHz (blue at 40, 80, and 160\,mJy/beam) and 20\,GHz (black at 50, 100, and 200\,mJy/beam). The separation between the radio and optical centroids (2.5, 2.6, and 2.7\,arsec at 4.8, 8.6 and 20\,GHz respectively) is too large for a confident ID to be made.}
 \label{fig:J012714-481332_overlay}
\end{figure}

This source has no optical counterpart in the SuperCOSMOS database. There is a galaxy 2.7\,arcsec from the 20\,GHz radio centroid as shown in Figure \ref{fig:J012714-481332_overlay}.The AT20G 4.8 and 8.6\,GHz radio centroids are 2.5 and 2.6\,arcsec distant from the galaxy so the association is uncertain. Further work is needed if this galaxy is to be accepted as the correct ID.

This source is present in the PMN catalog and was observed by \citet{healey_crates:all-sky_2007} and \citet{sadler_extragalactic_2008}, as summarised in table \ref{tab:data_J012714-481332}. This source has varied by 17\% at 4.85\,GHz over 14 years. The 8.6\,GHz flux density of \citet{healey_crates:all-sky_2007} has no quoted error, but assuming an uncertainty of up to 10\%, this corresponds to a variability of 18\% in just a year. When the measurements of \citet{sadler_extragalactic_2008} are taken into account the revised 20\,GHz variability index becomes 11\% over three years. The 95\,GHz variability is \<6\% over two years.

The lack of an optical ID suggests that this source is a distant galaxy, however a spectral turnover of 89\,GHz would suggest a nearby young galaxy. The low frequency variability is larger than normal for a GPS source, however the 20 and 95\,GHz variability is consistent with a genuine GPS galaxy.

\begin{table}
\centering
\caption{Data summary for J012714-481332}
\label{tab:data_J012714-481332}
\begin{tabular}{cccc}
\hline
\hline
Data Source & $\nu$ & Flux Density & Obs \\
            & GHz   & mJy          & Year\\ 
\hline
PMN                               & 4.85& $96\pm 10$  & 1990 \\
AT20G                             & 4.8 & $140\pm 7$  & 2004.83 \\
\hline
\citet{healey_crates:all-sky_2007}& 8.6 & $121.9$     & 2005.99 \\
AT20G                             & 8.6 & $181\pm 9$  & 2004.83 \\
\hline
AT20G                             & 20  & $237\pm 12$ & 2004.75 \\
\citet{sadler_extragalactic_2008} & 20  & $200\pm 11$ & 2005.5 \\
This paper                        & 20  & $174\pm 12$ & 2007.83 \\
\hline
\citet{sadler_extragalactic_2008} & 95  & $251\pm 30$ & 2005.5 \\
\citet{sadler_extragalactic_2008} & 95  & $232\pm 27$ & 2005.5 \\
This paper                        & 95  & $187\pm 16$ & 2007.83 \\
\hline
\end{tabular}
\end{table}

%%%%%%%%%%%%%%%%%%%%%%%%%%%%%%%%%%%%%%%%%%%%%%%%%%%%%%%
\begin{table}
\centering
\caption{Data summary for J012820-564939}
\label{tab:data_J012820-564939}
\begin{tabular}{cccc}
\hline
\hline
Data Source & $\nu$ & Flux Density & Obs \\
            & GHz   & mJy          & Year\\ 
\hline
\citet{drake_radio-excess_2003}   & 4.79& $121.0\pm 1.2$ & 2003?\\
PMN                               & 4.85& $141\pm 11$    & 1990 \\
AT20G                             & 4.8 & $130\pm 7$     & 2005.83 \\
\hline
\citet{drake_radio-excess_2003}   & 8.6 & $199.3\pm 1.7$ & 2003?\\
AT20G                             & 8.6 & $141 \pm 7$    & 2005.83 \\
\hline
AT20G                             & 20  & $189\pm 10$ & 2005.75 \\
This paper                        & 20  & $131\pm  9$ & 2007.83 \\
\hline
\end{tabular}
\end{table}

\subsection{AT20G J012820-564939 - a $z=0.066$ GPS galaxy}
This source is identified as a B$_J$ = 16.5\,mag galaxy within the SuperCOSMOS database. The optical colours are B-R$_C$ = 0.85\,mag and R$_C$-I$_C$= 0.81\,mag, which is consistent with a low redshift E or S0 galaxy with excess blue emission from an AGN or star formation. This source is in the sample of Radio-Excess IRAS Galaxies (REIG) by \citet{drake_radio-excess_2003} who measure a redshift of 0.066 but do not show a spectrum.

This source is in the PMN catalog and was also observed at 4.79 and 8.6\,GHz by \citet{drake_radio-excess_2003} as is summarised in table \ref{tab:data_J012820-564939}. There is no detectable variability at 4.8\,GHz but there is a change of 17\% at 8.6 and 20\,GHz over two years. The variability is not so high as to exclude a GPS classification of this source.
%%%%%%%%%%%%%%%%%%%%%%%%%%%%%%%%%%%%%%%%%%%%%

\subsection{AT20G J180859-832526 - a $z\sim0.5$ GPS galaxy}
This source is identified as a B$_J$ = 21.6\,mag galaxy within the SuperCOSMOS database. Although it is not the faintest source in the sample it does not have any measured R$_F$ or I$_C$ magnitudes so no determination of colour can be made. The redshift of this source has been estimated to be z$\sim$0.5.

This source was observed in the AT20G survey in 2005 at $95\pm 5$\,mJy and again in 2006 at $95\pm 5$\,mJy giving a revised variability of \<6\% over 2 years. The radio spectrum of this source has an optically thick fitted spectral index of $\alpha=+0.16$, which takes it outside of the original $\alpha>+0.2$ spectral index selection cutoff. There is nothing to exclude this source from being a genuine GPS source.
%%%%%%%%%%%%%%%%%%%%%%%%%%%%%%%%%%%%%%%%%%%%%%%%

\begin{figure}
\centering
 \epsfig{file=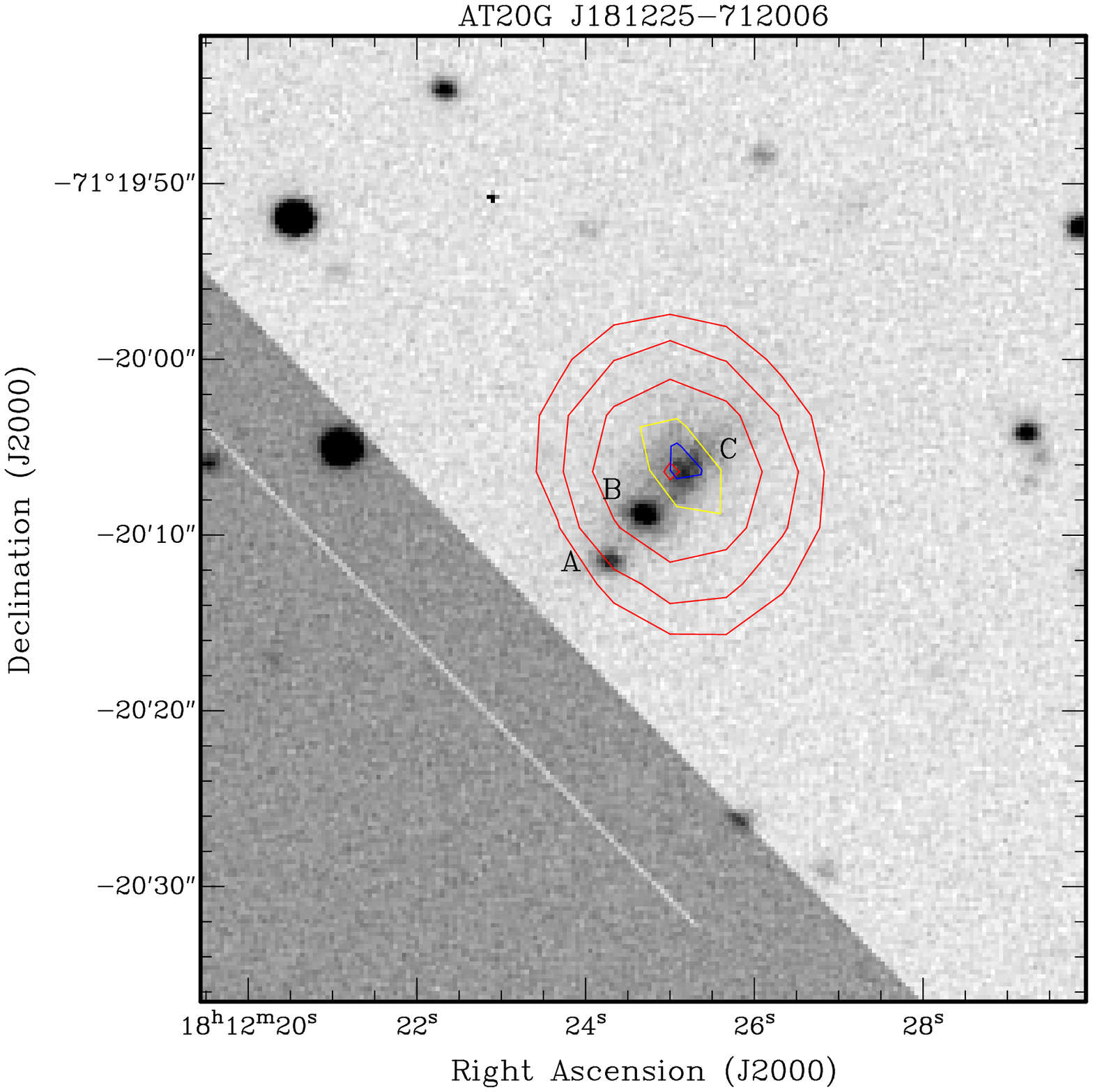, width=0.8\linewidth}
 \caption[Radio optical overlay for AT20G\,J181225-712006]{The source AT20G\,J181225-712006 with radio contours overlaid on an NTT optical R band image. Three optical galaxies are identified in the NTT acquisition image, with radio contours from the AT20G overlaid. Red contours are 20\,GHz at 5, 10, 20 and 40\,mJy/beam. The yellow and blue contours are 4.8 and 8.6\,GHz respectively and are both at 30\,mJy/beam.}
 \label{fig:J181225-712006_detail}
\end{figure}
\subsection{AT20G J181225-712006 - a $z=0.199$ GPS galaxy, possibly interacting}
\label{sec:J181225-712006}
A SuperCOSMOS blue image of this object is indicative of a B$_J$ = 18.4\,mag edge-on spiral galaxy and is identified as such within the catalog. The optical colours are B-R$_C$ = 1.65\,mag and R$_C$-I$_C$= 0.51\,mag, which is consistent with a low redshift E or S0 galaxy. 

\begin{figure}
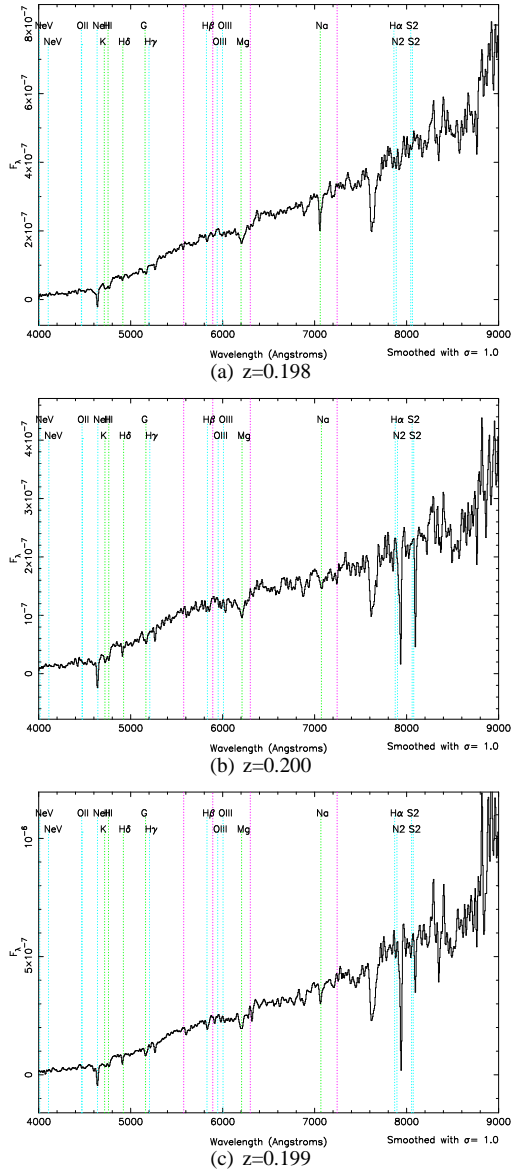

 \centering
 \subfigure[z=0.198]{\epsfig{file=C1392/J1812-7120a.ps,bb=80 30 580 740, totalheight=0.8\linewidth, angle=-90,clip=}}
 \subfigure[z=0.200]{\epsfig{file=C1392/J1812-7120b.ps,bb=80 30 580 740, totalheight=0.8\linewidth, angle=-90,clip=}}
 \subfigure[z=0.199]{\epsfig{file=C1392/J1812-7120c.ps,bb=80 30 580 740, totalheight=0.8\linewidth, angle=-90,clip=}}
 \caption[Spectra of the three components of AT20G\,J181225-712006]{Spectra of the three components of AT20G\,J181225-712006. In each case Ca -K, -H, Mg and Na are seen in absorption giving an average redshift of 0.199. The apparent absorption lines near the H$\alpha$ and [SII] lines in the lower two plots are the result of cosmic rays in the subtracted sky template. The unidentified absorption line near 7630\AA is atmospheric oxygen.}
 \label{fig:J181225-712006_spectra}
\end{figure}

The acquisition images that were taken during the ESO NTT observations show a more complex morphology. As is shown in Figure \ref{fig:J181225-712006_detail} there are three components to this object. As the three components are aligned linearly, it was possible to include each of them in the slit of the spectrograph and obtain separate spectra. The spectra are shown in Figure \ref{fig:J181225-712006_spectra}(a)-(c). Only absorption features are present and the average redshift is 0.199. An angular separation of approximately 2.5\,arcsec at this redshift corresponds to a projected linear separation of 8.1\,kpc between the central and outer components which, given the common distance to each, indicates that they are not only nearby to each other, but are probably interacting. With a redshift of z = 0.199, the optical colours suggest that there is some AGN or star formation activity which could be due to a recent interaction.

A radio optical overlay of this source is shown in Figure \ref{fig:J181225-712006_detail}. In each of the 5, 8, and 20\,GHz observations from the AT20G the radio emission is seen to be coincident with the faintest and most diffuse of the three components (C in fig. \ref{fig:J181225-712006_detail}). If we are witnessing the merger of two galaxies then it is tempting to suggest that the radio emission is the result of the merger process, however if this were the case then the central, brighter, galaxy would normally be expected to host the AGN. This source is present within the 2 Micron All Sky Extended source catalog \citep[2MASX][]{jarrett_2mass_2000} but little else is known about this object.

%%%%%%%%%%%%%%%%%%%%%%%%%%%%%%%%%%%%%%%%%%%%%%%%

\subsection{AT20G J181857-550815 - a $z=0.073$ GPS galaxy}
\begin{figure}
 \centering
 \subfigure[]{\epsfig{file=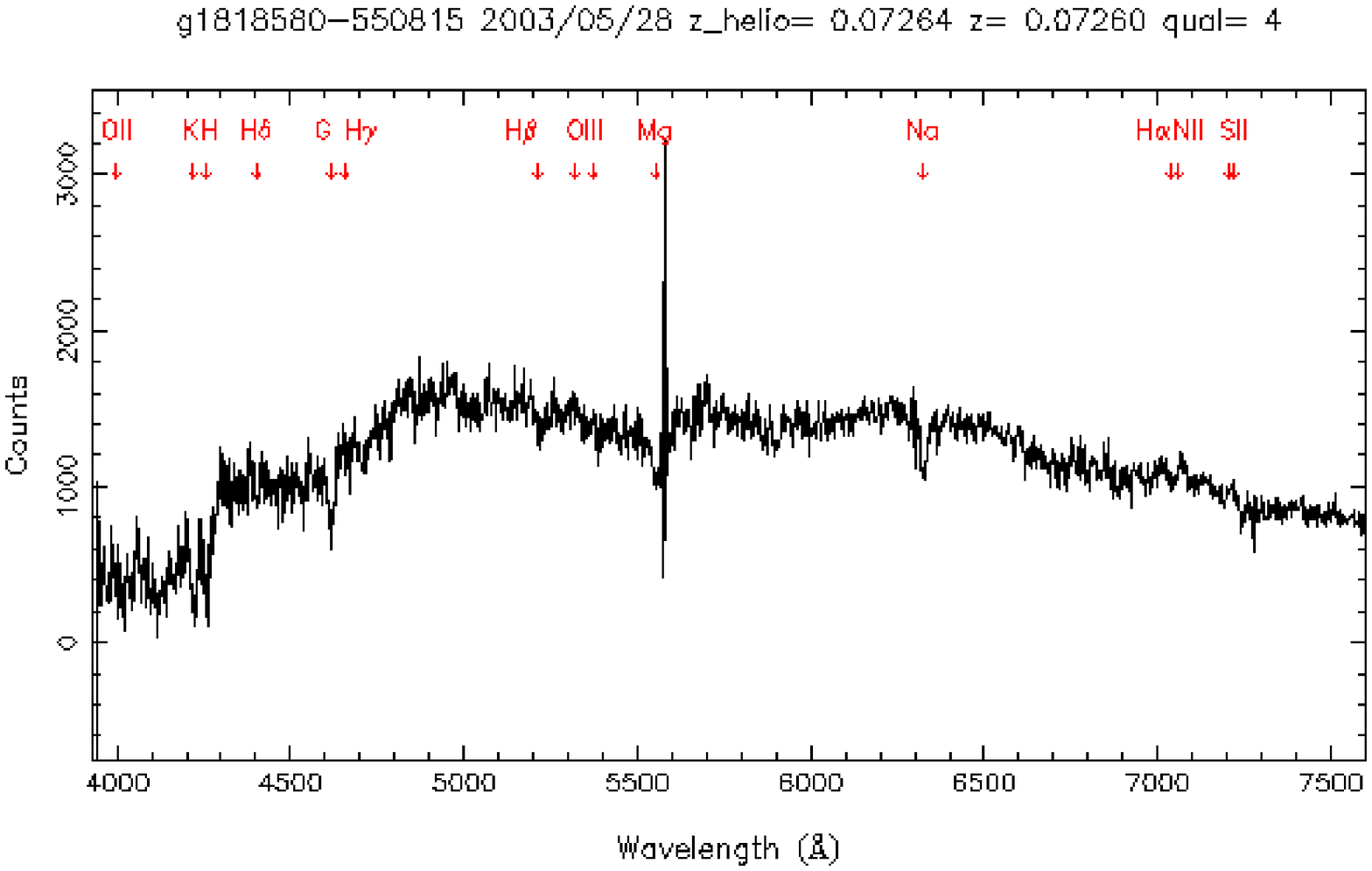, bb= -10 205 585 545,width=0.95\linewidth,clip=}}
 \subfigure[]{\epsfig{file=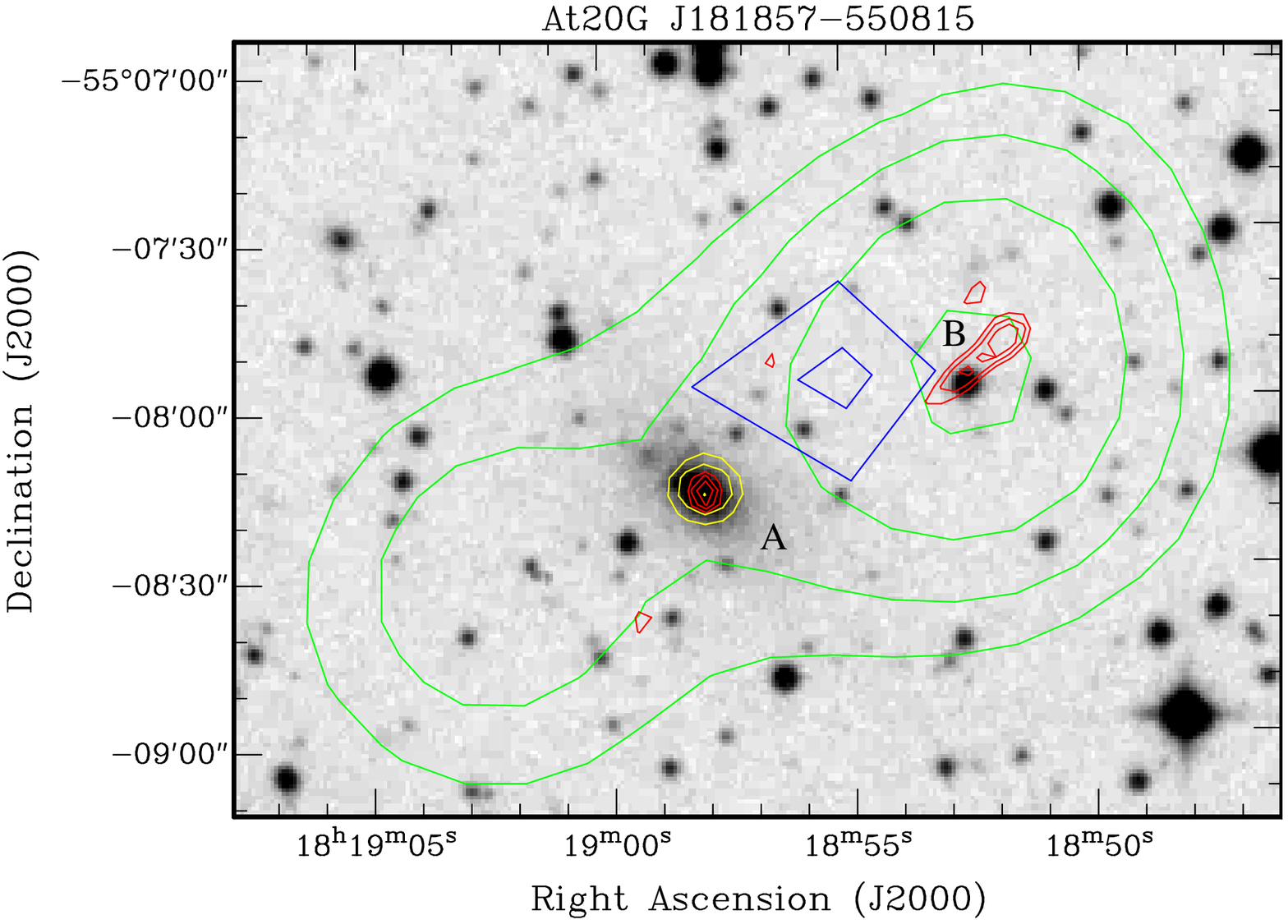, bb=0 65 540 450, width=0.8\linewidth,clip=}}
 \caption[Spectrum and radio optical overlay for AT20G J181857-550815]{AT20G J181857-550815. {\bf(a)} 6dFGS spectrum showing Na D, and the G-band in absorption and a strong H-K break. The redshift is 0.073. {\bf(b)} SuperCOSMOS blue image overlaid with PMN 4.85\,GHz in blue at 201 and 206\,mJy/beam, AT20G 4.8\,GHz in red at 25, 30, and 35\,mJy/beam, AT20G 20\,GHz in yellow at 16, 32, and 64\,mJy/beam, and SUMSS 843\,MHz in green at 40, 80 160, and 320\,mJy/beam. The radio and optical identifications are at the core component marked A. The PMN source is offset by 24 arcsec and is at the center of the blue contours. The extended feature seen in the red contours at position B is the faint jet. The SUMSS contours are consistent with component A being the host galaxy, and component B being the jet.}
 \label{fig:J181857-550815_pmn}
\end{figure}

This source is identified as a B$_J$ = 15.9\,mag galaxy within the SuperCOSMOS database.  The optical colours are B-R$_C$ = 0.95\,mag and R$_C$-I$_C$= 0.81\,mag which is consistent with a low redshift E or S0 galaxy, with excess blue emission due to AGN or star forming activity. The 6dFGS spectrum of this galaxy is shown in Figure \ref{fig:J181857-550815_pmn}(a), and shows stellar absorption lines typical of early-type galaxies but no obvious optical emission lines. The redshift is 0.073.

This source was observed in 2008 with e-VLBI resolution at 4.8\,GHz \citep{hancock_e-vlbi_2009}. This source is likely a restarted radio galaxy with an inverted core. Figure \ref{fig:J181857-550815_pmn} shows SUMSS, AT20G, and PMN radio contours on a SuperCOSMOS blue image. The 4.8\,GHz image from the AT20G shows significant emission that is consistent with a hot spot within the brighter of the SUMSS radio lobes (marked B in Figure \ref{fig:J181857-550815_pmn}). The PMN flux is higher than that measured by the AT20G, \citet{hancock_e-vlbi_2009}, or this paper, which is likely due to contributions from both the core and lobe. At a redshift of 0.073 the linear scale of the 4.8\,GHz jet is 1.3 kpc/arcsec. The separation between the core and jet is then 79\,kpc. The projected linear size of the SUMSS lobes is $\sim$110\,kpc.  The lack of 20\,GHz emission from the lobe suggests that they may no longer be being powered and could be a remnant of previous activity.

The core shows no significant variability at 4.8\,GHz, and is 14\% variable at 20\,GHz over one year. The radio spectrum of this source indicates that the spectral turnover is $\sim95$\,GHz or higher which is consistent with a young radio source which has recently become active. This source shows many of the characteristics of a genuine GPS source, despite the fact that the spectral turnover is yet to be observed.

%%%%%%%%%%%%%%%%%%%%%%%%%%%%%%%%%%%%%%%%%%%%%%%%%%

\subsection{AT20G J203540-694407 - a $z=0.875$ QSO}
\label{sec:J203540-694407}
\begin{figure}
 \centering
 \subfigure[]{\epsfig{file=C1392/AT20G2035-6944a.ps, bb=80 30 580 720, height=0.95\linewidth,angle=-90, clip=}}
 \subfigure[]{\epsfig{file=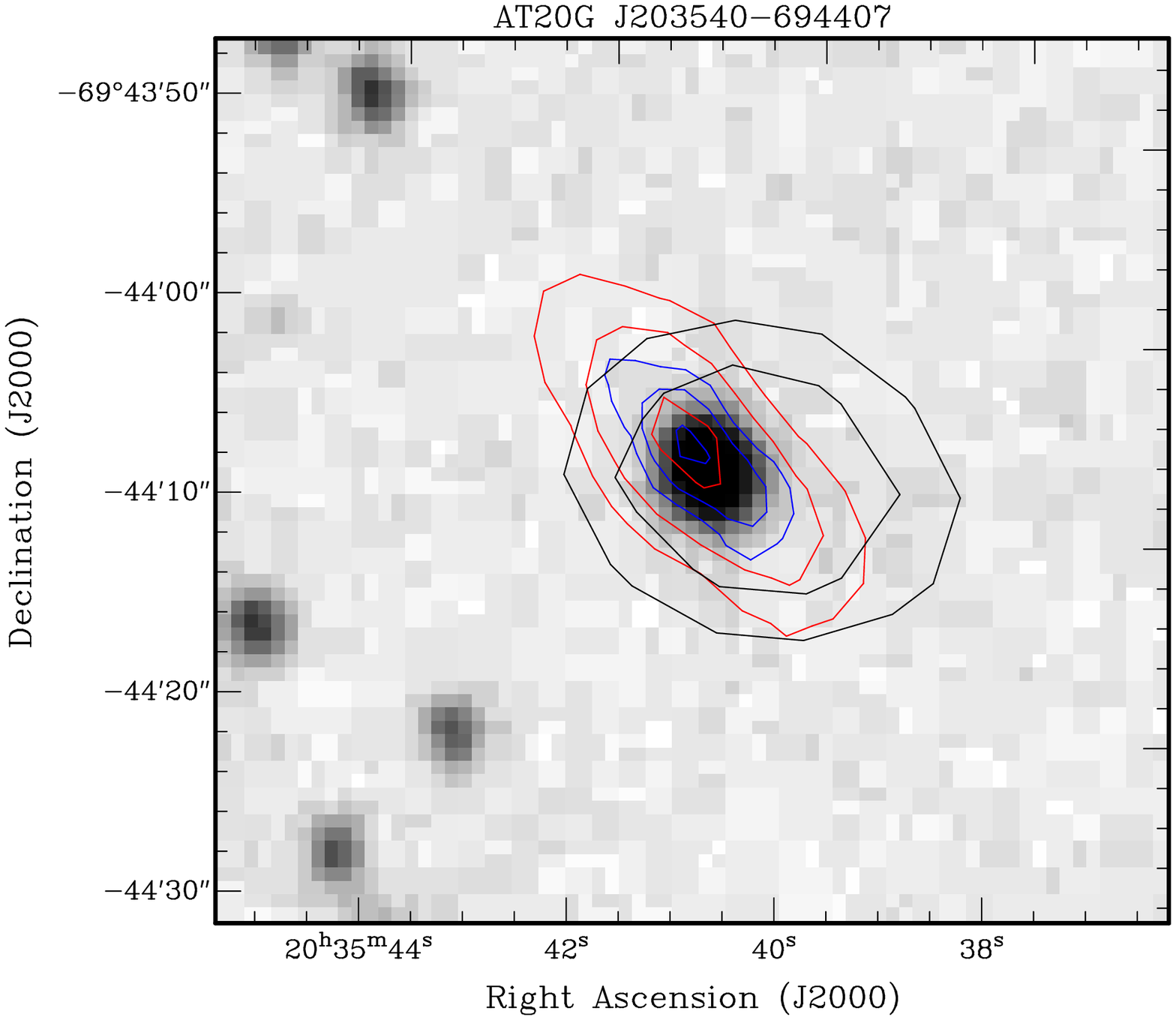, width=0.8\linewidth}}
 \caption[Spectrum and radio optical overlay for AT20G J203540-694407]{{\bf (a)} SSO 2.3m spectrum of J203540-694407. Flux units are arbitrary. MgII is seen in emission. The redshift is 0.875. The spectrum rises towards the blue which indicates that we are seeing a quasar. The wide absorption feature at 7630\AA\,is atmospheric absorption from oxygen. {\bf (b)} SuperCOSMOS blue image of AT20G J203540-694407 overlaid with AT20G radio contours at 4.8\,GHz (red at 40, 80, and 160\,mJy/beam), 8.6\,GHz (blue at 50, 100, and 200\,mJy/beam), and 20\,GHz (black at 35, 70, and 140\,mJy/beam). The elongation in the 4.8 and 8.6\,GHz contours is due to an elliptical beam shape.}
 \label{fig:J203540-694407_overlay}
 \label{fig:specJ203540-694407}
\end{figure}

This source is identified as being star like within the SuperCOSMOS database and has B$_J$ = 17.2 mag. The optical counterpart is 2.9\,arcsec distant from the 20\,GHz radio contours, however the association is accepted based on the 4.8 and 8.6\,GHz positions being within 2.5\,arcsec as depicted in Figure \ref{fig:J203540-694407_overlay}(b). The optical colours are B-R$_C$ = 0.95\,mag and R$_C$-I$_C$= 0.31\,mag, which is much bluer than expected for a normal E or S0 galaxy at the measured redshift of 0.875. An optical spectrum of this source is shown in Figure \ref{fig:specJ203540-694407}(a). The spectrum rises towards the blue and exhibits strong MgII emission, both of which are typical of quasars.

This source is in the PMN catalog with a 4.85\,GHz flux of $99\pm 8$\,mJy, giving a variability index of 30\% over 15 years, when compared to the AT20G flux. \citet{healey_crates:all-sky_2007} observed this source in 2005 with a 8.6\,GHz flux of $189.5$\,mJy 14 arcsec away from the AT20G position . The 8.6\,GHz data has no quoted uncertainty, but assuming an uncertainty as large as 10\%, the variability is 8\% over 2 years. The large amount of variability at and above the spectral peak (5\,GHz) is an indication that this is not a genuine GPS source. \citet{healey_cgrabs:all-sky_2008} identify this source as having associated gamma-ray emission but with a Figure of merit of 0.071 suggesting that the association is uncertain. This source is a quasar and not a genuine GPS galaxy.

%%%%%%%%%%%%%%%%%%%%%%%%%%%%%%%%%%%%%%%%%%%%%%%%

\subsection{AT20G J203958-720655 - a $z=1.0$ QSO}
\begin{figure}
 \centering
 \epsfig{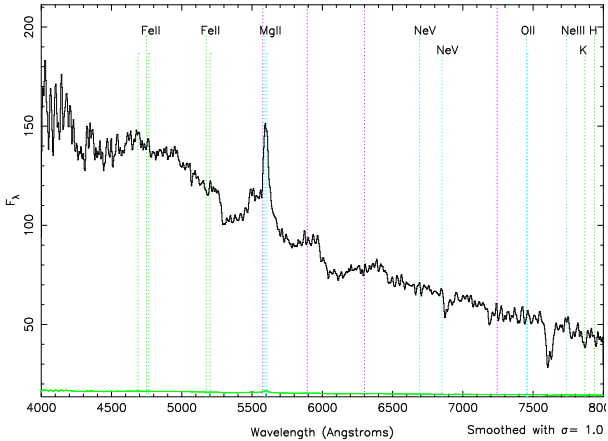}
 \caption[Spectrum of AT20G J203958-720655]{SSO 2.3m spectrum of J203958-720655. Flux units are arbitrary. MgII is seen in emission and FeII, H, and K are seen in absorption. The redshift is z=1.00. The spectrum rises towards the blue which indicates that we are seeing a quasar. The wide absorption feature at 7630\AA  is atmospheric absorption from oxygen.}
 \label{fig:specJ203958-720655}
\end{figure}

This source is identified as being star like within the SuperCOSMOS database and has B$_J$ = 17.1\,mag. The optical colours are B-R$_C$ = 0.55\,mag and R$_C$-I$_C$= 0.41\,mag which is much bluer than even the low redshift normal E or S0 galaxies. The optical spectrum is shown in Figure \ref{fig:specJ203958-720655} where a rising blue spectrum and strong MgII emission lines are indicative of a QSO at a redshift of 1.0.

This source is in the PMN catalog with a 4.85\,GHz flux of $122\pm 9$\,mJy. This source is unresolved in all AT20G radio images. This source shows significant variability at both 4.8\,GHz (20\% over 15 years) and 20\,GHz (9.7\% over 2 years). This source is most likely a quasar and not related to the genuine GPS sources.

%%%%%%%%%%%%%%%%%%%%%%%%%%%%%%%%%%%%%%%%%%%%%%%%%%%%

\subsection{AT20G J205503-635207 - a $z\sim 0.3$ GPS galaxy}
This source is identified as being star like within the SuperCOSMOS database and has B$_J$ = 20.2\,mag. The optical colours are B-R$_C$ = 1.25\,mag and R$_C$-I$_C$= 0.81\,mag, which is close to that expected for a low redshift, normal E or S0 galaxy. The estimated redshift is $\sim0.3$, which suggests an excess of blue emission.

This source is not in any of the on-line databases accessible via NED or VizieR. The 20\,GHz variability is less than 6\% over a year and is unresolved at all frequencies within the AT20G. There is not enough data to rule out this source as a GPS candidate.
%%%%%%%%%%%%%%%%%%%%%%%%%%%%%%%%%%%%%%%%%%%%%%%%%%%%%%%

\subsection{AT20G J212222-560014 - a $z=0.0518$ GPS galaxy in a cluster}
\begin{figure}
 \centering
 \subfigure[]{\epsfig{file=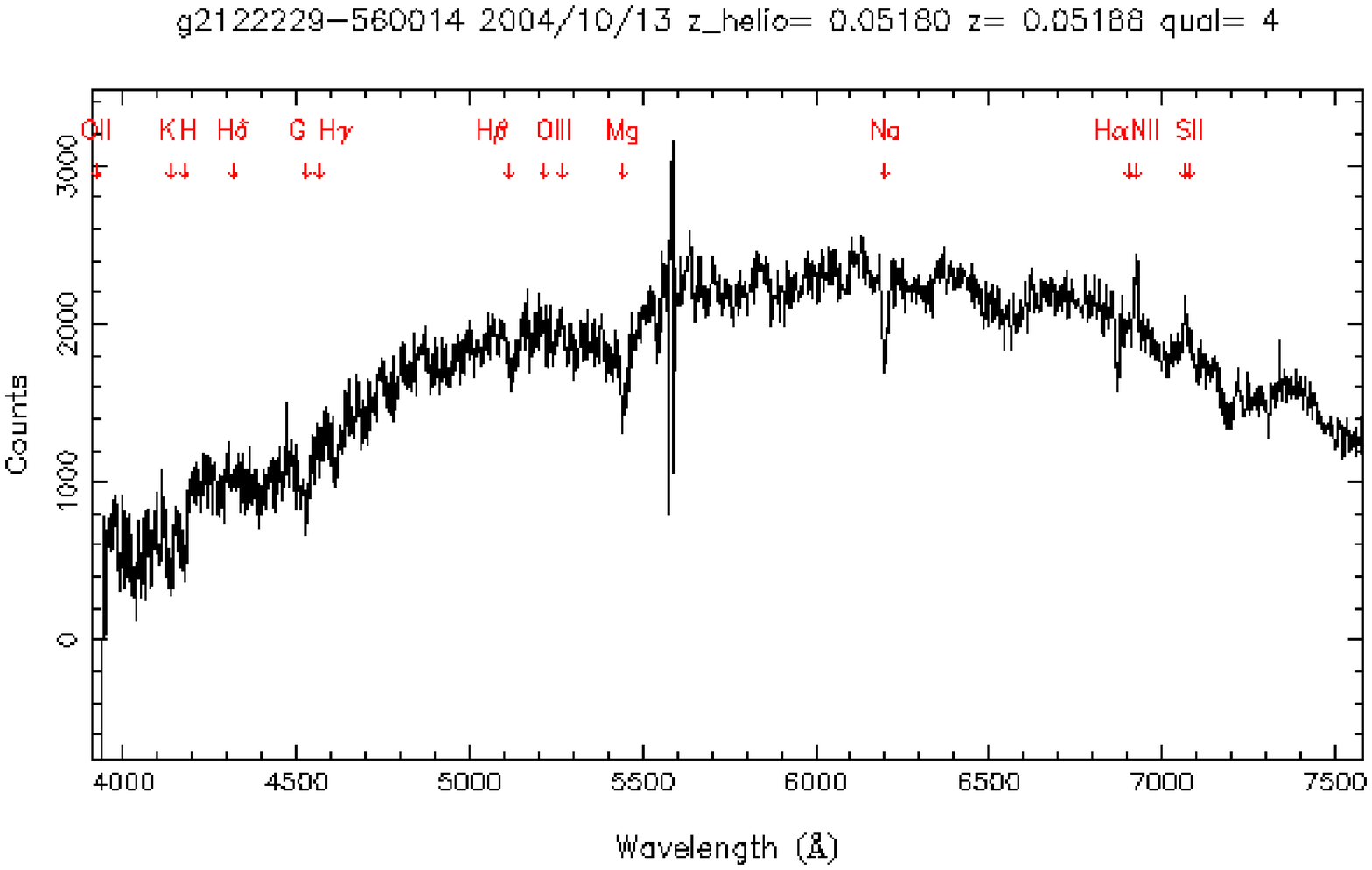, bb= -10 205 585 545, width=0.95\linewidth, clip=}}
 \subfigure[]{\epsfig{file=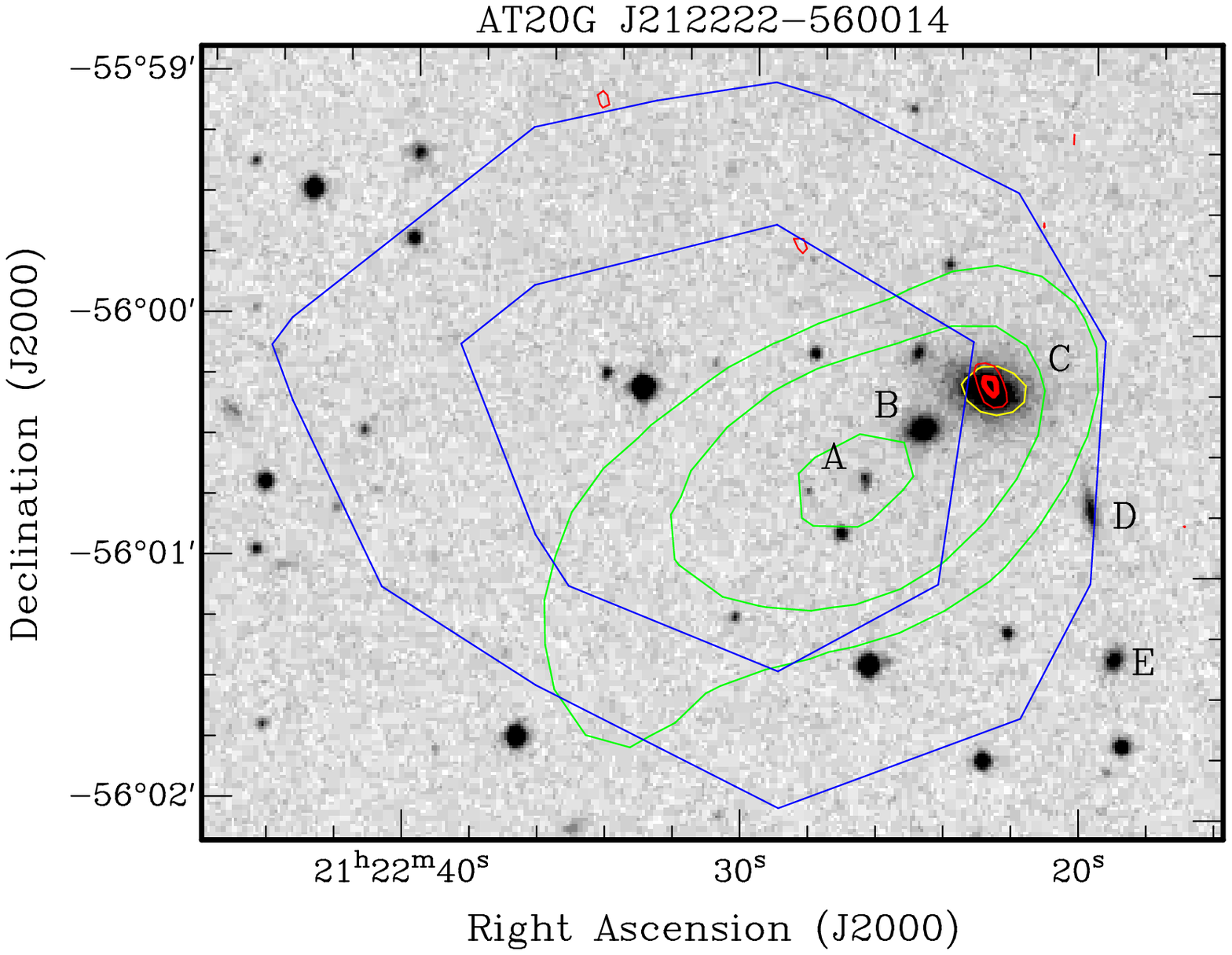, bb=55 220 540 600, width=0.85\linewidth,clip=}}
 \caption[Spectrum and radio optical overlay of AT20G J212222-560014]{AT20G J212222-560014. {\bf(a)} 6dF spectrum showing emission in H$\alpha$ and SII, and absorption in K, H, Na, and Mg. The redshift is 0.0518.  {\bf(b)} SuperCOSMOS blue image overlaid with radio contours. Contours are: Blue - PMN 4.8\,GHz at 50 and 60\,mJy/beam, Red - AT20G 4.8\,GHz at 15 and 25\,mJy/beam, Green - SUMSS 843\,MHz at 10, 20, and 40\,mJy/beam, and Yellow - AT20G 20\,GHz at 20\,mJy/beam. Identified objects are: A - APMUKS(BJ) B211848.98-561327.1, B - 2MASX J21222482-5600243, C - 2MASX J21222292-5600143 (also AT20G ID), D - APMUKS(BJ) B211842.26-561333.0, E - APMUKS(BJ) B211841.42-561410.4. }
 \label{fig:J212222-560014_pmn}
\end{figure}

This source is identified as a B$_J$ = 16.1\,mag galaxy within the SuperCOSMOS database. The optical colours are B-R$_C$ = 1.15\,mag and R$_C$-I$_C$= 0.71\,mag, which is only slightly more blue than a low redshift normal E or S0 galaxy. The 6dF spectrum depicted in Figure \ref{fig:J212222-560014_pmn}(a) shows a redshift of 0.0518, making this the nearest object within this sample. This galaxy is part of a galaxy cluster of at least 5 members. The optical ID is labeled as `C' in Figure \ref{fig:J212222-560014_pmn}(b).

The radio spectrum of this source remains inverted at 95\,GHz, although the upper limit on the 95\,GHz flux suggests that it may be beginning to turn over.

This source shows no significant variability at 20\,GHz.

The $73\pm 9$\,mJy source PMN J2122-5600 is located 1.17 arcmin to the east of J212222-560014 and is considered by \citet{healey_crates:all-sky_2007} to be associated. This association then includes J212222-560014 within the CRATES catalog of flat spectrum radio sources. 

Figure \ref{fig:J212222-560014_pmn}(b) shows the SuperCOSMOS blue survey image overlaid with AT20G 4.8 and 20\,GHz, PMN 4.85\,GHz, and SUMSS 843\,MHz contours. The offset between the AT20G and PMN contours is large but still within the astrometric uncertainties of the PMN survey. The AT20G flux is clearly coming from the optical galaxy. The PMN flux does not have a clear origin but the presence of the extended SUMSS emission and some low level extended flux in the AT20G image suggests that there is some diffuse low-frequency emission present. The fact that there is more PMN flux than AT20G flux ($73\pm 9$ vs $28\pm 2$\,mJy at 4.8\,GHz) is consistent with the idea that there is diffuse emission being generated by the galaxy cluster as well as more compact emission from the galaxy itself. It is also possible that this is a head-tail source within a cluster. There is no reason to exclude this source from being a genuine GPS galaxy.

%%%%%%%%%%%%%%%%%%%%%%%%%%%%%%%%%%%%%%%%%%%%%%%%

\subsection{AT20G J212402-602808 - a possible QSO}
This source is identified as being star like within the SuperCOSMOS database and has B$_J$ = 17.3\,mag. The optical colours are B-R$_C$ = 0.45\,mag and R$_C$-I$_C$= 0.51\,mag, and the estimated redshift is $\sim 1$, which indicates that there is excess blue emission due to an AGN or star formation. This source is visible at UV wavelengths and is considered by \citet{ageros_ultraviolet_2005} to be a candidate QSO.

The observation of the 4.8 and 8.6\,GHz fluxes were carried out almost a year after the 20\,GHz flux observations, which makes them almost simultaneous with the 2007 observations of this paper. A 20\,GHz variability of 20\% over a year is almost double the amount expected from a genuine GPS source. 

Given the spectral variability, and the UV detection it is likely that this source is a quasar and not a genuine GPS galaxy.
%%%%%%%%%%%%%%%%%%%%%%%%%%%%%%%%%%%%%%%%%%%%%%%%%%%%%%%%%%

\subsection{AT20G J213622-633551 - a possible QSO}
This source is identified as being star like within the SuperCOSMOS database and has B$_J$ = 19.8\,mag. This source was not detected in I$_C$ so no determination of optical colours can be made.

\begin{figure}
 \centering
 \epsfig{file=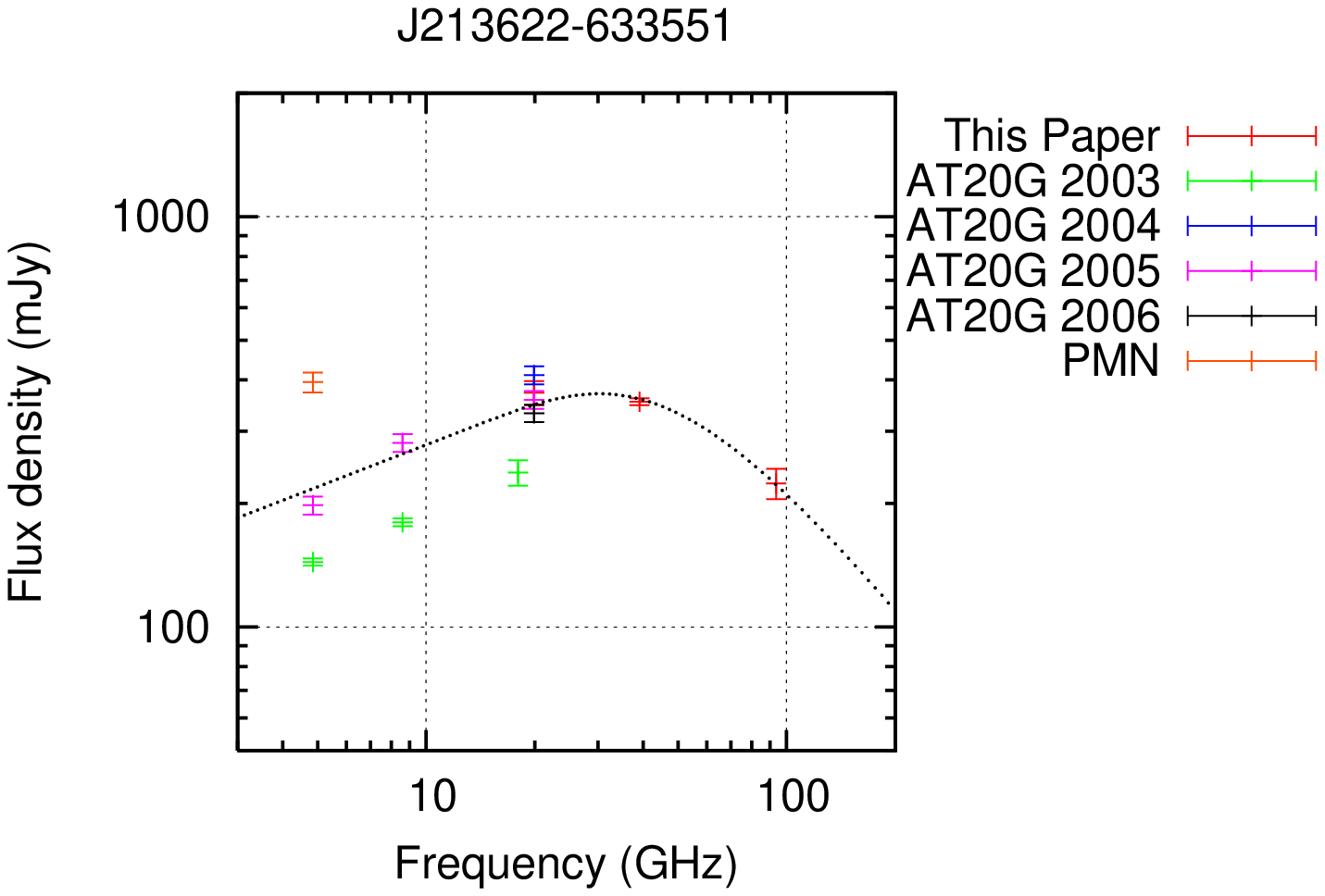, bb= 275 55 555 470, height=0.95\linewidth, angle=-90, clip=}
 \caption{Radio spectrum of J213622-633551 with data from the multiple AT20G observations and PMN catalog included. Despite the 17\% variability seen at 20\,GHz, the shape of the spectrum has not changed between 2003 and 2005. }
 \label{fig:J213622-633551_spec}
\end{figure}

This object was observed multiple times at multiple frequencies over the course of the AT20G pilot and survey. A summary of the observations is given in table \ref{tab:213622-633551}. Figure \ref{fig:J213622-633551_spec} plots each of the observations. The 2003 and 2005 observations (green and pink in the Figure) show some variability in total flux but not in spectral shape. Using the 20\,GHz data from 2004-2007 as well as the 18\,GHz flux from 2003 corrected with a spectral index of $\alpha =+0.38$ gives a revised variability index of 17\% over four years. The 4.8\,GHz variability is 20\% over 15 years. The amount of variability is not unheard of in genuine GPS samples, however this object is likely a quasar and not a genuine GPS galaxy.

\begin{table}
\centering
\caption{Data summary for J213622-633551}
\begin{tabular}{cccc}
\hline
\hline
Data Source & $\nu$ & Flux Density & Obs \\
            & GHz   & mJy          & Year\\ 
\hline
PMN                            & 4.85  & $395\pm 22$  & 1990\\
\citet{sadler_properties_2006} & 4.8   & $144\pm 3$   & 2003.92\\
AT20G                          & 4.8   & $198\pm 10$  & 2005.83\\
\hline
\citet{sadler_properties_2006} & 8.6   & $180\pm 4$   & 2003.92\\
AT20G                          & 8.6   & $281\pm 14$  & 2005.83\\
\hline
\citet{sadler_properties_2006} & 18    & $238\pm 17$  & 2003.83\\
\hline
AT20G                          & 20    & $411\pm 21$  & 2004.75\\
AT20G                          & 20    & $358\pm 18$  & 2005.75\\
AT20G                          & 20    & $332\pm 16$  & 2006.75\\
This paper                     & 20    & $372\pm 25$  & 2007.83\\
\hline
This paper                     & 40    & $354\pm  7$  & 2007.83\\
\hline
This paper                     & 95    & $224\pm 19$  & 2007.83\\
\hline
\end{tabular}
\label{tab:213622-633551}
\end{table}

%%%%%%%%%%%%%%%%%%%%%%%%%%%%%%%%%%%%%%%%%%

\subsection{AT20G J214447-694654 - a $z\sim 0.5$ GPS galaxy}
This source is identified as being star like within the SuperCOSMOS database and has B$_J$ = 21.4\,mag. This source was not detected in I$_C$ so no determination of optical colours can be made. The redshift for this source is estimated as $\sim 0.5$. This source does not appear within in NED or VizieR in any catalog aside from SUMSS. There is a PMN source $\sim 20$\,arcsec from this source with a flux of $30\pm 7$\,mJy but there is more than one possible optical ID for the PMN source so the association is uncertain. The lack of an optical counterpart suggests that this source is a faint or distant galaxy, however a radio spectral peak of 82\,GHz would put a young radio source at low redshift. There is nothing to suggest that this source is not a genuine GPS source.
%%%%%%%%%%%%%%%%%%%%%%%%%%%%%%%%%%%%%%%%%%5

\subsection{AT20G J230737-354828 - a variable radio galaxy or candidate QSO}
This source is identified as being a B$_J$ = 20.9\,mag galaxy within the SuperCOSMOS database. The optical colours are B-R$_C$ = 0.95\,mag and R$_C$-I$_C$= -0.09\,mag, which are possibly affected by variability (see below). The redshift of this source is estimated to be $\sim 1$, indicating that there is a large excess of blue emission. No optical spectrum is available for this source.

\citet{flesch_all-sky_2004} identify this source as having associated X-ray emission and as likely being a quasar. 

\begin{figure}
 \centering
 \epsfig{file=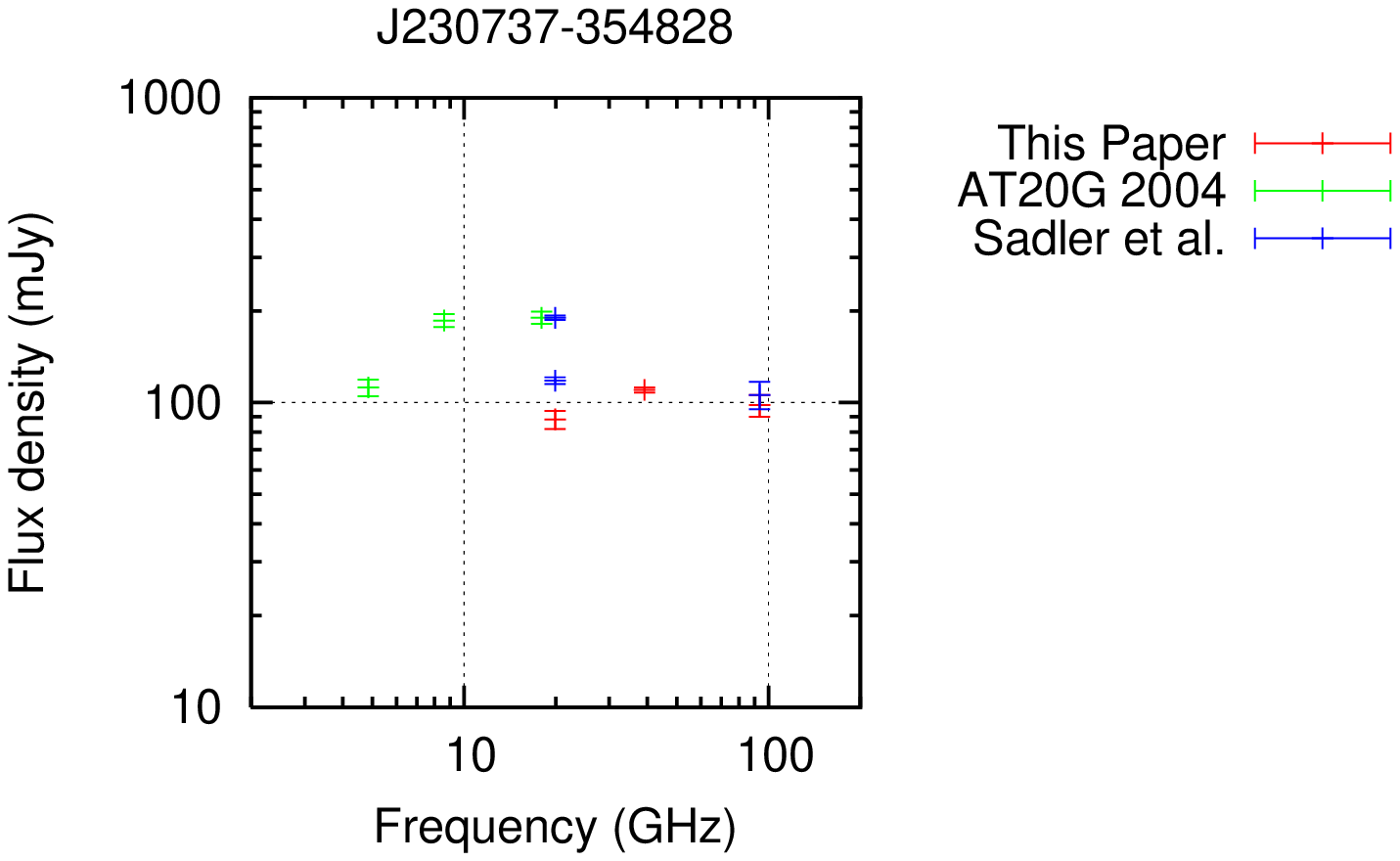, bb=300 55 555 470, height=0.95\linewidth, angle=-90,clip=}
 \caption{Radio spectrum of J230737-354828 incorporating the data from the AT20G, \citet{sadler_extragalactic_2008}, and this paper.}
 \label{fig:J230737-354828_spec}
\end{figure}

Table \ref{tab:230737-354828} shows the observations that have been made over the course of the AT20G survey and related projects. Figure \ref{fig:J230737-354828_spec} shows the radio spectrum with all the listed data included. The 36\% variability at 20\,GHz measured in this paper is supported by the 2006 observations of \citet{sadler_extragalactic_2008} which were on consecutive days. The observations of the AT20G and of this paper both show convex, GPS like spectra, but with peaks at different frequencies and with different flux densities. If this amount of variability extends to the optical then the optical colours could be contaminated as the observation of the B-R-I magnitudes were separated by as much as 20 years. The large amount of variability in both total flux and spectral shape make it likely that this source is a quasar and not a genuine GPS galaxy.

\begin{table}
\centering
\caption{Data summary for J230737-354828}
\begin{tabular}{cccc}
\hline
\hline
Data Source & $\nu$ & Flux Density & Obs \\
            & GHz   & mJy          & Year\\ 
\hline
AT20G                             & 4.8 & $112\pm 7$ & 2004.83\\
\hline
AT20G                             & 8.6 & $186\pm 9$ & 2004.83\\
\hline
AT20G                             & 18  & $190\pm 9$ & 2004.75\\
\hline
\citet{sadler_extragalactic_2008} & 20  & $118\pm 3$ & 2006.79\\
\citet{sadler_extragalactic_2008} & 20  & $190\pm 3$ & 2006.79\\
This paper                        & 20  & $88 \pm 6$ & 2007.83\\
\hline
This paper                        & 40  & $110\pm 2$ & 2007.83\\
\hline
\citet{sadler_extragalactic_2008} & 95  & $106\pm 11$& 2006.79\\
This paper                        & 95  & $98\pm 8$  & 2007.83\\
\hline
\end{tabular}
\label{tab:230737-354828}
\end{table}

%%%%%%%%%%%%%%%%%%%%%%%%%%%%%%%%%%%%%%%%%%%%%%%%%%%%%%%%%

\subsection{AT20G J233159-381147 - a $z=1.2$ QSO}
This source is a well known quasar identified as early as 1971 \citep{shimmins_identification_1971} and is identified as such in the SuperCOSMOS database. The source has B$_J$ = 17.3\,mag. The optical colours are B-R$_C$ = 0.55\,mag and R$_C$-I$_C$= 0.51\,mag, and the measured redshift is 1.20.

\citet{wright_variability_1984} use a variability measure similar to that of this paper to calculate a variability of 1\% for this source at 2.7\,GHz and 19\% variability at 4.8\,GHz over the period 1970-1979. More recent measurements by the PMN, and AT20G show a 5\,GHz variability index of \<6\% over 15 years and the 20\,GHz observations of the AT20G and this paper puts the variability at \<6\% over a period of 3 years. It is likely that this quasar has long periods of quiescence with short periods of flare activity.

This quasar has been detected in polarized flux both at 8.6\,GHz \citep[9.7\%][]{simard-normandie_linear_1981} and in optical V \citep[0.44\%][]{lamy_optical_2000}. \citet{wolter_x-ray_2001} detect a 0.3-3.5 keV X-ray flux of $1.5\times 10^{-13}\mathrm{erg cm}^{-2}\mathrm{s}^{-1}$. VLBI observations of this source at 2.3 and 8.4\,GHz were used to define the International Celestial Reference Frame \citep{johnston_radio_1995,ma_international_1998}.
%%%%%%%%%%%%%%%%%%%%%%%%%%%%%%%%%%%%%%%%%%%%%%%%%%%%%%

\subsection{AT20G J234743-494627 - a $z=0.643$ QSO}
This source is identified as being star like in the SuperCOSMOS database and has B$_J$ = 19.8\,mag. The optical colours are B-R$_C$ = 1.65\,mag and R$_C$-I$_C$= -0.81\,mag. \citet{healey_cgrabs:all-sky_2008} identify this source as a flat spectrum radio quasar at a redshift of 0.643.

\citet{healey_cgrabs:all-sky_2008} measure a 8.6\,GHz flux of $344$\,mJy. It is within the PMN catalog with a 4.85\,GHz flux of $367\pm 20$\,mJy. The variability at 4.85 and 8.6\,GHz is 26\% and 13\% respectively (assuming 10\% uncertainty at 8.6\,GHz), and 16\% at 20\,GHz. This level of variability at three frequencies is enough to bring the shape of the spectrum from being peaked to being flat. The identification of this source as FSRQ is not altered by the addition of new observations.

\bibliographystyle{mn2e}
\bibliography{high_frequency}
\label{lastpage}
\end{document}